\documentclass{article}

\usepackage[a4paper,top=3cm,bottom=3cm,left=3cm,right=3cm]{geometry}
\usepackage{authblk}
\usepackage{amsthm,amsmath,amsfonts,amssymb}
\usepackage[authoryear]{natbib}
\usepackage[colorlinks,allcolors=blue]{hyperref}
\usepackage{graphicx}
\usepackage[ruled,vlined]{algorithm2e}
\usepackage{bbm}
\usepackage{thmtools}
\usepackage[nameinlink]{cleveref}
\usepackage{subfigure}
\usepackage[normalem]{ulem}
\usepackage{xparse}
\usepackage{xcolor}
\usepackage{tabularx,booktabs}
\usepackage[section]{placeins}
\usepackage{multirow}

\theoremstyle{plain}

\newtheorem{theorem}{Theorem}[section]
\newtheorem{proposition}[theorem]{Proposition}

\theoremstyle{remark}

\newcommand{\iid}{\stackrel{\mbox{\scriptsize iid}}{\sim}}
\newcommand{\ind}{\stackrel{\mbox{\scriptsize ind}}{\sim}}
\newcommand{\virgolette}[1]{``#1''}
\newcommand{\e}{\operatorname{e}}
\newcommand{\Ghatb}{\hat{G}^{(b)}}
\newcommand{\Ghatn}{\hat{G}^{(n)}}
\newcommand{\BEhat}{\widehat{BE}}
\newcommand{\NEhat}{\widehat{NE}}
\newcommand{\Lone}{L_1}

\renewcommand{\url}[1]{\href{#1}{\small\texttt{#1}}}
\let\tilde\widetilde
\let\bm\boldsymbol

\DeclareMathOperator*{\invGamma}{InvGamma}
\DeclareMathOperator*{\alr}{alr}
\DeclareMathOperator*{\invalr}{alr^{-1}}
\DeclareMathOperator*{\argmax}{argmax}
\DeclareMathOperator*{\Var}{Var}
\DeclareMathOperator*{\E}{\mathbb{E}}
\DeclareMathOperator*{\Eadj}{E^{\,\text{adj}}}
\DeclareMathOperator{\dd}{\text{d}\!}
\NewDocumentCommand\w{e{^_}}{%
    \IfNoValueTF{#1}{%
        \IfNoValueTF{#2}{\bm{w}}{\bm{w}_{(#2)}}
    }{%
        \IfNoValueTF{#2}{\bm{w}^{(#1)}}{\bm{^{(#1)}_{(#2)}}}
    }
}
\NewDocumentCommand\wtilde{e{^_}}{%
    \IfNoValueTF{#1}{%
        \IfNoValueTF{#2}{\tilde{\bm{w}}}{\tilde{\bm{w}}_{(#2)}}
    }{%
        \IfNoValueTF{#2}{\tilde{\bm{w}}^{(#1)}}{\tilde{\bm{w}}^{(#1)}_{(#2)}}
    }
}
\makeatletter
    \RenewDocumentCommand\eqref{sm}{%
      \IfBooleanTF#1%
      {\textup{\tagform@{\ref*{#2}}}}%
      {\textup{\tagform@{\ref{#2}}}}%
    }
    
\makeatother

\title{\textbf{Bayesian nonparametric boundary detection\\for multiple areal data}}
\author[1]{Matteo Gianella}
\author[2]{Mario Beraha}
\author[1]{Alessandra Guglielmi}
\affil[1]{\small\textit{Department of Mathematics, Politecnico di Milano -- Milan (ITALY)}}
\affil[2]{\small\textit{Department of Economics, Management and Statistics, University of Milano-Bicocca -- Milan (ITALY)}}
\date{15 May 2026}

\begin{document}

\maketitle

\begin{abstract}
    We consider the problem of boundary detection for areal data, focusing on situations where for each areal unit multiple observations are available. We propose a Bayesian nonparametric mixture model for the area-specific population densities, with spatially dependent weights and a random number of components. Contrary to previously proposed methods for boundary detection, which consider one observation per areal unit, ours does not require external information such as area-specific covariates or dissimilarity metrics. Instead, by exploiting information from multiple samples per area, it is able to identify boundaries between areas that exhibit different densities. Crucially, the number of mixture components needs to be learned from data to obtain meaningful boundary detection, due to the non-identifiability of overfitted mixtures. Therefore, we assume it random by placing a prior on it. The motivating application is the analysis of economic inequality in the greater Los Angeles
    region, which typically yields social inequality and unrest.
    Efficient posterior computation is facilitated by a transdimensional Markov Chain Monte Carlo sampler which exploits the recently introduced \emph{optimal auxiliary priors} to improve the mixing. The methodology is validated via extensive simulations and applied to the income data in the greater Los Angeles region. We identify several boundaries in the income distributions, which can be explained \textit{ex-post} in terms of the percentage of the population without health insurance, though not in terms of the total number of crimes, showing the usefulness of such an analysis to policymakers.
\end{abstract}

\section{Introduction}\label{sec:introduction}
In this paper, we consider the problem of \textit{boundary detection} for areal data, focusing on situations in which multiple observations are available for each areal unit.
We propose a Bayesian nonparametric mixture model for area-specific population densities, with spatially dependent weights and a random number of components, capable of identifying boundaries between areas that exhibit different densities.
Boundary analysis methods are routinely used to identify borders (or zones) that distinguish different spatial regions. The typical main assumption underneath such methods is a strong geographical correlation or dependence in the variable of interest across nearby areas. If, for some neighbouring areas, such a correlation is not observed in the data, this is used as evidence of the presence of a \textit{boundary}. This procedure has been proposed, for instance, to highlight different patterns in disease mapping \citep[see, e.g.,][]{lee.mitchell2012,li2015bayesian,gao2022bayesian,aiello2023detecting} or in environmental applications \citep{qu2021boundary}.
 We formulate boundary detection within the Bayesian structural learning framework, where the adjacency graph that encodes spatial structure is random and inferred from the data. 

\subsection{Personal income data in the greater Los Angeles region}\label{subsec:introapplication} 
Our study is motivated by the analysis of economic inequality in the greater Los Angeles region, including Los Angeles (LA), Ventura, and Orange counties. Working on data from the Public Use Microdata Sample (PUMS) of the American Community Survey (ACS) \citep[see][]{uscb.aboutacs}, we aim to identify boundaries that delineate areas with significant differences in income distributions. Specifically, we consider the personal income of survey respondents as a proxy for their economic status. The ACS data, collected annually, encompass information on incomes, jobs, and education. However, a limited number of geographic summaries are available in the PUMS dataset for confidentiality reasons. In particular, the finest unit of geography contained there is the so-called Public Use Microdata Area (PUMA), which are spatial non-overlapping areas that partition each state into contiguous geographic units containing roughly 100,000 people. PUMAs, having been created for demographic reporting, do not represent a useful partition for urban planners and policymakers, as they may not adequately represent the true spatial heterogeneity within larger metropolitan regions. This limitation is especially pertinent for densely populated areas, where multiple small PUMAs may coexist, and for less urbanised regions, where a single PUMA might cover a vast geographical area. In contrast, policymakers would need to detect larger zones where the social indicators are homogeneous while identifying possible \virgolette{danger areas} as the borders between two areas with significantly different economic and social indicators.

\begin{figure}[t]
    \centering
    \subfigure{\includegraphics[width=0.48\textwidth]{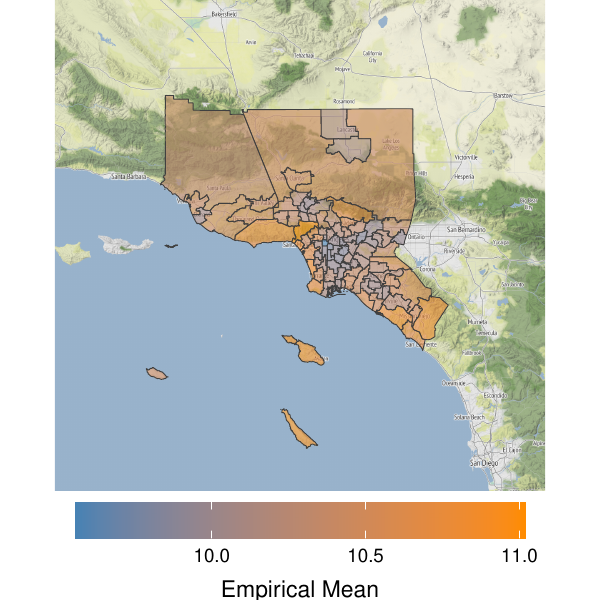}}%
    \hfill
    \subfigure{\includegraphics[width=0.48\textwidth]{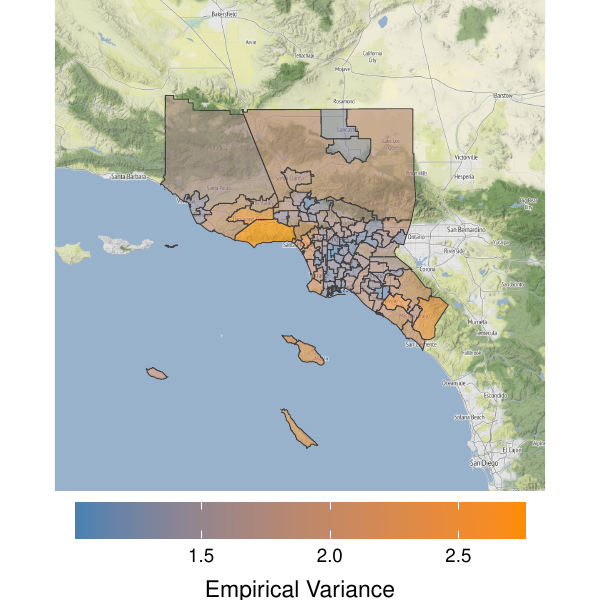}}%
    \caption{California census income data in the log scale. Each area is coloured according to the empirical mean (left) and variance (right) of the log-income.}
    \label{fig:CaliCensusData_exploration1}
\end{figure}

\Cref{fig:CaliCensusData_exploration1} reports the sample means and variances of the log-income of survey respondents 
in each PUMA. These simple summary plots motivate modelling the spatial dependence, as geographically contiguous districts tend to exhibit similar distributions (e.g., see the northern part of the map, where the data show a smooth colour change across geographically contiguous PUMAs). High-income means appear to be located in coastal and specific suburban corridors. This figure justifies modelling spatial dependence between geographically contiguous areas. More generally, it is well-known that models incorporating spatially structured random effects that capture residual spatial autocorrelation allow information to be \virgolette{borrowed} across neighbouring regions, leading to more robust inference and realistic uncertainty quantification than models assuming conditional independence among areas; see, for instance, \cite{besag1991bayesian}. 
Nevertheless, in \Cref{fig:CaliCensusData_exploration1}, there are cases where two adjacent areas appear to have very different summary statistics. See, for instance, the sharp difference between the greater Los Angeles region in the south and the PUMAs in LA downtown.

Typically, for boundary detection problems,  only one observation is available for each areal unit, e.g., the average response per area, 
and, depending on the approach, boundaries may be informed either by pairwise dissimilarity metrics between neighbouring areas in localised CAR models \citep{lee2013carbayes}, or assessed post-hoc from posterior contrasts in a latent/residual spatial surface after optionally adjusting for area-level covariates in Bayesian wombling frameworks \citep{lu2005areal}. In our context, instead, we have multiple observations (i.e., personal incomes of survey respondents) for each area, and we address boundary detection by identifying boundaries that separate areal units with significantly different area-specific income densities from their neighbours.

\Cref{fig:CaliCensusData_eda2} (left panel) show the kernel density estimates of the individual personal income in each area in the greater LA region. In \Cref{subsec:data_exploration} we will show that there is substantial heterogeneity in these densities, reflected in shifts in location (mean/median income), differences in spread and modal structure, and variation in the lower-income tails. These features suggest that geographically contiguous areas may differ markedly in their socio-economic composition. The $\Lone$ distances between the kernel density estimates of datapoints for each pair of neighboring PUMAs,  although mean and variance heatmaps indicate spatial correlation in log-incomes, reveal adjacent areas with pronounced distributional discrepancies; see \Cref{fig:CaliCensusData_eda2} (right panel). This exploratory analysis supports the use of a boundary detection approach based on full density differences rather than simple difference-in-means tests, as it better captures the heterogeneous variability present in the data.

As a baseline, we explored simpler strategies that reduce each area to low-di\-men\-sio\-nal summaries. Specifically, we considered a parametric Bayesian MCAR model for boundary detection and the SKATER regionalisation algorithm \citep{assunccao2006efficient} applied to vectors of empirical quantiles. These approaches can perform well when relevant heterogeneity is captured by the chosen summaries, but they may miss distributional discontinuities when neighbouring areas have similar quantiles yet meaningfully different density shapes; in our data, this limitation is consequential (full details are deferred to the Supplementary Material).

A further practical complication is that we do not have harmonised area-level covariates at the PUMA resolution: PUMAs are designed for disclosure control and do not align cleanly with other administrative partitions for which socioeconomic indicators are routinely available. 
As a result, covariate-driven dissimilarity metrics are not a viable input for our analysis. This motivates a model-based approach that uses the within-area samples directly: we treat area-specific income densities as random objects and declare a \emph{boundary} when two geographically contiguous units exhibit negligible spatial dependence in their estimated densities, i.e., when the local borrowing of strength breaks down across their shared edge. Intuitively, the estimated boundary underlines where the income gap, as represented by a random density, is more marked. This information holds potential for policymakers aiming to devise interventions that address social and economic inequalities.

\subsection{Our contributions and outline of the article}\label{subsec:ourcontributions} 
We ground the boundary detection problem in a Bayesian structural learning framework \citep{lauritzen1996graphical}, assuming a prior distribution for the adjacency graph that represents the spatial structure. In each PUMA $i$, we assume that observations are independent and identically distributed from the density $f_i$, and we model the density $f_i$ of the (log) income via a finite mixture of Gaussian distributions with a random number of components, a well-established method for approximating any density \citep[][Section 2.3.3]{ghosal2017fundamentals}. To induce dependence across neighbouring areas, we build on the spatial mixture model introduced in \cite{beraha2021spatially} by assuming a logistic multivariate CAR prior for the weights of the mixture, and this prior incorporates the random adjacency graph. In particular, this allows us to address boundary detection by identifying boundaries for which the associated geographically contiguous areal densities are estimated as being very different without resorting to dissimilarity metrics or covariates, as previously mentioned.

A key difference is that, while \cite{beraha2021spatially} fixes the number of components $H$ to a \virgolette{large} value and proposes a sparse prior for the mixture weights, following the classical procedure to deal with overfitted mixtures, we treat $H$ as a random variable. This approach is necessary because the non-identifiability of overfitted mixtures \citep[see][]{rousseau.mengersen2011} severely limits the ability to identify boundaries, as explained in greater detail below. Treating $H$ as random introduces well-known difficulties in the algorithm for posterior inference.  We design a suitable transdimensional Markov chain Monte Carlo (MCMC) sampler akin to \cite{green1995reversible}. However, unlike the classical split-merge reversible jump MCMC of \cite{green1995reversible}, our algorithm is based on the \emph{optimal proposals} of \cite{norets2021optimal}, where new mixture components are selected from an auxiliary prior that depends on the data, making our approach potentially more efficient. Note that, despite the similarities in modelling the area-specific densities with \cite{beraha2021spatially}, the statistical problems addressed in these two papers are different.

In principle, distributional boundaries could also be explored using empirical distances between within-area samples, such as the $\Lone$ norm between estimated densities or the empirical Wasserstein distance. However, such approaches do not provide a principled mechanism for spatial borrowing of strength, and they do not directly distinguish sampling noise from genuine distributional discontinuities when sample sizes vary across areas. Our framework addresses these limitations by assuming a generative model for the data in each area and learning the dependence structure through the random adjacency graph. This yields a principled probabilistic mechanism for identifying boundaries, while simultaneously stabilising estimation of area-specific densities by sharing information across neighbouring units within homogeneous regions.

While our approach does not require covariates or dissimilarity metrics to detect boundaries, additional information can be included in the model in a straightforward way. For instance, it is possible to include area-level covariates and use them to model the mean spatial surface either with a linear or non-linear regression model, as well as including spatial dissimilarity metrics to inform the presence/absence of boundaries. We discuss two possible strategies in \Cref{sec:discussion}. Crucially, our method enables inference when area-level covariates are unavailable, either because of privacy concerns (common in the case of public health data), or because they are not harmonized with the spatial partition of interest. Examples of the latter case include our ACS data and educational data, where school districts are not consistent with counties or other legal/statistical areas \citep{geverdt2024edge_sdgrf}.

The rest of this article is organised as follows. \Cref{sec:model} introduces our area-depen\-dent mixture model for boundary detection through the random adiacency graph. \Cref{sec:algorithm} describes essential details of the MCMC algorithm. In \Cref{sec:simstudies_summary} we provide a summary and key findings of the simulation studies and stress tests we performed to show the abilities of our model and MCMC algorithm.
The application of our boundary detection model to the California census income dataset is discussed in \Cref{sec:real_case_scenario}. \Cref{sec:discussion} concludes the paper with a discussion and directions for future research.

The (online) Supplementary Material (SM) contains insights about the borrowing of strength induced by our logistic multivariate CAR prior in \Cref{sec:borrow}. Full details of the MCMC algorithm can be found in \Cref{sec:MCMC} of the SM. \Cref{sec:simstudies,sec:stress_tests} of the SM contain the simulation studies described in full details
and the stress tests implemented to assess the robustness of our model when boundary detection can be challenging, respectively.
In particular, the simulation study in \Cref{subsec:de_simstudy} of the SM focuses on joint spatial density estimation, the one in \Cref{subsec:sl_simstudy} of the SM on structural learning under misspecification and the one in \Cref{subsec:bd_simstudy} of the SM is an
extensive simulation study and sensitivity analysis focusing on boundary detection under missspecification. The same task is then performed in two more challenging settings: when the data show identical central tendencies and differs only in the tails (\Cref{sec:bd_simstudy_samemedian} of the SM) and when the number of observations per area is very large (\Cref{subsec:bd_simstudy-highdim} of the SM).
The SM also includes a comparison with alternative models or empirical techniques for boundary detection in \Cref{sec:comparisons}. However, these models and algorithms typically perform boundary detection only when there is a single response per area; for this reason, we applied them to a vector of empirical quantiles of all the data in each area. In particular, the competitor models are described in \Cref{subsec:competitor_models}, while comparison of posterior inference for the simulated data and the California census income dataset are reported in \Cref{subsec:comparison_synthetic} and \Cref{subsec:comparison_calicensusdata} of the SM.
In \Cref{sec:interpret_boundaries} in the SM, as a double-check, we investigate whether the estimated boundaries, at least for LA county, can be explained by available exogenous covariates. The SM also reports additional plots and tables on posterior inference for
the California census income dataset in its \Cref{sec:extra_plots_and_tables}.

The MCMC algorithm, coded in \texttt{C++}, is publicly available in an \texttt{R} package called \texttt{SPMIX} and available at the following link: \url{https://github.com/TeoGiane/SPMIX}. The code required to reproduce plots and tables of this paper and the SM is publicly available at the following link: \url{https://github.com/TeoGiane/SPMIX-applications}.

\section{A model for boundary detection}\label{sec:model}
Consider observations $\bm{y} = \left(\bm{y}_1,\dots,\bm{y}_I\right)$, where $\bm{y}_i = \left(y_{i,1},\dots,y_{i,N_i}\right)$ for $i = 1,2,\dots,I$ and $j = 1,2,\dots,N_i$, where $y_{i,j}$ is the observation of individual $j$ in area $i$. Each vector $\bm{y}_i$ is associated with an area $i$, and $N_i$ is the number of observations in such an area, which might differ through areas. In the application discussed in \Cref{sec:real_case_scenario}, we will use PUMAs as the areal unit. With notation $i \sim k$, we mean that areas $i$ and $k$ are geographically contiguous, meaning that they share at least a border. We represent the spatial dependence across areas through a random $I \times I$ binary matrix (or adjacency graph) $G$. This graph is defined as follows: $(a)$ $G_{i,i} = 0$  for every $i = 1, \dots, I$; $(b)$ $G_{i,k} = 0$ if the corresponding areas are not geographically contiguous (i.e., if $i \not\sim k$); $(c)$ $G_{i,k}$ is a binary random variable, i.e., $G_{i,k} \in \{0,1\}$, for all edges $(i,k)$ such that $i \sim k$. We define the set $\{(i,k): i \sim k\}$ as the set of \textit{admissible edges} $\Eadj$. Then, if two geographically contiguous areas $i$ and $k$ are such that $G_{i,k} = 1$, they are called \textit{neighbouring areas} and the associated graph edge $(i,k)$ is a \textit{neighbouring edge}. Similarly, if $G_{i,k} = 0$ for two geographically contiguous areas $i$ and $k$, they are called \textit{boundary areas} and the associated edge $(i,k)$ is a \textit{boundary edge}.

Given these definitions, let us describe how boundary detection is performed. As part of the Bayesian model, we assume $G$ random with some (marginal) prior. All the inference on the boundary detection problem is based on the (marginal) posterior distribution of $G$, obtained via the MCMC algorithm we introduce below and detailed in \Cref{sec:MCMC} of the SM. Then, it is straightforward to estimate $\mathbb P(G_{i, k} = 1 \mid \bm y)$ for any $(i, k) \in \Eadj$. An \emph{estimated boundary edge} is an edge $(i, k) \in \Eadj$ such that $\mathbb P(G_{i, k} = 1 \mid \bm y) < \gamma$, i.e., for which the posterior probability $\mathbb P(G_{i, k} = 1 \mid \bm y)$  that there is an edge between areas $i$ and $k$ is smaller than a threshold $\gamma \in (0, 1)$. Conversely, an \textit{estimated neighbouring edge} is an edge $(i,k) \in \Eadj$ such that $\mathbb P(G_{i, k} = 1 \mid \bm y)\geq \gamma$. The \textit{estimated boundary graph} $\Ghatb$
is therefore the graph whose edges consist of all the estimated boundary edges,
while the \textit{estimated neighbouring graph} $\Ghatn$ is 
the graph whose edges consist of all the estimated 
neighbouring edges. Strictly speaking, $\Ghatb$ and $\Ghatn$ are disjoint graphs such that 
$\Ghatb \cup \Ghatn = \Eadj$. Finally, if two geographically contiguous areas $i$ and $k$ are such that $(i,k) \in \Ghatn$, they are called \textit{estimated neighbouring areas}, while if $(i,k) \in \Ghatb$, they are called \textit{estimated boundary areas}. In this model, we assume that an isolated areal unit $i$ (e.g., an island) does not have admissible edges, i.e., there exists no area $k$ such that $i \sim k$. As a consequence, no boundaries can be found between an island $i$ and other areas. Of course, in different applications, assuming that the set of admissible edges of an island $i$ is non-empty might be relevant.

Since we are interested in detecting differences in income distribution across geographically contiguous areas, the first building block is a model for spatially-dependent density estimation. To this end, we model the density in each area via a Gaussian mixture model, given their well-known ability to approximate any density (under mild conditions; see \cite{ghosal2017fundamentals}, Section 2.3.3). We assume
\begin{equation}
    \label{eqn:modeldata}
    y_{i,j} \mid \w_i, \bm{\tau}, H \ind f_i(\cdot;H):=\sum_{h=1}^H w_{i,h} \, \mathcal{N}\left(\,\cdot \mid \tau_h\right), \text{ for } j = 1, \dots, N_i,\ i=1,\dots,I,
\end{equation}
where $\w_i = \left(w_{i,1}, \dots, w_{i,H}\right)$ is a vector in the $H$-dimensional simplex $S^{H}$, i.e., $w_{i,h} \geq 0$ for all $h$ and $\sum_{h} w_{i,h} = 1$, and $\mathcal{N}\left(\cdot \mid \tau_{h}\right)$ denotes the Gaussian density with parameters $\tau_{h} = (\mu_{h}, \sigma^{2}_{h})$, being $\mu_{h}$ the mean and $\sigma^{2}_{h}$ the variance. Observe that the parameters $\tau_h = 1, \ldots, H$, in \eqref{eqn:modeldata} are common across different areas: this assumption is often made when modelling related densities since it allows the adoption of a more parsimonious model but preserves the flexibility we need for density estimation \citep[see, e.g.,][]{quintana2022dependent}. Moreover, note that data within each area $i$, $ y_{i,j}, j = 1, \dots, N_i$, are assumed conditionally independent and identically distributed from $f_i$.

To make likelihood in \eqref{eqn:modeldata} more tractable from a computational point of view, we introduce, for each observation $y_{i,j}$, the corresponding latent cluster allocation variable $s_{i,j}$ for $j=1,\dots,N_i$ and $i=1,\dots,I$. Each latent variable indicates the component of the mixture to which the corresponding observation is allocated. This representation will be useful to derive a Gibbs sampler for our model. We rewrite \eqref{eqn:modeldata} as:
\begin{align}
    y_{i,j} \mid s_{i,j} = h, \tau_{h}, H &\ind \mathcal{N}\left(\,\cdot \mid \tau_h\right) \quad j = 1, \dots, N_i \text{ and } i=1,\dots,I, \label{eqn:data_in_clust}\\
    \mathbb{P}\left(s_{i,j} = h \mid \w_i, H\right) &= w_{i,h} \quad h=1,\dots,H. \label{eqn:prior_clust_allocs}
\end{align}
A component in the mixture is said to be empty if no observations have been allocated to such a component. In the following, we will denote any allocated component as a cluster, and the number of clusters is the number of allocated components.

We take the Bayesian approach and complete the likelihood \eqref{eqn:modeldata} with a joint prior for parameters $\bm \tau=(\tau_1,\ldots,\tau_H)$, $\bm w_{(i)}$'s and $H$.  As it is standard in Bayesian nonparametric models, we assume
\begin{equation}
    \tau_h=(\mu_h,\sigma_h^2)\iid P_0,\ h=1,\ldots,H,
    \label{eqn:P_0}
\end{equation}
where $P_0$ is the normal-inverse-gamma density, i.e., $P_0(\dd \mu_h, \dd \sigma_h^2)={\mathcal N}(\dd \mu_h; \mu_0; \sigma_h^2 / \lambda) \times \invGamma\-(\dd \sigma_h^2; c, d)$. The conditional prior of $\bm w_{(i)}$, given $H$ and the adjacency graph $G$, assumes the strong spatial correlation across the observations we expect to see in the data. To incorporate this prior knowledge into the model, given that the $\tau_h$'s are shared, we induce spatial dependence through the prior on the weights by assuming a logistic multivariate conditionally autoregressive (CAR) prior \citep{beraha2021spatially} as the marginal joint prior for $(\w_1, \dots, \w_I)$. That is, we let:
\begin{align}
    w_{i, h}&= \frac{\e^{\tilde{w}_{i, h}}}{1 + \sum_{h=1}^{H-1}\e^{\tilde{w}_{i,h}}} \quad h=1,\dots,(H-1), & w_{i,H} &= \frac{1}{1 + \sum_{h=1}^{H-1}\e^{\tilde{w}_{i,h}}}.
    \label{eqn:alr}
\end{align}
and assume a multivariate CAR distribution with parameters $\rho$, $\sigma^2$ and the graph $G$ for $(\wtilde_1, \dots, \-\wtilde_I)$:
\begin{equation}
    vec(\wtilde_1, \dots, \wtilde_I) \mid \sigma^2, G, H \sim \mathcal{N}_{I(H-1)}\left(\bm{0}, \left(\left(F - \rho G\right) \otimes \dfrac{1}{\sigma^2}\mathbf{I}_{H-1}\right)^{-1}\right),
    \label{eqn:logMCAR_def}
\end{equation}
where $\mathbf{I}_{H-1}$ is the identity matrix of dimension $(H-1)$ and $F = diag( \rho\sum_k G_{1,k} + (1-\rho)$, $\dots, \rho\sum_k G_{I,k} + (1-\rho))$, $\rho$ and $\sigma^2$ positive. Note that \eqref{eqn:logMCAR_def} is defined in terms of a multivariate CAR model, generalising the univariate CAR model in \cite{leroux2000estimation}. See \cite{beraha2021spatially} for further details and properties. Parameter $\rho$ describes the global level of spatial correlation between the areas, where $\rho = 0$ represents independence and values of $\rho$ close to 1 stand for strong spatial association. Then, the random graph $G$ represents the local correlation structure between the areas. In the rest of the paper, we fix $\rho$ to a value close to 1 to encourage spatial association. Note that fixing $\rho$ to a large value is common practice in CAR models for boundary detection (when $G$ is random) since the spatial correlation structure can be determined locally by $G$ rather than globally by $\rho$ \citep[see, e.g.,][]{lee.mitchell2012, lee2013carbayes}. We will assume $G$ random (see below) since $G$ is the relevant parameter for boundary detection, as underlined at the beginning of this section. However, also assuming  $\rho$ random clearly introduces non-identifiability in \eqref{eqn:logMCAR_def}. The marginal prior we assume for $\sigma^2$ is standard, i.e., 
\begin{equation}
\label{eqn:prior_sigma}
\sigma^2\sim \operatorname{InvGamma}(\alpha/2, \beta/2), \ \alpha,\beta >0,
\end{equation}
where the prior mean of $\sigma^2$ is $\beta/(\alpha-2)$ when $\alpha > 2$. For the graph $G$ we assume:
\begin{equation}
  \begin{split}
    \label{eqn:edgeprior}
    G_{i,k} \mid p &\iid \operatorname{Be}(p) \quad \text{ for all } (i,k) \in \Eadj; \\
    p &\sim \operatorname{Beta}\left(a, b\right), \ a,b>0.
  \end{split}
\end{equation}
From the above equation, it is clear that parameter $p$ identifies the (random) prior probability of edge inclusion, i.e., $p = \mathbb{P}(G_{i,k} = 1)$, for each admissible edge. Then, according to our definition of boundary edge,  $1 - p = \mathbb{P}(G_{i,k} = 0)$, $(i,k) \in \Eadj$, represents the prior probability of having a boundary edge between geographically contiguous areas. The Beta marginal prior is a default choice when modelling the probability of success.

Notice that our model for the collection of densities $(f_1, \ldots, f_I)$ can be seen as a Gaussian mixture based on a finite-dimensional instance of a multivariate species sampling model \citep{franzolini2025multivariate}. A fundamental object in this context is the partially exchangeable partition probability function (pEPPF), which characterises the law of the random partition induced by the almost surely discrete mixing probability measures underlying the densities in \eqref{eqn:modeldata} across the $I$ groups. The pEPPF provides a formal characterization of the partition structure and it's crucial for understanding the borrowing of strength across different areas.
In our setting, \textit{borrowing of strength} occurs when observations from different areas are assigned to the same mixture component. To formalize this, let $s_{i,j} \in \{1, \dots, H\}$ be the latent cluster allocation variable such that $y_{i,j} \mid s_{i,j}, \boldsymbol{\tau}, H \ind \mathcal{N}\left(\ \cdot \mid \tau_{s_{i,j}}\right)$, where $H$ denotes the total number of available mixture components (atoms) and $\tau_h$ represents the $h$-th atom (see \eqref{eqn:data_in_clust}). Then, two observations in different groups are assigned to the same mixture component or, equivalently, there is a tie across two groups $i$ and $k$, when $\tau_{s_{i,j}} = \tau_{s_{k,j'}}$.
\begin{equation*}
    \mathbb{P}\left(\tau_{s_{i,j}} = \tau_{k,j'} \mid G, H, \rho, \sigma^2\right).
\end{equation*}%
While the pEPPF remains analytically intractable due to the logistic multivariate CAR prior, it is nevertheless possible to characterise the pairwise tie probabilities across areas. We provide an analytical expression for $\mathbb{P}\left(\tau_{s_{i,j}} = \tau_{k,j'} \mid G, H, \rho, \sigma^2\right)$ in \Cref{prop:borrowing_of_strength} of the Supplementary Material, and briefly discuss its numerical evaluation.

Summing up so far, the random parameters of our model are $\bm \tau$, $\w_i$, $H$, $G$ and $\sigma^2$. We have multiple observations in each area, assumed i.i.d. according to the mixture model \eqref{eqn:modeldata}. Our model addresses boundary detection by identifying boundaries $(i,k)$ for which the associated vector of mixture weights $\w_i$ and $\w_k$ are estimated as being very different using the logistic multivariate CAR prior \eqref{eqn:logMCAR_def} containing $G$. Since the area-specific densities differ only through the mixture weights, we identify as boundaries those edges which separate areal units with significantly different area-specific densities from their neighbours.

The final building block of our model is the marginal prior for the number of components $H$. Indeed, as already mentioned in the Introduction, fixing $H$ leads to undesired consequences for boundary detection. This deals with the well-known non-identifiability of mixture models \citep{rousseau.mengersen2011}. In particular, within model \eqref{eqn:modeldata}, if $\w_i \approx \w_j$, then $f_i \approx f_j$, but the reverse is not true if $H$ is too large.

\begin{figure}[t]
    \centering
    \includegraphics[width=0.66\textwidth]{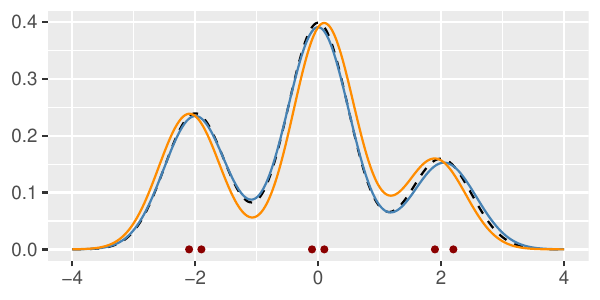}
    \caption{Example of non-identifiability with overfitted mixtures. The black dashed line is $f_0$. The blue and orange lines are the mixtures $\tilde{f}(\bm{w}_1)$ and $\tilde{f}(\bm{w}_2)$ of six Gaussian kernels.}
    \label{fig:CounterExample1}
\end{figure}

To understand the issue, we provide two examples. Consider, first, two areas and suppose that data, in both areas, are distributed according to a mixture of $H_0$ Gaussian kernels. If, under our model, we fix $H = H_0$, the model is identifiable, and no issue is expected. On the other hand, if $H > H_0$, for a common set of atoms $\bm{\tau}$, we can have multiple and different configurations of weights $\bm w_1$ and $\bm w_2$ that approximate well the data-generating process. In particular, assume that the true density is
\begin{equation*}
    f_{0}\left(\cdot\right) = \sum_{h = 1}^{3} w_{0,h}\,\mathcal{N}\left(\cdot \mid \mu_{0,h}, \sigma_0^2\right),
\end{equation*}
with $\bm{w}_{0} = (0.3, 0.5, 0.2)$, $\bm{\mu}_{0} = \left(-2, 0, 2\right)$ and $\sigma_0^2 = 0.5$.
If we set $H=6$ and choose a common set of atoms $\bm{\mu} = (-2.1,-1.9,-0.1, 0.1, 1.9, 2.2)$
in the density
\begin{equation*}
    \tilde{f}\left(\bm{w}\right) = \sum_{h=1}^H w_{h} \, \mathcal{N}\left(\,\cdot \mid \mu_h, \sigma_0^2\right),
\end{equation*}
which has the same structure as the likelihood in \eqref{eqn:modeldata}, there exist at least two configurations, $\bm{w}_1 = (0.15, 0.15, 0.25, 0.25, 0.1, 0.1)$ and $\bm{w}_2 = (0.29, 0.01, 0.01, 0.49, 0.199, 0.001)$, for which both $\tilde{f}(\bm{w}_1)$ and $\tilde{f}(\bm{w}_2)$ approximate $f_0$ well enough; see \Cref{fig:CounterExample1}. As a result, if two geographically contiguous areas share the same density, model \eqref{eqn:data_in_clust}-\eqref{eqn:edgeprior} would not be able to detect such similarity based solely on the mixture weights. The same argument also holds for geographically contiguous areas with different densities. This issue might lead to poor boundary detection because the full-conditional of the graph $G$ depends on the weights; see \Cref{sec:MCMC} in the Supplementary Material.

The issue described above is not purely hypothetical, but, on the contrary, it is often encountered in practice. Let us consider a simulated dataset with  $I = 36$ of areas in a unit-squared domain. Depending on the area, we simulate $100$ i.i.d. data points either $(i)$ from a Student's $t$ distribution with $6$ degrees of freedom, centred in $4$ and with standard deviation equal to $1.5$
or $(ii)$ from a Skew Normal distribution with location  $\xi = 4$, scale  $\omega = 1.3$ and shape  $\alpha = -3$ parameters. This implies that, if we define $\delta = \alpha / \sqrt{1+\alpha^2}$, the mean of the distribution is $\xi + \omega\delta\sqrt{2/\pi} \approx 3.016$ and the variance is $\omega^2(1 - 2\delta^2 / \pi) \approx 0.722$.
We apply the model \eqref{eqn:data_in_clust}-\eqref{eqn:edgeprior} and run the MCMC sampler for a total of 10,000 iterations, discarding the first half as burn-in. We compute $\mathbb{P}\left(G_{i,j} = 1 \mid \bm{y}\right)$ for every admissible edge when we fix the number of components to a reasonably small value ($H = 3$) and to a larger number, i.e., $H = 10$. As the cut-off threshold for boundary detection, we choose $\gamma = 0.5$, i.e., compute the posterior median (boundary) graph. We observe that, when $H = 3$, we achieve perfect boundary detection: the true boundary and the posterior median boundary graphs coincide. This is not true when $H = 10$, as probabilities $\mathbb{P}( G_{i,j} = 1 \mid \mathbf{y} )$ are all between $0.9992$ and $1$, implying that the boundary graph is empty for every reasonable value of $\gamma$ ($\leq 0.5$). This indicates that boundary detection cannot be achieved when the number of components is fixed too large.

These two examples show that fixing $H$ might be problematic for boundary detection. Parameter $H$ must be learned from the data, and we assume as its marginal prior a shifted Poisson distribution, and we write:
\begin{equation}
\label{eqn:prior_H}
    H - 1 \sim Poi(\Lambda)
\end{equation}
for $\Lambda > 0$. Notice that, if $H=1$ a.s., though the model is well defined, there is only one mixture weight in each area, which equals $1$. In this case, no boundaries will be found, since all estimated densities will be identical.

\section{Posterior computation} 
\label{sec:algorithm}

From the conditional distribution of the data in \eqref{eqn:data_in_clust}-\eqref{eqn:prior_clust_allocs} and the prior \eqref{eqn:P_0}-\eqref{eqn:prior_H}, we implement a transdimensional MCMC algorithm to obtain the joint posterior distribution of the number of components $H$, the vector of parameters $\bm{\theta}_{H} = (\bm{W}, \bm{\tau}, \bm{s}_{1},\dots,\bm{s}_{I})$, whose dimension depends on $H$, and the remaining global parameters $(\sigma^2, G)$. Here, $\bm{W}$ is the $I \times H$ matrix whose $i,h$-th element is the mixture weight $w_{i,h}$, while $\bm{s}_i = \left(s_{i,1},\dots,s_{i,N_i}\right)$, for $i = 1,2,\dots,I$, denotes the latent cluster allocation variables associated to every observation in area $i$ and introduced in \eqref{eqn:data_in_clust}-\eqref{eqn:prior_clust_allocs}. Note that the support of each $s_{i,j}$ is $\{1,\dots,H\}$. 
A transdimensional sampling algorithm (as the reversible jump MCMC) usually consists of two major steps: $(i)$ a \textit{between models move} which consists in the joint update of $H$ and the corresponding parameter vector, and $(ii)$ a \textit{within model move} which implements, conditionally to $H$, a sampling scheme to update the vector of parameters $\bm{\theta}_{H}$, $\sigma^2$ and $G$.

For us, the within model move is straightforward. On the other hand, the between models move is not standard. Below, we describe how to adapt the approach of \cite{norets2021optimal} that sidesteps the need of split-merge moves, but is based on \emph{auxiliary priors} that maximize the overall chain mixing. To the best of our knowledge, this is the first time that the approach by \cite{norets2021optimal} is applied to mixture models.

\subsection{Between-models move}
The reversible jump MCMC sampler \citep{green1995reversible} provides a general framework for transdimensional simulation schemes. It can be viewed as an extension of the Metropolis-Hastings algorithm. As it happens in standard Metropolis-Hastings, given the current state of the chain $(H,\bm{\theta}_{H})$, the next state $(H',\bm{\theta}_{H'})$ is sampled from a proposal distribution $q[(H,\bm{\theta}_{H}),(H',\bm{\theta}_{H'})]$ and accepted with probability
\begin{equation*}
    \alpha \left[\left(H,\bm{\theta}_{H}\right), \left(H',\bm{\theta}_{H'}\right) \right] = \min\left\{ 1, \frac{\pi\left( H', \bm{\theta}_{H'} \mid \bm{y}\right) \, q\left[ (H',\bm{\theta}_{H'}) , \left(H,\bm{\theta}_{H} \right) \right]}{\pi\left( H, \bm{\theta}_{H} \mid \bm{y} \right) \, q\left[ \left(H,\bm{\theta}_{H}\right), (H',\bm{\theta}_{H'}) \right]} \right\}.
\end{equation*}
Usually, the proposal distribution is defined in two steps. If $\bm{\theta}_{H} \in \mathbb{R}^{n_{H}}$ and $\bm{\theta}_{H'} \in \mathbb{R}^{n_{H'}}$, with $n_{H'}>n_{H}$ and $d=n_{H'}-n_{H}$, first a random vector $\bm{u}\in\mathbb{R}^{d}$ is sampled from a distribution $q_{d}(\bm{u})$ and then $\bm{\theta}_{H'}$ is defined as $g_{H \rightarrow H'}(\bm{\theta}_{H}, \bm{u})$ for a suitable mapping function $g_{H\rightarrow H'}$ that maps $\mathbb{R}^{n_H}$ into $\mathbb{R}^{n_H'}$. Since both the proposal distribution $q_{d}(\bm{u})$ and the mapping function $g_{H \rightarrow H'}$ are arbitrary, the definition of an efficient between-models move might be a difficult task.

The approach we follow is based on \cite{norets2021optimal}, where the author defines optimal auxiliary priors and proposals for generic nested models indexed by an integer $H$ in $\{1,2,\dots\}$ with unknown parameter $\bm{\theta}_{H}$ and prior of the form $\pi\left(\bm{\theta}_{H} \mid H\right)\pi\left(H\right)$. Since the models are nested, the unknown parameters are nested as well, i.e., if $H' > H$, the first $H$ elements of $\bm{\theta}_{H'}$ correspond to vector $\bm{\theta}_{H}$. Given the current state $(H, \bm \theta_H)$, consider moving to $(H', \bm \theta_{H'})$ with $H' = H+1$. We denote the parameter vector associated with the extra component by $\bm{\theta}$. The joint distribution for $\left(\bm{y}, \bm{\theta}_{H'}, H\right)$ is given by:
\begin{equation*}
    f\left(\bm{y},\bm{\theta}_{H'}, H\right) = \tilde{\pi}_{H}\left(\bm{\theta} \mid \bm{\theta}_{H}, \bm{y}\right)\mathcal{L}\left(\bm{y} \mid H, \bm{\theta}_{H}\right)\pi(\bm{\theta}_{H} \mid H)\pi(H),
\end{equation*}
where $\tilde{\pi}_{H}\left(\bm{\theta} \mid \bm{\theta}_{H}, \bm{y}\right)$ needs to be defined. We choose such prior as the conditional posterior distribution of the extra component, i.e., 
\begin{align*}
    \tilde{\pi}_{H}\left(\bm{\theta} \mid \bm{\theta}_{H}, \bm{y}\right) &= \pi(\bm{\theta} \mid \bm{y}, H+1, \bm{\theta}_{H}),\\
    &\propto \mathcal{L}\left(\bm{y} \mid H+1, \bm{\theta}_{H+1} \right)\pi\left(\bm{\theta}_{H+1} \mid H+1\right).
\end{align*}
This choice guarantees optimal conditions in terms of overall chain mixing and minimisation of the estimated variance, as shown in \cite{norets2021optimal}. Note that mixture models are not generally considered as nested models. This is typically due to the usual assumption of a Dirichlet prior for the weights. Instead, our logistic multivariate CAR prior lends itself naturally to the nested model framework much like the mixture of experts models analysed in \cite{norets2021optimal}. Moreover, the proposal that adds one mixture component while leaving the others unchanged is along the same spirit of modern MCMC algorithms for Bayesian (nonparametric) mixture models. Think, for instance, of Algorithm 8 in \cite{neal2000markov}, the birth-death sampler in \cite{stephens2000bayesian}, the slice sampler in \cite{kalli2011slice}, and the conditional sampler in \cite{argiento2022infinity}. The main difference is that with our approach, the new component is selected by taking the data into account, while in the other papers mentioned, the new components are proposed from the prior distribution, making our approach potentially more efficient.

Nonetheless, the optimal posterior $\pi(\bm{\theta} \mid \bm{y}, H+1, \bm{\theta}_{H})$ is not known in closed form, so we use its Laplace approximation. Such approximation is justified by the Bernstein–von Mises theorem and the asymptotic behaviour of the (conditional) posterior distribution we consider: see, e.g., \cite{walker1969asymptotic} and the references therein. Note that sampling the extra component $\bm{\theta}$ from its (approximated) conditional posterior distribution $\pi(\bm{\theta} \mid \bm{y}, H+1, \bm{\theta}_{H})$ avoids the artificial construction of proposal distributions and mapping functions. Moreover, since we have marginalised w.r.t. the latent cluster allocation variables, the dimension of the proposal distribution is sensibly reduced. This mitigates the well-known problem of poor mixing for multidimensional Metropolis-Hastings simulation schemes \citep[see, e.g.,][]{robert2014metropolis}. \Cref{alg:rjmove_pseudocode} provides a detailed description of the reversible jump step in pseudo code.

\begin{algorithm}[t]
    \caption{Between Models Move for the model for boundary detection}
    \label{alg:rjmove_pseudocode}
    \SetAlgoLined
    sample $H'$ in $\{H-1;H+1\}$ with probability $\{1/2;1/2\}$\;
    \eIf{$H'=H+1$}{
        compute the parameters of the approximated optimal posterior (see \eqref{eqn:optmean} and \eqref{eqn:optcov} in the SM)\;
        sample $\left(\wtilde^{H+1}, \tau_{H+1}\right) \sim \mathcal{N}\left(\bm{\mu}^{*},\bm{V}^{*}\right)$\;
        compute $A_{H,H+1}$ according to \eqref{eqn:increasearate} in the SM\;
        sample $U \sim \mathcal{U}[0;1]$\;
        \eIf{$U \leq A_{H,H+1}$}{
            accept the move and enlarge the state of the chain\;
        }{
            reject the move and let the state of the chain unchanged\;
        }
    }{
        randomly select the component to drop\;
        compute the parameters of the approximated optimal posterior (see \eqref{eqn:optmean} and \eqref{eqn:optcov} in the SM)\;
        compute $A_{H,H-1}$ according to \eqref{eqn:reducearate}  in the SM\;
        sample $U \sim \mathcal{U}[0;1]$\;
        \eIf{$U \leq A_{H,H-1}$}{
            accept the move and reduce the state of the chain\;
        }{
            reject the move and let the state of the chain unchanged\;
        }
    }
\end{algorithm}

\subsection{Within-model move}
The update of the parameter vector $\bm{\theta}_H$, $\sigma^2$ and $G$, given $H$, is rather standard. The within-model move is obtained repeatedly sampling parameters as follows:
\begin{enumerate}
    \item For any $i = 1,\dots, I$ and $j = 1, \dots, N_i$, independently update the cluster allocation variables from its full conditional distribution, using a Gibbs sampler update;
    \item Independently update the atoms $\bm{\tau}$ of the mixture from its full conditional distribution, via a Gibbs sampler update;
    \item For each $i = 1, \dots, I$ and $h = 1, \dots, H - 1$, sample the transformed weights $\tilde{w}_{i,h}$ via an augmented Gibbs sampler update. This update is made introducing a latent Polya-Gamma random variable  $\omega_{i,h}$ and sampling sequentially from the full conditional $\pi( \tilde{w}_{i,h} \mid \tilde{\bm{W}}_{-(i,h)}, \bm{s}_i, \sigma^2, \omega_{i,h}; \rho )$, now available in closed form. We denote with $\tilde{\bm{W}}_{-(i,h)}$ the matrix $\tilde{\bm{W}}$ without its $(i,h)$-th element. Details of this step can be found in \cite{beraha2021spatially}, Section 5.
    \item Sample $\sigma^2$ from its full conditional distribution;
    \item For any admissible edge $\left(i,j\right)$ such that $i \sim j$, sample the corresponding graph edge $G_{i,j}$ from its full conditional. This allows for the update of multiple graph edges in a single MCMC iteration, unlike other types of sampling strategies: see, for instance, the BDMCMC algorithm for Gaussian graphical models in \cite{mohammadi2015bayesian}.
\end{enumerate}
For a detailed description of the transdimensional Gibbs sampler, with explicit computation of the full conditional distributions and the reversible jump proposal, please refer to \Cref{sec:MCMC} in the Supplementary Material.

\section{Simulation studies}
\label{sec:simstudies_summary}
We perform extensive simulation studies
to evaluate the performance of our \textit{SPMIX} model with respect to $(i)$ joint spatial density estimation, $(ii)$ structural learning of the adjacency graph, and $(iii)$ boundary detection under model misspecification. We provide in this section a summary of these simulations, while more details are reported in \Cref{sec:simstudies} of the Supplementary Material.
    
The first simulation study, described in \Cref{subsec:de_simstudy} of the SM, is built to assess the model's ability to borrow information across areas when local sample sizes are small ($N_i = 100$). In these setting, \textit{SPMIX} demonstrates high accuracy in recovering spatially varying densities, even when the underlying data-generating process involves complex, spatially dependent weights. These findings confirm that, in the case of well-specified models, we retrieve the correct number of components across all areas and capture nuanced distributional differences.

Structural learning %
is analysed in \Cref{subsec:sl_simstudy} of the SM by examining the sensitivity of the graph estimation process to various hyperprior specifications under a misspecified regime, i.e., when the generating process does not coincide with the likelihood.
We found that the choice of the prior for the edge inclusion probability $p$ is particularly critical; a sparse specification, such as $p \sim \operatorname{Beta}(2, I)$, penalizes redundant edges and prevents the detection of spurious boundaries. Furthermore, the global spatial strength parameter $\rho$ and the between-area variance $\sigma^2$ act as the primary mechanisms for identifying distributional discrepancies.

To assess whether the proposed model is robust and reliable under misspecification, we then perform a large-scale study involving 50 independent replicates of a 36-area domain. This simulation study, detailed in \Cref{subsec:bd_simstudy} of the SM, 
provides a robust assessment of the boundary detection performance, using metrics such as Precision, Sensitivity, and AUC, while evaluating the sensitivity
of the results to the number of mixture components $H$ and the spatial association parameter $\rho$. 
Across all replicates, the model with a random number of components and high spatial association ($\rho \geq 0.95$) consistently outperforms the fixed-component counterparts. In particular, we observe that while finite mixtures can lead to a drop in precision, especially in case of overfitted mixtures, the proposed transdimensional MCMC maintains high sensitivity and specificity. Threshold-independent metrics, such as the area under the ROC Curve (AUC), underscore the model's robustness in discriminating between contiguous areas with distinct data-generating processes when the graph threshold $\gamma = 0.5$ is used.
Interestingly, the model with random $H$ shows superior MCMC mixing and convergence properties.

In \Cref{sec:stress_tests} in the SM, we assess the robustness of boundary detection provided by our model for two demanding stress tests.
In the first scenario, we simulate data in different areas from different distributions but with identical median values. In the second scenario, we consider a extremely high data resolution in each area, for a total of $108,000$ observations. In both cases, the model with random $H$ and high global spatial association ($\rho = 0.95$) correctly identifies the true boundary graph. 

One last remark must be made on the computational efficiency of our model with this large dataset: the reversible jump step is clearly the bottleneck of the MCMC algorithm. In fact, we observe an average of $11$ MCMC iterations per second in the simulated scenarios with $I = 36$ areas and $N_i = 100$ observations per area, which is a reasonable computational time for a model of this complexity. Clearly, the computational time drastically increases with the number of areas and observations: to fit the model for this large dataset 
we needed four days as the total computational time to run $40,000$ MCMC iterations. The implementation of more efficient algorithms and sampling strategies is beyond the scope of this work, but is addressed in the Discussion section.

\section{California census income dataset} \label{sec:real_case_scenario}
Here we analyse the personal income data of California citizens, extracted by the public use microdata sample from the 2020 ACS. After describing the dataset and performing a preliminary exploratory analysis that validates our modelling assumptions in \Cref{subsec:data_exploration}, we report posterior inference in \Cref{subsec:post_inf_cali}. 
\Cref{subsec:comparison_calicensusdata}
in the Supplementary Material shows the estimates for the California census income dataset under three alternative models for boundary detection available in the literature. To the best of our knowledge, existing models and algorithms perform boundary detection only in the case of a single response per area, while our model achieves the same goal in the case of multiple responses in each geographical unit. For this reason, we have applied these methods to summary statistics of the data in each area.
The section also shows sensitivity analysis with respect to $\rho$, the global spatial association parameter.  
Finally, \Cref{sec:interpret_boundaries} in the Supplementary Material illustrates the estimated boundaries, focusing in particular on the area of downtown Los Angeles, in light of other data sources related to crime and public health.

\subsection{Data description and exploration}\label{subsec:data_exploration}

\begin{figure}[t]
    \centering
    \subfigure{\includegraphics[width=0.48\textwidth]{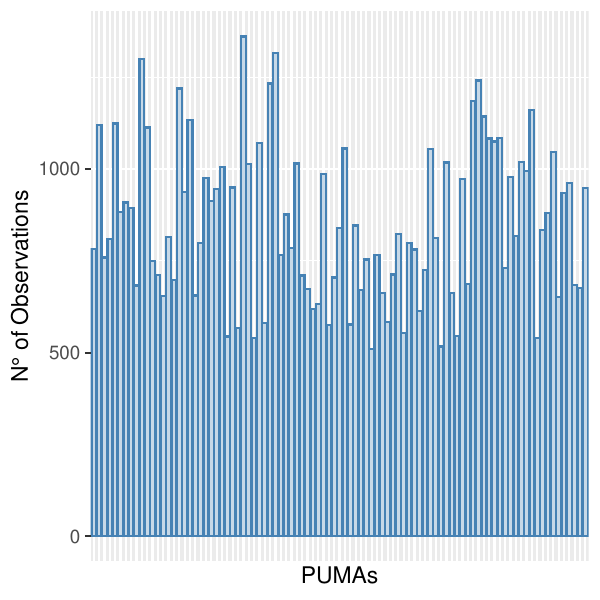}}
    \hfill
    \subfigure{\includegraphics[width=0.48\textwidth]{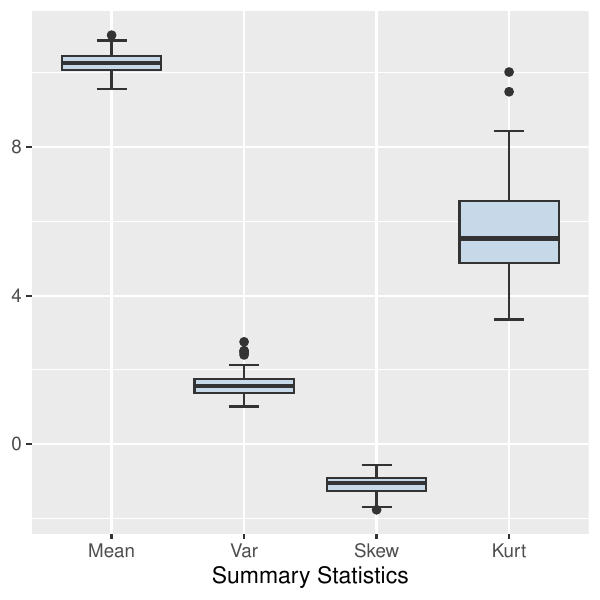}}
    \caption{Summary statistics of the California census income dataset: barplot of the number of observations in each PUMA (left); boxplots of the empirical mean, variance, skewness and kurtosis in each PUMA (right).}
    \label{fig:CaliCensusData_eda1}
\end{figure}

We focus on personal income data in California in the greater Los Angeles region, which consists of LA, Ventura, and Orange counties, as explained in the Introduction. We model the logarithm of the income of a person $j$ in PUMA $i$ as $y_{i,j}$, for a total of $I=93$ PUMAs. See also \cite{beraha2023normalised} for an analysis of personal income data in all California PUMAs, although with a different focus from ours. \Cref{fig:CaliCensusData_exploration1} reports the sample means and variances of data $\{y_{i,j}\}$ in each PUMA. Each PUMA has between $600$ and $1,300$ observations,
for a total of $79,319$ individual records. \Cref{fig:CaliCensusData_eda1} (left panel) 
shows the number of observations in each PUMA. Averaging over all PUMAs, the empirical mean and variance of the logarithm of the areal income data are 10.28 and 1.58, respectively, while  the variance of the empirical variances in all the PUMAs is 0.21. High-income means appear to be located in coastal and specific suburban corridors; see \Cref{fig:CaliCensusData_exploration1}. 

\Cref{fig:CaliCensusData_eda1}, right panel, reports relevant summary statistics of the data distribution in each PUMA. The data exhibits a pronounced negative skewness: namely, if we compute the empirical skewness in each PUMA, we obtain a mean value of $-1.08$, with values varying significantly across areas (from $-1.78$ to $-0.56$). This indicates a persistent left-tail density across all PUMAs. We also observe a large variability in the empirical kurtosis in each PUMA: this suggests that key differences between empirical distributions can be observed more in the tails of the distributions, rather than in their bulks. However, we can also observe significant spatial non-stationarity in both the first and second moments of the distributions. Specifically, the mean log-income across areal units ranges from $9.56$ to $11.02$, while the intra-areal standard deviation varies from $1.01$ to $1.66$. This suggests that the underlying distribution of log-incomes in each PUMA can not be described only by summary statistics describing the centre of the distribution but is also influenced by asymmetries and areal-specific variabilities that requires a flexible density estimation procedure.

\begin{figure}[t]
    \centering
    \subfigure{\includegraphics[width=0.48\textwidth]{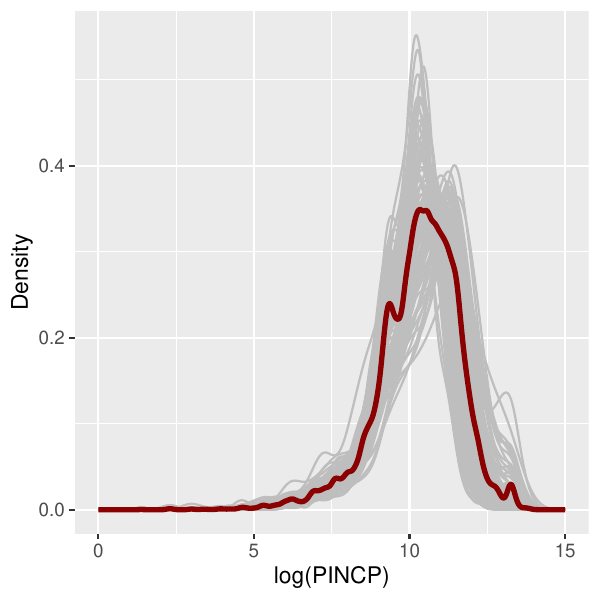}}
    \hfill
    \subfigure{\includegraphics[width=0.48\textwidth]{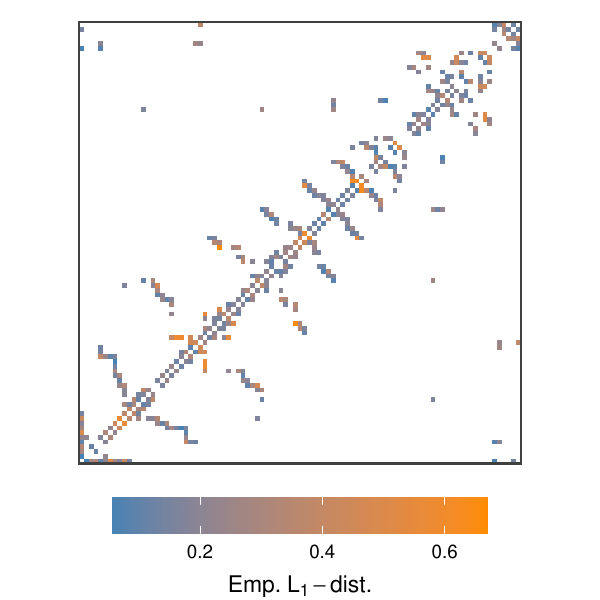}}
    \caption{Exploratory data analysis on the California census income dataset: plot of the kernel density estimate of the $\log$-income data in each PUMA, with the mean curve highlighted in red (left); $\Lone$ distances between kernel density estimates in each pair of geographically contiguous areas (right).}
    \label{fig:CaliCensusData_eda2}
\end{figure}

Since our approach to boundary detection is based on differences in the areal densities, we compute the empirical densities, i.e., the kernel density estimate of the $\log$-income data for each PUMA. As illustrated in the right panel of \Cref{fig:CaliCensusData_eda2}, the greater Los Angeles region is characterised by substantial differences in the empirical densities. The variability observed is heterogeneous: horizontal shifts in the density peaks reflect differences in mean/median income; variations in the range and in the peaks of the distributions reveal that geographically continuous areas might possess different socio-economic structures, regardless of their proximity. Finally, the thickness of the lower-income tail is highly sensitive to geography, which might provide a crucial feature to identify boundaries between affluent enclaves and economically disadvantaged sectors.

In order to quantify the variability observed in the area-specific empirical densities,
we compute, for each pair of geographically contiguous areas (i.e., for every $(i,k) \in \Eadj$) the $\Lone$ distance between the kernel density estimates and report their values in the right panel of \Cref{fig:CaliCensusData_eda2}. Even if the heatmaps of the empirical mean and variance in each PUMA show spatial correlation in the mean log-incomes, this plot highlights areas where geographically contiguous PUMAs exhibit high $\Lone$ distances. This exploratory data analysis confirms that a boundary detection approach based on discrepancies between areal-specific densities would be more robust than traditional difference-in-means tests, as it would capture the heterogeneous variability we observe in the log-incomes of the greater Los Angeles region.

\subsection{Posterior inference}
\label{subsec:post_inf_cali}
We apply our model \eqref{eqn:data_in_clust}-\eqref{eqn:prior_H}, as described in \Cref{sec:model}, to the dataset. We fix the hyperparameters of $P_0$ as follows: $\mu_0 = 10, \lambda = 0.1, c = 4, d = 4$; this implies that $\mu_0$, which expresses the prior mean of the data under \eqref{eqn:modeldata}, is close to the overall sample mean and that, a priori, $\sigma^2_h$ concentrates on values in the range $(0.5; 3.8)$ with $\E(\sigma^2_h) \simeq 1.7$ and $\Var(\sigma^2_h) \simeq 0.9$, in such a way that the prior mean and variance are then not too far from the corresponding empirical estimates. For the common between-areas variance $\sigma^2$ in \eqref{eqn:prior_sigma}, we fix $\alpha = 6$ and $\beta = 4$, yielding both prior mean and variance equal to 1. We fix $\rho = 0.95$ in \eqref{eqn:logMCAR_def} to encourage spatial association between areas. Moreover, we set $p$ in \eqref{eqn:edgeprior} as $p \sim \operatorname{Beta}\left(2, I\right)$, where $I = 93$ is the total number of PUMAs. These values \citep[see][]{paci2020structural} make the prior for $G$ sparse, as we set a priori a small probability $p$ of edge inclusion (see \Cref{subsec:bd_simstudy} of the Supplementary Material for further details about this choice for the prior).

\begin{figure}[t]
    \centering
    \subfigure{\includegraphics[width=0.48\textwidth]{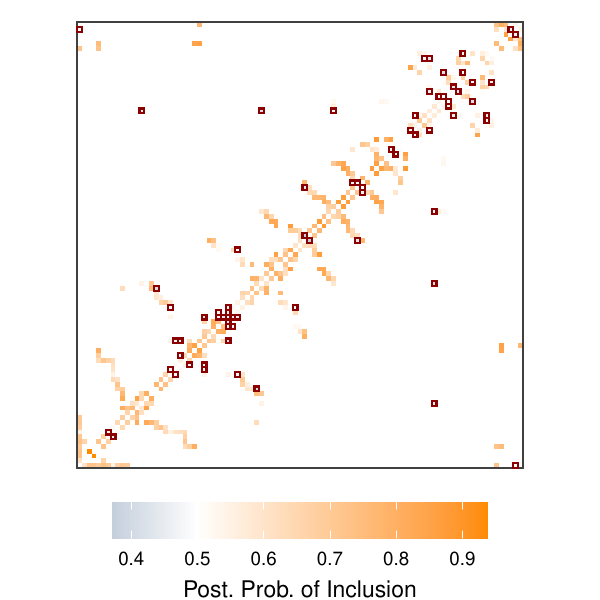}}%
    \hfill
    \subfigure{\includegraphics[width=0.48\textwidth]{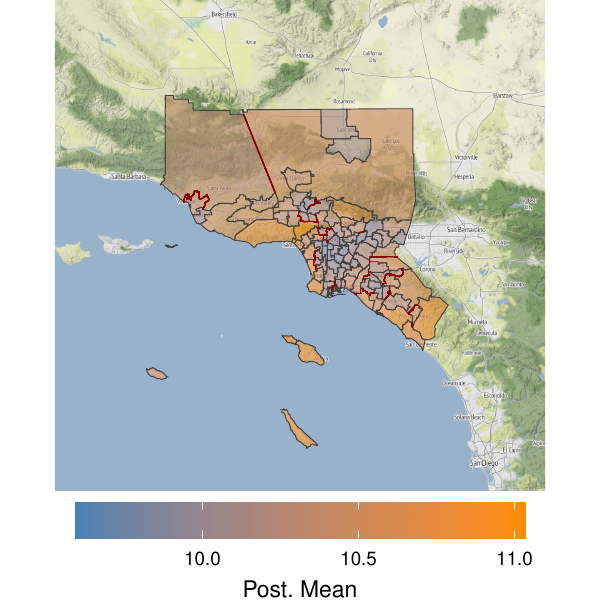}}%
    \caption{Posterior probabilities of edge inclusion $\mathbb{P}(G_{i,k} = 1 \mid \bm{y})$ and boundary graph $\Ghatb$ ($\gamma=0.5$) in red (left); posterior means of the estimated densities on the map with estimated boundary edges highlighted in red (right).}
    \label{fig:CaliCensusData_BDresults}
\end{figure}

We need a long run of our algorithm to retrieve an accurate posterior sample, due to the high dimensionality of the proposal distribution in the between-models move and the high number of observations. Hence, we run the sampler in \Cref{sec:algorithm} for a total of $40,000$ iterations, discarding the first $35,000$ as burn-in, with a final sample size of $5,000$ iterations. The MCMC algorithm returns the posterior mode of $H$  equal to $8$.
\Cref{fig:CaliCensusData_BDresults} (left panel) reports the  matrix of posterior probabilities $\mathbb{P}(G_{i,k} = 1 \mid \bm{y})$ of edge inclusion, for each link $(i,j) \in \Eadj$ as discussed in \Cref{sec:model}. In the figure, non-admissible edges (i.e., the non-bordering areas) are represented as white spots, while boundary edges, identified by $\mathbb{P}(G_{i,k} = 1 \mid \bm{y}) < \gamma = 0.5$, are in red. To make a clearer picture, we also report the estimated boundary graph over the map of the $93$ PUMAs; see  \Cref{fig:CaliCensusData_BDresults} (right panel). The heatmap on the right displays the means of the estimated densities for each area. \Cref{fig:CaliCensusData_BDresults-supp} in the Supplementary Material shows the area-specific variance of the estimated densities. From these plots, it is clear that many detected boundaries can be explained in terms of posterior means and/or posterior variances between geographically contiguous areas. In particular, we observe a sharp boundary that separates the bay area of LA County from PUMAs associated to Los Angeles.
This boundary represents one of the most compelling findings of our analysis; see \Cref{sec:interpret_boundaries} in the Supplementary Material, where we validate the boundary by economic and social indicators.

\begin{figure}[t]
    \centering
    \subfigure{\includegraphics[width=0.48\textwidth]{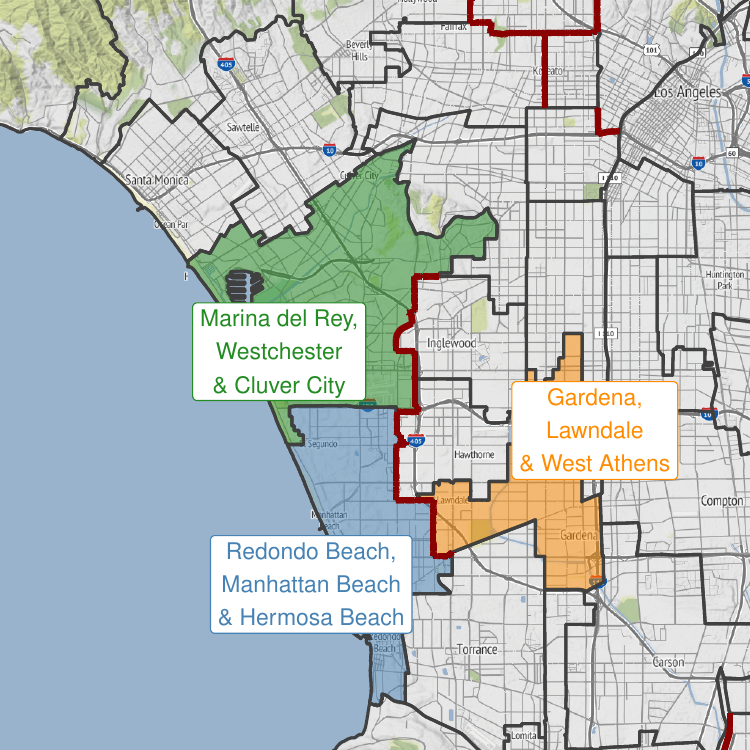}}%
    \hfill
    \subfigure{\includegraphics[width=0.48\textwidth]{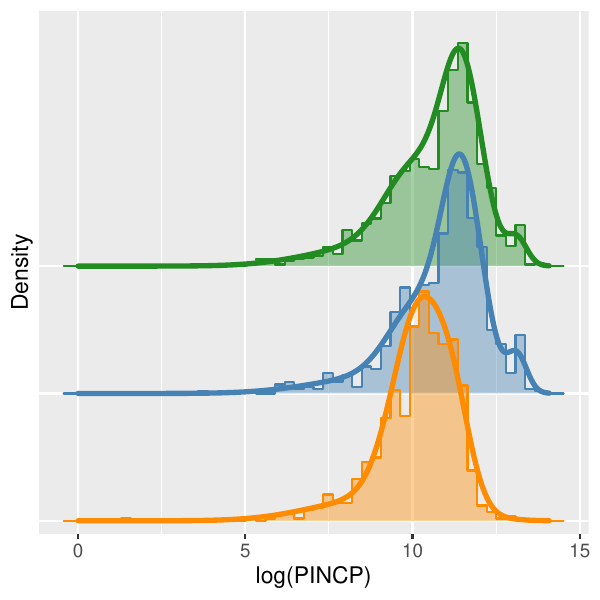}}%
    \caption{Location (left panel) and posterior predictive densities (right panel) for three PUMAs: \textit{Redondo Beach, Manhattan Beach \& Hermosa Beach} (in blue), \textit{Marina del Rey, Westchester \& Culver City} (in green) and \textit{Gardena, Lawndale \& West Athens} (in orange). Estimated boundary edges are highlighted in red.}
    \label{fig:CaliCensusData_densities}
\end{figure}

We select three PUMAs on the estimated boundary between the LA bay area and the inner part of the city. We consider the \textit{Redondo Beach, Manhattan Beach \& Hermosa Beach} PUMA and two of its adjacent areas: \textit{Gardena, Lawndale \& West Athens} and \textit{Marina del Rey, Westchester \& Culver City} PUMAs. Our method has estimated a boundary edge between \textit{Redondo Beach, Manhattan Beach \& Hermosa Beach} and \textit{Gardena, Lawndale \& West Athens} PUMAs while the \textit{Marina del Rey, Westchester \& Culver City} PUMA has been estimated as a neighbouring area of \textit{Redondo Beach, Manhattan Beach \& Hermosa Beach}; see \Cref{fig:CaliCensusData_densities}, left panel. We also report the estimated densities and the associated data histograms of the three areas in the right panel of \Cref{fig:CaliCensusData_densities}. Note that the density estimates precisely match the empirical histograms, indicating good performance of our model in density estimation.

Moreover, while the estimated densities of \textit{Redondo Beach, Manhattan Beach \& Hermosa Beach} and \textit{Marina del Rey, Westchester \& Culver City} PUMAs share several common features (for instance, the same mode and variance, a secondary mode on the left tail of the distribution and a small peak at the far right of the density), there are evident differences between the estimated densities of \textit{Redondo Beach, Manhattan Beach \& Hermosa Beach} and \textit{Gardena, Lawndale \& West Athens} PUMAs: for instance, notice that the density associated with \textit{Gardena, Lawndale \& West Athens} PUMA has a lower mode, a lighter left tail of the distribution with no evident secondary modes and extremely light right tail, with no peak as exhibited by the other two areas we have considered. To quantify the differences between the estimated densities, we have computed the associated $\Lone$ distances, obtaining the value $\simeq 0.128$ for the distance between \textit{Redondo Beach, Manhattan Beach \& Hermosa Beach} and \textit{Marina del Rey, Westchester \& Culver City}, detected as neighbouring areas, and $\simeq 0.621$ for the distance between \textit{Redondo Beach, Manhattan Beach \& Hermosa Beach} and \textit{Gardena, Lawndale \& West Athens}, instead detected as separated by a boundary.

\begin{figure}[t]
    \subfigure[\label{subfig:L1_distances_global}]{\includegraphics[width=0.48\textwidth]{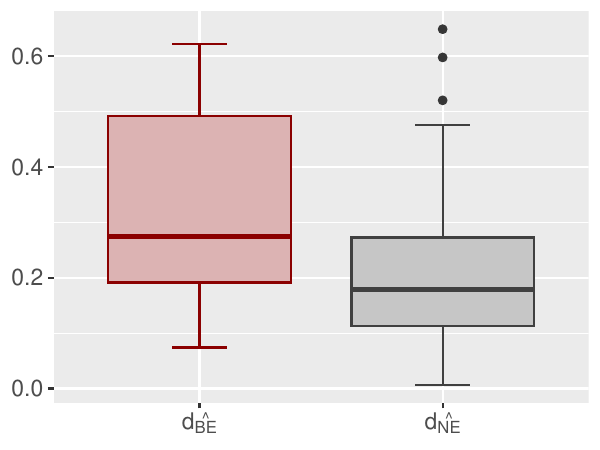}}
    \subfigure[\label{subfig:L1_distances_local}]{\includegraphics[width=0.48\textwidth]{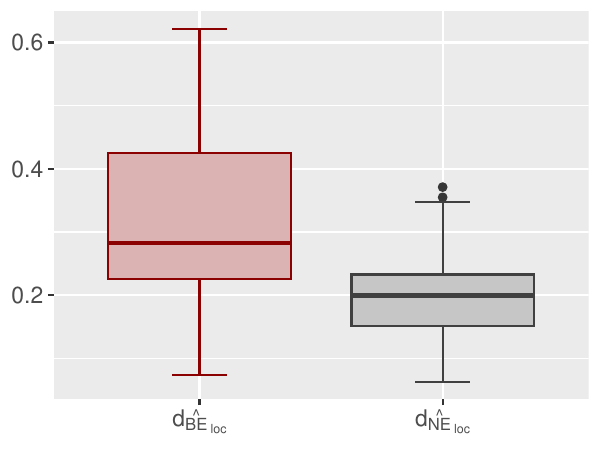}}
    \caption{California Census income dataset: global and local density comparisons in the $\Lone$ metric: boxplots of $d_{\BEhat}$ and $d_{\NEhat}$ (left); boxplots of $d_{\BEhat_{\mbox{\tiny{loc}}}}$ and $d_{\NEhat_{\mbox{\tiny{loc}}}}$ (right).}
    \label{fig:L1_distances}
\end{figure}

As a further check of the goodness of fit of our model, we compute the $\Lone$ distance between pairs of densities associated to estimated boundary edges and compare it with the $\Lone$ distance between pairs of densities associated to estimated neighbouring edges. Such a comparison can be done both \emph{globally} and \emph{locally}. To this end, recall that $\Ghatn$ is the estimated neighbouring graph, $\Ghatb$ is the estimated boundary graph and that $\Ghatb \cup \Ghatb = \Eadj$. To compare the estimated densities globally, we define the sets $\NEhat = \{(i,k) : (i,k) \in \Ghatn\}$ and $\BEhat = \{(i,k) : (i,k) \in \Ghatb\}$. Hence, $\NEhat$ is the set of edges of $\Ghatn$ (i.e. the set of all estimated neighbouring edges), while $\BEhat$ denotes the set of edges of $\Ghatb$ (i.e., the set of all estimated boundary edges). Then, we compute the $\Lone$ distances $d_{\Lone}(\cdot,\cdot)$ between the estimated densities $\hat{f}_i(\cdot)$ and $\hat{f}_k(\cdot)$ for all $(i,k) \in \NEhat$, obtaining the set $d_{\NEhat} := \{d_{\Lone}(\hat{f}_i,\hat{f}_k) : (i,k) \in \NEhat\}$, i.e., the set of all $\Lone$ distances between density pairs associated with estimated neighbouring edges. With a similar argument applied over $(i,k) \in \BEhat$, we obtain the set of $\Lone$ distances between all density pairs associated with estimated boundary edges $d_{\BEhat} := \{d_{\Lone}(\hat{f}_i,\hat{f}_k) : (i,k) \in \BEhat\}$. \Cref{subfig:L1_distances_global} summarises the set of $\Lone$ distances $d_{\NEhat}$ and  $d_{\BEhat}$ in separated boxplots.

The local comparison, on the other hand, goes as follows: first, we define $\NEhat_i = \{k : i \sim k, (i, k) \in \Ghatn\}$ as the set of estimated neighbouring areas of area $i$ and $\BEhat_i = \{k: i \sim k, (i,k) \in \Ghatb\}$ as the set of estimated bordering areas of area $i$, for $i=1,\ldots,I$. Then, for each PUMA, we compute the average $\Lone$ distance between the posterior estimate $\hat{f}_i$ of the density in the $i$-th area and the estimated densities $\{\hat{f}_k\}_{k \in \NEhat_i}$; similarly, we compute the average $\Lone$ distance between the posterior estimate $\hat{f}_i$ and $\{\hat{f}_k\}_{k \in \BEhat_i}$ for all $i$. We then obtain, as in the global comparison, two sets of $\Lone$ distances, called $d_{\NEhat_{\mbox{\tiny{loc}}}}$ and  $d_{\BEhat_{\mbox{\tiny{loc}}}}$. More precisely, $d_{\NEhat_{\mbox{\tiny{loc}}}}$ is then defined as the set  $\{ \lvert \NEhat_i \rvert^{-1} \sum_{k \in \NEhat_i} d_{\Lone}(\hat{f}_i, \hat{f}_k), i \in 1,\dots,I \wedge i \text{ s.t. } \NEhat_i \neq \emptyset \}$, while the definition of $d_{\BEhat_{\mbox{\tiny{loc}}}}$ is similar. \Cref{subfig:L1_distances_local} summarises the set of $\Lone$ distances  $d_{\NEhat_{\mbox{\tiny{loc}}}}$ and  $d_{\BEhat_{\mbox{\tiny{loc}}}}$ in separated boxplots.

To wrap up, \Cref{fig:L1_distances} reports the boxplots of the $\Lone$ distances between posterior density estimates among adjacent and boundary areas in case of global and local comparisons. In both cases, we see that the $\Lone$ distances between neighbouring areas are smaller than between boundary areas. Such difference is more evident in the local comparison, where the two interquartile ranges do not intersect. This shows that our model is able to discriminate between neighbouring and boundary areas through the estimated densities.

\section{Discussion}\label{sec:discussion}
In this paper, we have introduced a flexible Bayesian model for the joint estimation of spatially-dependent densities and performing boundary detection among areas that exhibit strikingly different densities. Because the model draws on multiple observations within each spatial unit, it detects boundaries intrinsically, without relying on ad-hoc dissimilarity measures, thereby widening its applicability across a broad range of applied contexts. The model takes a logistic multivariate CAR prior for the weights of the assumed mixtures, and this prior incorporates the random adjacency graph.

Our methodology proves valuable in identifying areas with significant disparities in population income in the greater Los Angeles region. For policymakers, urban planners, and politicians committed to fostering more equitable policies, such information is instrumental in strategically planning interventions to mitigate social and economic inequalities.
Specifically, our analysis reveals a clear division in the city of Los Angeles, with a wealthier bay area, encompassing affluent neighbourhoods like Beverly Hills, and a less affluent region covering downtown LA and the southern part of the city. This division aligns with findings from various studies and reports by sociologists, economists, and policymakers, affirming the robustness of the inference our model provides. 

As mentioned in the Introduction, there exists a vast literature on CAR priors or their generalisations for boundary detection, all defined in the case of a single response per area and often in the presence of dissimilarity metrics between area-specific covariates. The logistic Multivariate CAR prior in \eqref{eqn:logMCAR_def} adapts and extends CAR priors to the context of multiple observations per area, while prior \eqref{eqn:edgeprior} controls boundary detection without any kind of covariates. Alternative distributions to CAR priors in case of areal data include the DAGAR priors, both univariate \citep{datta2019spatial} and multivariate. Recently, \cite{aiello2023detecting} have extended this prior to boundary detection in the case of multivariate areal data. DAGAR priors might provide a computational advantage  over CAR priors, since the undirected graph $G$ is replaced by a directed acyclic graph (DAG), which makes posterior inference faster. However, the definition of this DAG depends on an arbitrary topological ordering of the areas. In the case of irregular grids, as the one we consider here, there are no theoretical guarantees that posterior inference is invariant w.r.t. the ordering of the nodes. Moreover, the adoption of CAR priors yields an efficient MCMC sampling scheme for all parameters without introducing any approximation in $G$, whereas a DAGAR-like prior in our context would have required a much more complex update for the graphical structure. We proved that our model detects boundaries when the associated estimated areal densities for the variable of interest (the log-income) are different, even when summary statistics are similar. Simpler Bayesian models applied to area-specific empirical quantiles can only detect boundaries driven by those summary statistics themselves or driven by dissimilarities based on extra covariates, which might not be available in general.

As mentioned earlier, our model does not require area-specific covariates to inform the boundary detection. However, additional information from covariates can be easily incorporated into the model. For instance, integrating area-specific covariates can inform the boundary detection process by replacing \eqref{eqn:edgeprior} with a probit or logit regression model
\begin{equation*}
    \mathbb{P}(G_{i,k} = 1) = g(\bm \beta^T \Phi(\bm x_i, \bm x_k)),
\end{equation*}
where $\Phi(\bm x_i, \bm x_k)$ is a vector-valued dissimilarity function based on covariates $\bm x_i$, $\bm x_k$ of areas $i$ and $k$ (e.g., $\Phi(\bm x_i, \bm x_k)_\ell = \lvert x_{i, \ell} - x_{k, \ell} \rvert$) and $\bm{\beta}$ is a (random) parameter; see, for instance, \cite{lee.mitchell2012}. Individual and area-specific covariates can be easily incorporated into the model by modifying the mixture kernel as follows
\begin{equation*}
    y_{i,j} \mid \bm w_{(i)}, \bm \beta, \bm \sigma, H \iid \sum_{h=1}^H w_{i, h} \mathcal{N}(\cdot \mid \beta_h^T x_{i, j}, \sigma^2_h).
\end{equation*}
This mixture model can be considered as a finite-dimensional version of the \virgolette{single-weight} dependent Dirichlet process mixture \citep{quintana2022dependent}. This adaptability underscores our model's versatility in accommodating diverse data sources for robust boundary detection.

Posterior inference requires reversible jump MCMC moves, whose computational challenges are well-known. The high dimensionality of the proposal distribution implies that the mixing of the chain for $H$ is not extremely good. However, we propose an algorithm that does not get stuck in local modes of $H$ and, regardless of the initial value, always yields the same posterior distribution. Moreover, by combining the \emph{optimal proposal} of \cite{norets2021optimal} and an efficient \texttt{C++} code implementation, available via the \texttt{R} package \texttt{SPMIX}, we were able to scale inference up to a dataset with almost $80,000$ observations while keeping a reasonable runtime.

We need to point out that, to obtain an accurate MCMC approximation for posterior inference with such a number of observations, we need a very large number of iterations of our MCMC algorithm, and this requires an extended run-time. Consequently, scaling our approach to massive datasets with possibly millions of observations or high-dimensional parameters seems non-trivial. In particular, we believe that adapting ideas from \cite{miller2018mixture} and \cite{argiento2022infinity} can lead to the development of other spatially-dependent priors for which posterior inference is more efficient (in terms of chain mixing) and less demanding (but still likely unfeasible for millions of data). Another alternative is to resort to Consensus Monte Carlo algorithms. The key idea of these algorithms is to split the data into subsets (usually called shards), perform posterior inference through MCMC in each subset in parallel and then combine the posterior inferences. This is an approximate method that avoids full posterior simulation with the full dataset; since the MCMC algorithms in each shard use only a portion of the whole dataset, they show a better mixing. The resulting algorithm scales much better w.r.t. the size of the dataset, since the most challenging parts can be executed in parallel and with no communication between cores. However, there is little or no work for Consensus Monte Carlo algorithms in the context of spatial data in the literature so far.

\section*{Acknowledgments}
The authors acknowledge the support by MUR, grant Dipartimento di Eccellenza 2023-2027. Alessandra Guglielmi has also been supported by MUR - Prin 2022 - Grant no. 2022CLTYP4, funded by the European Union - Next Generation EU. Mario Beraha received funding from the European Research Council (ERC) under the European Union's Horizon 2020 research and innovation programme under grant agreement No 817257.

\section*{Supplementary Material}
The (online) Supplementary Material (SM) contains insights about the borrowing of strength induced by our logistic multivariate CAR prior in \Cref{sec:borrow}. Full details of the MCMC algorithm can be found in \Cref{sec:MCMC} of the SM. \Cref{sec:simstudies,sec:stress_tests} of the SM contain the simulation studies described in full details and the stress tests implemented to assess the robustness of our model when boundary detection can be challenging, respectively.
\Cref{sec:comparisons} of the SM reports a comparison with alternative models or empirical techniques for boundary detection: competitor models are described in \Cref{subsec:competitor_models}, while comparison of posterior inference for the simulated data and the California census income dataset are reported in \Cref{subsec:comparison_synthetic} and \Cref{subsec:comparison_calicensusdata} of the SM. In \Cref{sec:interpret_boundaries} in the SM, we investigate whether the estimated boundaries, at least for LA county, can be explained by available exogenous covariates. \Cref{sec:extra_plots_and_tables} of the SM reports additional plots and tables on posterior inference for the California census income dataset%

\bibliographystyle{apalike}
\bibliography{SPMIX}

@PREAMBLE{ {\providecommand\noopsort[1]{#1}} }

@article{beraha2021spatially,
  title={{Spatially dependent mixture models via the Logistic Multivariate CAR prior}},
  author={\noopsort{Beraha}, Mario and Pegoraro, Matteo and Peli, Riccardo and Guglielmi, Alessandra},
  journal={Spatial Statistics},
  volume={46},
  pages={100548},
  year={2021},
  publisher={Elsevier}
}

@article{green1995reversible,
  title={{Reversible jump Markov chain Monte Carlo computation and Bayesian model determination}},
  author={\noopsort{Green}, Peter J},
  journal={Biometrika},
  volume={82},
  number={4},
  pages={711--732},
  year={1995},
  publisher={Oxford University Press}
}

@article{lu2005areal,
  title   = {Bayesian areal wombling for geographical boundary analysis},
  author  = {\noopsort{Lu}, Haolan and Carlin, Bradley P.},
  journal = {Geographical Analysis},
  volume  = {37},
  number  = {3},
  pages   = {265--285},
  year    = {2005},
  doi     = {10.1111/j.1538-4632.2005.00624.x}
}

@article{stephens2000bayesian,
  title={{Bayesian analysis of mixture models with an unknown number of components - an alternative to reversible jump methods}},
  author={\noopsort{Stephens}, Matthew},
  journal={The Annals of Statistics},
  volume={28},
  pages={40--74},
  year={2000},
  publisher={JSTOR}
}

@article{argiento2022infinity,
  title={{Is infinity that far? A Bayesian nonparametric perspective of finite mixture models}},
  author={\noopsort{Argiento}, Raffaele and De Iorio, Maria},
  journal={The Annals of Statistics},
  volume={50},
  number={5},
  pages={2641--2663},
  year={2022},
  publisher={Institute of Mathematical Statistics}
}

@article{miller2018mixture,
  title={{Mixture models with a prior on the number of components}},
  author={\noopsort{Miller}, Jeffrey W and Harrison, Matthew T},
  journal={Journal of the American Statistical Association},
  volume={113},
  number={521},
  pages={340--356},
  year={2018},
  publisher={Taylor \& Francis}
}

@article{neal2000markov,
  title={{Markov chain sampling methods for Dirichlet process mixture models}},
  author={\noopsort{Neal}, Radford M},
  journal={Journal of computational and graphical statistics},
  volume={9},
  number={2},
  pages={249--265},
  year={2000},
  publisher={Taylor \& Francis}
}

@article{norets2021optimal,
  title={{Optimal auxiliary priors and reversible jump proposals for a class of variable dimension models}},
  author={\noopsort{Norets}, Andriy},
  journal={Econometric Theory},
  volume={37},
  number={1},
  pages={49--81},
  year={2021},
  publisher={Cambridge University Press}
}

@article{lee.mitchell2012,
  title={Boundary detection in disease mapping studies},
  author={\noopsort{Lee}, Duncan and Mitchell, Richard},
  journal={Biostatistics},
  volume={13},
  number={3},
  pages={415--426},
  year={2012},
  publisher={Oxford University Press}
}

@article{lee2013carbayes,
  title = {{CARBayes}: An {R} Package for {B}ayesian Spatial Modeling with Conditional Autoregressive Priors},
  author = {Duncan \noopsort{Lee}},
  journal = {Journal of Statistical Software},
  year = {2013},
  volume = {55},
  number = {13},
  pages = {1--24}
}

@article{paci2020structural,
  title={Structural learning of contemporaneous dependencies in graphical VAR models},
  author={\noopsort{Paci}, Lucia and Consonni, Guido},
  journal={Computational Statistics \& Data Analysis},
  volume={144},
  pages={106880},
  year={2020},
  publisher={Elsevier}
}

@article{rousseau.mengersen2011,
  title={Asymptotic behaviour of the posterior distribution in overfitted mixture models},
  author={\noopsort{Rousseau}, Judith and Mengersen, Kerrie},
  journal={Journal of the Royal Statistical Society: Series B (Statistical Methodology)},
  volume={73},
  number={5},
  pages={689--710},
  year={2011},
  publisher={Wiley Online Library}
}

@article{quintana2022dependent,
  title={The dependent Dirichlet process and related models},
  author={\noopsort{Quintana}, Fernando A and M{\"u}ller, Peter and Jara, Alejandro and MacEachern, Steven N},
  journal={Statistical Science},
  volume={37},
  number={1},
  pages={24--41},
  year={2022},
  publisher={Institute of Mathematical Statistics}
}

@article{li2015bayesian,
  title={Bayesian models for detecting difference boundaries in areal data},
  author={\noopsort{Li}, Pei and Banerjee, Sudipto and Hanson, Timothy A and McBean, Alexander M},
  journal={Statistica Sinica},
  volume={25},
  number={1},
  pages={385},
  year={2015},
  publisher={NIH Public Access}
}

@book{lauritzen1996graphical,
  title={Graphical models},
  author={\noopsort{Lauritzen}, Steffen L},
  year={1996},
  publisher={Clarendon Press}
}

@article{walker1969asymptotic,
  title={On the asymptotic behaviour of posterior distributions},
  author={\noopsort{Walker}, Andrew M},
  journal={Journal of the Royal Statistical Society: Series B (Methodological)},
  volume={31},
  number={1},
  pages={80--88},
  year={1969},
  publisher={Wiley Online Library}
}

@article{mohammadi2015bayesian,
  author = {A. \noopsort{Mohammadi} and E. C. Wit},
  title = {{Bayesian Structure Learning in Sparse Gaussian Graphical Models}},
  volume = {10},
  journal = {Bayesian Analysis},
  number = {1},
  pages = {109 -- 138},
  year = {2015}
}

@article{qu2021boundary,
  title={Boundary detection using a Bayesian hierarchical model for multiscale spatial data},
  author={\noopsort{Qu}, Kai and Bradley, Jonathan R and Niu, Xufeng},
  journal={Technometrics},
  volume={63},
  number={1},
  pages={64--76},
  year={2021},
  publisher={Taylor \& Francis}
}

@article{gao2022bayesian,
  title={Bayesian Models for Multivariate Difference Boundary Detection in Areal Data},
  author={\noopsort{Gao}, Leiwen and Banerjee, Sudipto and Ritz, Beate},
  journal={Biostatistics},
  volume={24},  
  number={4},
  pages={922--944},
  year={2023}
}

@inproceedings{leroux2000estimation,
  title={Estimation of disease rates in small areas: a new mixed model for spatial dependence},
  author={\noopsort{Leroux}, Brian G and Lei, Xingye and Breslow, Norman},
  booktitle={Statistical models in epidemiology, the environment, and clinical trials},
  pages={179--191},
  year={2000},
  organization={Springer}
}

@misc{uscb.aboutacs,   
  title = {About the American Community Survey},   
  author = {{\noopsort{United} States Census Bureau}},
  year = {2023},
  note = {URL: \url{https://www.census.gov/programs-surveys/acs/about.html}. Accessed: April 27, 2023} 
}

@misc{lac.opendata-portal,
    title = {{County of Los Angeles Open Data}},
    author = {\noopsort{County of Los Angeles}},
    year = {2023},
    note = {URL: \url{https://data.lacounty.gov/}. Accessed: April 27, 2023}
}

@article{hipp2007income,
  title={Income inequality, race, and place: Does the distribution of race and class within neighborhoods affect crime rates?},
  author={\noopsort{Hipp}, John R},
  journal={Criminology},
  volume={45},
  number={3},
  pages={665--697},
  year={2007},
  publisher={Wiley Online Library}
}

@article{kalli2011slice,
  title={Slice sampling mixture models},
  author={\noopsort{Kalli}, Maria and Griffin, Jim E and Walker, Stephen G},
  journal={Statistics and computing},
  volume={21},
  pages={93--105},
  year={2011},
  publisher={Springer}
}

@article{robert2014metropolis,
  title={The Metropolis--Hastings Algorithm},
  author={\noopsort{Robert}, Christian P},
  journal={Wiley StatsRef: Statistics Reference Online},
  pages={1--15},
  year={2014},
  publisher={Wiley Online Library}
}

@article{beraha2023normalised,
    author = {\noopsort{Beraha}, Mario and Griffin, Jim E},
    title = "{Normalised latent measure factor models}",
    journal = {Journal of the Royal Statistical Society Series B: Statistical Methodology},
    volume = {85},
    number = {4},
    pages = {1247-1270},
    year = {2023}
}

@book{ghosal2017fundamentals,
  title={Fundamentals of nonparametric Bayesian inference},
  author={\noopsort{Ghosal}, Subhashis and Van der Vaart, Aad},
  volume={44},
  year={2017},
  publisher={Cambridge University Press}
}

@article{polson2013bayesian,
  title={Bayesian inference for logistic models using P{\'o}lya--Gamma latent variables},
  author={\noopsort{Polson}, Nicholas G and Scott, James G and Windle, Jesse},
  journal={Journal of the American Statistical Association},
  volume={108},
  number={504},
  pages={1339--1349},
  year={2013},
  publisher={Taylor \& Francis}
}

@article{braveman2010socioeconomic,
  title={{Socioeconomic disparities in health in the United States: what the patterns tell us}},
  author={\noopsort{Braveman}, Paula A and Cubbin, Catherine and Egerter, Susan and Williams, David R and Pamuk, Elsie},
  journal={American Journal of Public Health},
  volume={100},
  number={S1},
  pages={S186--S196},
  year={2010},
  publisher={American Public Health Association}
}

@misc{aiello2023detecting,
  title={Detecting Spatial Health Disparities Using Disease Maps}, 
  author={Luca \noopsort{Aiello} and Sudipto Banerjee},
  year={2023},
  note={arXiv preprint arXiv:2309.02086}
}

@inproceedings{plummer2003jags,
  title={JAGS: A program for analysis of Bayesian graphical models using Gibbs sampling},
  author={\noopsort{Plummer}, Martyn and others},
  booktitle={Proceedings of the 3rd international workshop on distributed statistical computing},
  volume={124},
  pages={1--10},
  year={2003},
  organization={Vienna, Austria}
}

@article{datta2019spatial,
  title = {{Spatial Disease Mapping Using Directed Acyclic Graph Auto-Regressive (DAGAR) Models}},
  author = {Abhirup \noopsort{Datta} and Sudipto Banerjee and James S. Hodges and Leiwen Gao},
  journal = {Bayesian Analysis},
  volume = {14},
  number = {4},
  pages = {1221 -- 1244},
  year = {2019},
  publisher = {International Society for Bayesian Analysis}
}

@article{assunccao2006efficient,
  title={Efficient regionalization techniques for socio-economic geographical units using minimum spanning trees},
  author={\noopsort{Assun{\c{c}}{\~a}o}, Renato M and Neves, Marcos Corr{\^e}a and C{\^a}mara, Gilberto and da Costa Freitas, Corina},
  journal={International Journal of Geographical Information Science},
  volume={20},
  number={7},
  pages={797--811},
  year={2006},
  publisher={Taylor \& Francis}
}

@article{bivand2022r,
  title = {R Packages for Analyzing Spatial Data: A Comparative Case Study with Areal Data},
  author = {Roger \noopsort{Bivand}},
  journal = {Geographical Analysis},
  volume = {54},
  number = {3},
  pages = {488-518},
  year = {2022},
  publisher={Wiley Online Library}
}

@techreport{geverdt2024edge_sdgrf,
  title       = {Education Demographic and Geographic Estimates Program (EDGE): School District Geographic Relationship Files Technical Documentation},
  author      = {\noopsort{Geverdt}, Doug and Maselli, Annie},
  year        = {2024},
  institution = {U.S. Department of Education, National Center for Education Statistics},
  address     = {Washington, DC},
  urldate     = {2025-12-05}
}

@article{gelman1992inference,
  title={Inference from iterative simulation using multiple sequences},
  author={Gelman, Andrew and Rubin, Donald B},
  journal={Statistical science},
  volume={7},
  number={4},
  pages={457--472},
  year={1992},
  publisher={Institute of Mathematical Statistics}
}

@article{vehtari2021rank,
  title={Rank-normalization, folding, and localization: An improved $\hat{R}$ for assessing convergence of MCMC (with discussion)},
  author={Vehtari, Aki and Gelman, Andrew and Simpson, Daniel and Carpenter, Bob and B{\"u}rkner, Paul-Christian},
  journal={Bayesian analysis},
  volume={16},
  number={2},
  pages={667--718},
  year={2021},
  publisher={International Society for Bayesian Analysis}
}

@article{besag1991bayesian,
  title={Bayesian image restoration, with two applications in spatial statistics},
  author={Besag, Julian and York, Jeremy and Molli{\'e}, Annie},
  journal={Annals of the Institute of Statistical Mathematics},
  volume={43},
  number={1},
  pages={1--20},
  year={1991},
  publisher={Springer}
}

@article{franzolini2025multivariate,
  title={Multivariate species sampling models},
  author={Franzolini, Beatrice and Lijoi, Antonio and Pr{\"u}nster, Igor and Rebaudo, Giovanni},
  journal={arXiv preprint arXiv:2503.24004},
  year={2025}
}

\clearpage\appendix
\begin{center}
    \LARGE \textbf{Supplementary Material:\\Bayesian nonparametric boundary detection\\for multiple areal data}
\end{center}
\vspace{1cm}

\setcounter{section}{0}
\setcounter{equation}{0}
\setcounter{figure}{0}
\setcounter{table}{0}
\setcounter{page}{1}

\renewcommand{\thesection}{S\arabic{section}}
\renewcommand{\thefigure}{S\arabic{figure}}
\renewcommand{\thetable}{S\arabic{table}}
\renewcommand{\theequation}{S\arabic{equation}}

\makeatletter
    \renewcommand{\p@subfigure}{\thefigure}
    \renewcommand{\theHfigure}{S\arabic{figure}}
    \renewcommand{\theHtable}{S\arabic{table}}
    \renewcommand{\theHequation}{S\arabic{equation}}
\makeatother

\section{Borrowing of strength of the prior model}\label{sec:borrow}
\begin{proposition}
    \label{prop:borrowing_of_strength}
    Let $x_{i,j} := \tau_{s_{i,j}}$, where $s_{i,j}\in\{1,\ldots,H\}$ is the allocation
    variable of observation $j$ in area $i$. For $i \neq k$, define the pairwise tie probability as
    \begin{equation*}
        \tau_{i,k} := \mathbb{P}\left(x_{i,j}=x_{k,j'}\mid G, H,\rho,\sigma^2\right).
    \end{equation*}
    Let $A_G(\rho) = \left[\rho\bigl(\operatorname{diag}(G\mathbf 1)-G\bigr) + (1-\rho)I \right]^{-1}$ and define $a_{i,k}=[A_G(\rho)]_{i,k}$. For $H\geq 2$, let $V_{i,k}$ be the $(2H-2) \times (2H-2)$ matrix defined as
    \begin{equation*}
        V_{i,k} =
        \begin{pmatrix}
            a_{i,i}I_{H-1} & a_{i,k}I_{H-1} \\
            a_{k,i}I_{H-1} & a_{k,k}I_{H-1}
        \end{pmatrix}
    \end{equation*}
    Then,
    \begin{equation*}
        \resizebox{.99\textwidth}{!}{
        $\displaystyle\tau_{i,k} = \int_{\mathbb R^{2H-2}} \frac{1+\sum_{h=1}^{H-1}\exp(z_h+z'_h)}{\left(1+\sum_{h=1}^{H-1}\exp(z_h)\right)\left(1+\sum_{h=1}^{H-1}\exp(z'_h)\right)}\,\varphi_{2H-2}\left[
        \begin{pmatrix}
            z \\ z'
        \end{pmatrix}\bigg\lvert\ 
        0,\sigma^2 V_{i,k}\right]\,\mathrm{d}z\,\mathrm{d}z',$}
    \end{equation*}
    where $\varphi_d(\cdot \mid m,V)$ denotes the $d$-variate Gaussian density with mean
    $m$ and covariance matrix $V$. If $H = 1$, then $\tau_{i,k} = 1$.
\end{proposition}

\begin{proof}
    Note that by marginal exchangeability of the observations in the same population, the tie probability across groups $i$ and $k$, $\mathbb{P}\left(x_{i,j} = x_{k,j'} \mid G, H, \rho, \sigma^2\right)$ does not depend on the indices $\left(j,j'\right)$ and equals $\mathbb{P}\left(x_{i,1} = x_{k,1} \mid G, H, \rho, \sigma^2\right)$. Moreover, since the atoms are almost surely distinct, $x_{i,1}=x_{k,1}$ if and only if $s_{i,1}=s_{k,1}$. Hence, conditionally on the weights,
    \begin{equation*}
            \mathbb{P}\left(x_{i,1} = x_{k,1}\mid \w_i, \w_k, H\right) = \sum_{h=1}^{H}\,w_{i,h}\,w_{k,h}.
    \end{equation*}
    Taking expectation with respect to the logistic multivariate CAR prior gives the first identity. The integral expression follows by writing the weights as the inverse additive log-ratio transform of the Gaussian log-ratios. In particular,
    \begin{equation*}
        w_{i,h} = \frac{\exp(z_h)}{1+\sum_{\ell=1}^{H-1}\exp(z_\ell)}, \quad h=1,\ldots,H-1, \qquad w_{i,H} = \frac{1}{1+\sum_{\ell=1}^{H-1}\exp(z_\ell)},
    \end{equation*}
    and analogously for area $k$. Substituting these expressions into $\sum_{h=1}^H w_{i,h}\,w_{k,h}$ yields the displayed integrand.
\end{proof}

Equivalently, $\tau_{i,k}$ can be written as an expectation with respect to a pair of Gaussian log-ratio vectors. Let
\begin{equation*}
    \left(Z_i^\top,Z_k^\top\right)^\top \sim N_{2H-2} \left(0,\, \sigma^2 V_{i,k}\right),
\end{equation*}
and let $S(z)$ denote the inverse additive log-ratio map. Then
\begin{equation*}
    \tau_{i,k} = \mathbb{E}\left[S(Z_i)^\top S(Z_k)\right],
\end{equation*}
hence, $\tau_{i,k}$ can be studied by means of Monte Carlo simulation.
While this analytical representation clarifies which are the hyperparameters directly involved in tie probabilities across groups $\tau_{i,k}$, their effect on borrowing of strength is not analytically evident. Indeed, $\rho$ modifies the spatial covariance entries $a_{i,k}$ while $\sigma^2$ rescales both the marginal variability of the log-ratios and their cross-covariance. Hence, apart from limiting cases such as $\sigma^2 \to 0$, for which the weights concentrate around the uniform vector and $\tau_{i,k} \to 1/H$, monotonicity or explicit interpretations of the induced borrowing of strength are generally not available in closed form and are best assessed numerically.

\section{Sampling strategy in details}\label{sec:MCMC}
In this section, we give more details about the two major steps of the reversible jump MCMC algorithm, i.e., the \textit{between-models move} which consists in the joint update of $H$ and the corresponding parameter vector, and the \textit{within-model move} which implements, conditionally to $H$, a sampling scheme to update the vector of parameters $\bm{\theta}_{H}$, $\sigma^2$ and $G$, as introduced in \Cref{sec:algorithm} of the manuscript. 

\paragraph*{Within-model move}
The state of the MCMC, given the number of components $H$, is described by the latent allocation variables $\{s_{i,j}\}_{i,j}$ for $i=1,\dots,I$ and $j=1,\dots,N_{i}$, the common mixture atoms $\bm{\tau}=(\tau_{1},\dots,\tau_H)$, the transformed weights $(\wtilde_{1},\dots,\wtilde_{I})$ (where $\wtilde_{i}=\alr(\w_{i})$ for $i=1,\dots,I$), the graph $G$, the graph sparsity parameter $p$ and the common variance $\sigma^2$. The within-model move is a Gibbs sampler obtained by repeatedly sampling from the following conditional distributions:

\begin{itemize}
    \item Independently update the components of the common mixture atom vector from
    \begin{equation*}
        \pi\left(\tau_{h} \mid rest\right) \propto P_{0}\left(\tau_{h}\right) \textstyle\prod_{i,j\,:\,s_{i,j} = h}\,\mathcal{N}\left(y_{i,j} \mid \tau_{h}\right), \quad h=1,\dots,H. \\[5pt]
    \end{equation*}
    \item For $i = 1, \dots, I$, $j = 1, \dots, N_i$, independently update the cluster allocation variables from
    \begin{equation*}
        \pi\left(s_{i,j} = h \mid rest\right) \propto \invalr\left(w_{i,h}\right)\,\mathcal{N}\left(y_{i,j} \mid \tau_{h}\right), \quad h=1,\dots,H. \\[5pt]
    \end{equation*}
    \item For each $i=1,\dots,I$ and each $h = 1, \dots, H$ independently sample the transformed weight $\tilde{w}_{i,h}$ via the augmented Gibbs sampler. The augmentation technique goes as follows: we start from the full conditional for $\tilde{w}_{i,h}$,
    \begin{equation}
        \pi ( \tilde{w}_{i,h} \mid \tilde{\bm{W}}_{-(i,h)}, rest ) \propto \pi ( \tilde{w}_{i,h} \mid \tilde{\bm{W}}_{-(i,h)}, \rho, \sigma^2 ) \times \mathcal{L} ( \tilde{w}_{i,h} \mid \bm{s}_{i}, \tilde{\bm{w}}_{(i),-h} ),
        \label{eqn:wtilde_fullcond}
    \end{equation}
    where $\tilde{\bm{w}}_{(i),-h}$ denotes the vector $\wtilde_{i}$ once component $h$ has been removed. Now:
    \begin{equation*}
        \pi ( \tilde{w}_{i,h} \mid \tilde{\bm{W}}_{-(i,h)}, \rho, \sigma^2 ) \sim \mathcal{N}\left( \mu^{*}_{i,h}, \Sigma^{*}_{i,h}\right),
    \end{equation*}
    where
    \begin{align*}
        \mu^{*}_{i,h} &= \mu_{i,h} + \bm{\Sigma}_{h,-h}\bm{\Sigma}_{-h,-h}^{-1}(\tilde{\bm{w}}_{i,-h}-\bm{\mu}_{i,-h}),\\
        \Sigma^{*}_{i,h} &= (\rho\textstyle\sum_{k=1}^{I}G_{i,k} + 1 - \rho )^{-1} ( \Sigma_{h,h}-\bm{\Sigma}_{h,-h}\bm{\Sigma}_{-h,-h}^{-1}\bm{\Sigma}_{-h,h} ).
    \end{align*}
   The second  factor in the right-hand side of \eqref{eqn:wtilde_fullcond}
      can be written as
    \begin{align}
        \mathcal{L} ( \tilde{w}_{i,h} \mid \bm{s}_{i}, \tilde{\bm{w}}_{(i),-h} ) &= \frac{(\operatorname{e}^{\eta_{i,h}})^{N_{i,h}}}{(1+\operatorname{e}^{\eta_{i,h}})^{N_i}} \label{eqn:polyagammatrick} \\
        &= 2^{-N_i}\operatorname{e}^{(N_{i,h} - N_{i}/2)\eta_{i,h}} \int_{0}^{+\infty} \operatorname{e}^{-\omega\eta_{i,h}^2/2}\pi(\omega)\mathrm{d}\omega, \nonumber
    \end{align}
    where $\eta_{i,h} = \tilde{w}_{i,h} - \log\sum_{h' \neq h} \operatorname{e}^{\tilde{w}_{i,h'}}$, $N_i$ is the number of observations in area $i$ and $N_{i,h}$ is the observations in area $i$ assigned to component $h$. The second equivalence in \eqref{eqn:polyagammatrick} comes from \cite{polson2013bayesian} and $\pi(\omega)$ is the density of a Polya-Gamma distribution of parameters $(N_i, 0)$. Please, refer to the same paper for details about the definition and properties of Polya-Gamma random variables. Hence, introducing an auxiliary Polya-Gamma random variable $\omega_{i,h} \sim \operatorname{Polya-Gamma}(N_i, 0)$ for each $i$ and $h$, we can disintegrate the above measure w.r.t. $\omega_{i,h}$ and obtain closed-form expressions for the full conditionals of $\tilde{w}_{i,h}$ and $\omega_{i,h}$. Finally, the augmented Gibbs sampler step consists of the following steps: 
    \smallskip
    \begingroup
    \renewcommand{\theenumi}{\roman{enumi}}
    \begin{enumerate}
        \item sample the auxiliary Polya-Gamma random variable $\omega_{i,h}$ from
        \begin{equation*}
            \pi\left(\omega_{i,h} \mid \wtilde_{i} \right) = \operatorname{Polya-Gamma}\left(N_{i}, \tilde{w}_{i,h} - \log\textstyle\sum_{h' \neq h}\operatorname{e}^{\tilde{w}_{i,h'}}\right)
        \end{equation*}
        for each $i,h$;
        \smallskip
        
        \item sample the transformed weight $\tilde{w}_{i,h}$ from the augmented full conditional distribution
        \begin{equation*}
            \pi(\tilde{w}_{i,h} \mid \tilde{\bm{W}}_{-(i,h)}, \bm{s}_i, \sigma^2, \omega_{i,h} ) = \mathcal{N}(\hat{\mu}_{i,h}, \hat{\Sigma}_{i,h}),
        \end{equation*}
        where the parameters of the Gaussian random variables are the following:
        \begin{align*}
            \hat{\mu}_{i,h} &= \bigg(\frac{\mu^{*}_{i,h}}{\Sigma^{*}_{i,h}} + N_{i,h} - \frac{N_{i}}{2} + \omega_{i,h}\log{\textstyle\sum}_{h' \neq h}\operatorname{e}^{\tilde{w}_{i,h'}}\bigg) \bigg(\frac{1}{\Sigma_{i,h}^{*}} + \omega_{i,h}\bigg)^{-1}, \\
            \hat{\Sigma}_{i,h} &= \bigg(\frac{1}{\Sigma^{*}_{i,h}} + \omega_{i,h}\bigg)^{-1};
        \end{align*}
    \end{enumerate}
    \endgroup

    \item for any $(i,k) \in \Eadj$, independently sample edge $G_{i,k}$ from
    {\small
    \begin{align}
        \label{eqn:Gij_fullcond}
        \pi\left( G_{i,k} = 1 \mid rest \right) &\propto \exp \left\{ \log\left(\frac{p}{1-p}\right) + \frac{\rho}{2\sigma^2}\wtilde_i'\wtilde_k \right\}, &\pi\left(G_{i,k} = 0 \mid rest\right) &\propto 1;
    \end{align}
    }
    \item Sample the graph sparsity parameter $p$ from
    \begin{equation*}
        \pi \left( p \mid rest \right) = \textstyle\operatorname{Beta} \left( a + \sum_{(i,k) \in \Eadj} G_{i,k}, b + \lvert \Eadj \rvert - \sum_{(i,k) \in \Eadj} G_{i,k} \right);
    \end{equation*}

    \item sample the common variance $\sigma^2$ from
    \begin{equation*}
        \pi \left( \sigma^2 \mid rest \right) = \operatorname{Inv-Gamma}\left(\alpha_p / 2, \beta_p / 2\right),
    \end{equation*}
    where the posterior parameters are given by
    \begin{align*}
        \alpha_{p} &= \alpha + I(H-1), &
        \beta_{p} &= \beta + {\textstyle\sum}_{i,k=1}^{I} (F-\rho G)_{i,k}\wtilde_{i}'\wtilde_{k}.
    \end{align*}

\end{itemize}

\smallskip
\paragraph*{Between-models move}
The general idea behind the between-models move has already been presented in \Cref{sec:algorithm} of the paper. Here, we specify the details in the case of our model. Specifically, the conditional posterior distribution becomes the joint law $\pi( \wtilde^{H+1}, \tau_{H+1} \mid \bm{y}, H+1, \tilde{\bm{W}}, \bm{\tau})$. Such a distribution, in an implicit form, is proportional to
\begin{equation*}\resizebox{.98\hsize}{!}{$\displaystyle
    \prod_{i=1}^{I}\mathcal{L}\left(\bm{y}_{i} \mid \invalr\left(\wtilde_i, \tilde{w}_{i,H+1}\right), \left(\bm{\tau}, \tau_{H+1}\right), H+1 \right) \prod_{h=1}^{H+1} \pi\left(\wtilde^h \mid \sigma^2, G, H+1 \right) \pi\left(\tau_h \mid H+1\right).$}
\end{equation*}
Its Laplace approximation is a multivariate Gaussian distribution with parameters
\begin{align}
    \label{eqn:optmean}
    \bm{\mu}^* &= \argmax_{\left(\wtilde^{H+1}, \tau_{H+1}\right)} \log\pi(\wtilde^{H+1}, \tau_{H+1} \mid \bm{y}, H+1, \tilde{\bm{W}}, \bm{\tau}); \\
    \label{eqn:optcov}
    \bm{V}^* &= -\mathbf{H}^{-1}\left(\log\pi(\wtilde^{H+1}, \tau_{H+1} \mid \bm{y}, H+1, \tilde{\bm{W}}, \bm{\tau})\right)\left(\bm{\mu}^*\right),
\end{align}
where $\mathbf{H}\left(f\right)\left(\cdot\right)$ denotes the Hessian of function $f$ evaluated in $\cdot$.

The parameter vector $(\wtilde^{H+1}, \tau_{H+1})$, associated to the component we aim to propose is sampled from $\tilde{\pi}(\wtilde^{H+1}, \tau_{H+1} \mid \bm{\mu}^*, \bm{V}^*) \approx \mathcal{N}\left(\bm{\mu}^*, \bm{V}^*\right)$. While in the original reversible jump MCMC by \cite{green1995reversible}, the proposed state is generated by the composition of a random proposal distribution and a deterministic mapping function, in this work, we rely on a direct sample of the state of the new component.

We now focus on the computation of the reversible jump acceptance ratio with current state $(H, \bm{\theta}_H)$ and proposed state $(H', \bm{\theta}_{H'})$, with $H'$ equal to either $H+1$ or $H-1$. At each iteration, we either decide to increase or reduce the problem dimension with equal probability. To simplify notation, let $\tilde{\bm{W}}^+ = (\tilde{\bm{W}}, \wtilde^{H+1})$, $\bm{\tau}^+ = \left(\bm{\tau}, \tau_{H+1}\right)$. Moreover, when we write $\alr\left(\bm{A}\right)$ and $\bm{A}$ is a matrix, we mean that the additive log ratio transformation is applied to each row of matrix $\bm{A}$. Then, in case $H' = H+1$, the move is accepted with probability $\min\left(1, A_{H,H+1}\right)$, where
\begin{equation}\resizebox{.9\hsize}{!}{$\displaystyle
    A_{H,H+1} = \frac{\mathcal{L}\left(\bm{y} \mid \invalr(\tilde{\bm{W}}^+), \bm{\tau}^+, H+1 \right) \pi\left(\tilde{\bm{W}}^+, \bm{\tau}^+ \mid \sigma^2, G, H + 1\right)\pi\left(H+1\right)}{\mathcal{L}\left(\bm{y} \mid \invalr(\tilde{\bm{W}}), \bm{\tau}, H \right) \pi\left(\tilde{\bm{W}}, \bm{\tau} \mid \sigma^2, G, H \right)\pi\left(H\right)} \times \frac{1}{\tilde{\pi}\left(\wtilde^{H+1}, \tau_{H+1} \mid \bm{\mu}^*, \bm{V}^*\right)}.$}
    \label{eqn:increasearate}
\end{equation}
On the other hand, in case $H' = H - 1$, a random component $r$ between $1$ and $H-1$ is selected as a candidate for removal. Then, we define $\tilde{\bm{W}}^-$ as the matrix $\tilde{\bm{W}}$ after column $r$ has been removed and,  similarly, we define the vector $\bm{\tau}^-$. Then, the move is accepted with probability $\min\left(1, A_{H,H-1}\right)$, where
\begin{equation}\resizebox{.9\hsize}{!}{$\displaystyle
    A_{H,H-1} = \frac{\mathcal{L}\left(\bm{y} \mid \invalr(\tilde{\bm{W}}^-), \bm{\tau}^-, H-1 \right) \pi\left(\tilde{\bm{W}}^-, \bm{\tau}^- \mid \sigma^2, G, H-1\right)\pi\left(H-1\right)}{\mathcal{L}\left(\bm{y} \mid \invalr(\tilde{\bm{W}}), \bm{\tau}, H \right) \pi\left(\tilde{\bm{W}}, \bm{\tau} \mid \sigma^2, G, H \right)\pi\left(H\right)} \times \tilde{\pi}\left(\wtilde^{H}, \tau_{H} \mid \bm{\mu}^*, \bm{V}^*\right).$}
    \label{eqn:reducearate}
\end{equation}

\section{Simulation studies}
\label{sec:simstudies}
This section provides extensive simulation studies designed to evaluate the performance of the proposed model across three distinct tasks: joint spatial density estimation, structural learning of the underlying graph, and boundary detection. Namely, \Cref{subsec:de_simstudy} examines the model's ability to recover spatially varying densities focusing on scenarios with small sample sizes per area. \Cref{subsec:sl_simstudy} focuses on structural learning and investigates how different prior specifications for the graph sparsity and spatial strength affect the estimation of the adjacency structure, closely related to the boundary detection, ultimate goal of this model and manusctipt. Finally, \Cref{subsec:bd_simstudy} focuses on boundary detection and presents a large-scale simulation study across 50 independent replicates. This study provides a robust assessment of boundary detection performance, using metrics such as Precision, Sensitivity, and AUC, while evaluating the sensitivity of the results to the number of mixture components and spatial dependence parameters. This last section also  provides detailed MCMC diagnostics and a short discussion about the computational efficiency of the algorithm described in \Cref{sec:algorithm} of the manuscript and fully detailed in \Cref{sec:MCMC} here.

\subsection{Simulation Study - joint spatial density estimation}
\label{subsec:de_simstudy}
\begin{figure}[t]
    \centering
    \subfigure[$\w^1$\label{subfig:DEsimstudy_w1}]{\includegraphics[width=0.31\textwidth]{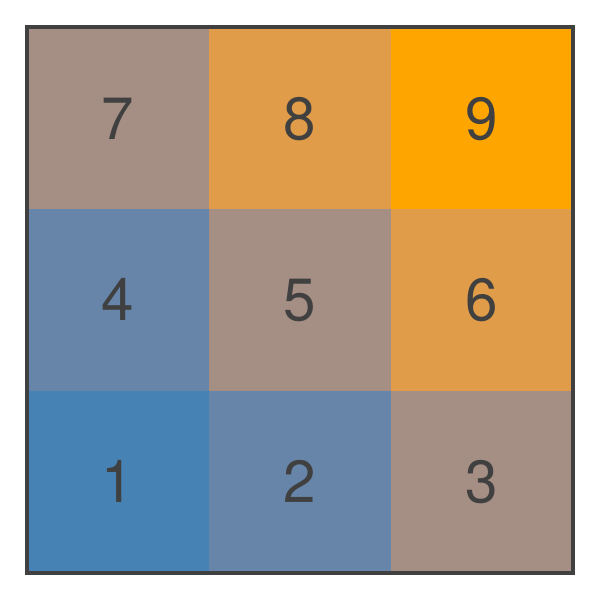}}%
    \hfill
    \subfigure[$\w^2$\label{subfig:DEsimstudy_w2}]{\includegraphics[width=0.31\textwidth]{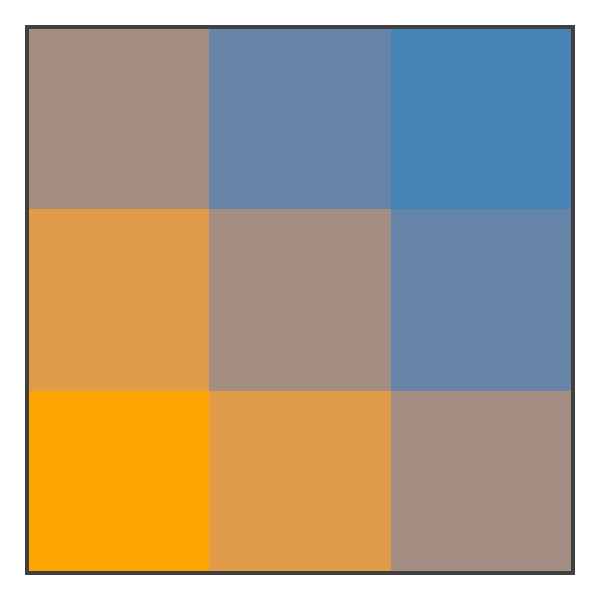}}%
    \hfill
    \subfigure[Adjacency graph\label{subfig:DEsimstudy_G}]{\includegraphics[width=0.31\textwidth]{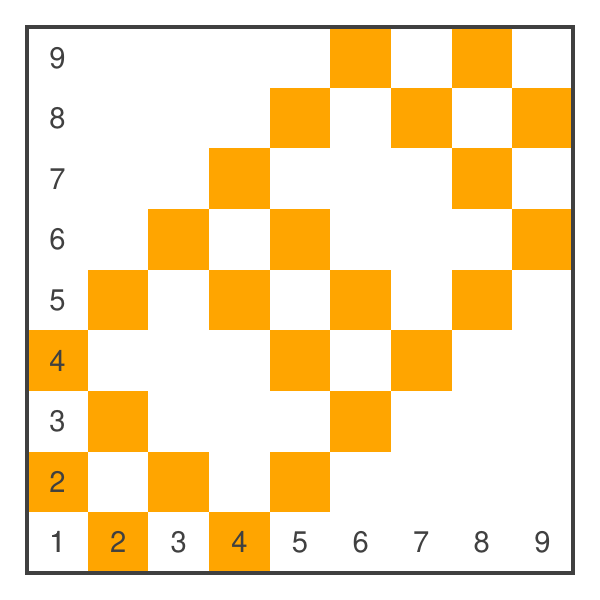}}%
    \caption{Simulation from spatially dependent weights in \Cref{subsec:de_simstudy}: (a) and (b) shows the values of $w_{i,1}$ and $w_{i,2}$ for each area. (c) represent the adjacency graph, where orange squares denote couples of geographically contiguous areas.}
    \label{fig:DEsimstudy_setup}
\end{figure}

We consider $I=9$ areas, obtained by splitting a square unit area domain into equal squared areas (see \Cref{subfig:DEsimstudy_w1}); for any area $i=1,\ldots,9$, observations are simulated independently as:
\begin{equation*}
    y_{i,j} \iid w_{i,1}\,\mathcal{N}\left(-5,1\right) + w_{i,2}\,\mathcal{N}\left(0,1\right) + w_{i,3}\,\mathcal{N}\left(5,1\right) \quad j = 1, \dots, 100.
\end{equation*}
Note that the number of samples $N_i$ in each area $i$ is relatively small ($N_i=100$), so the sharing of information between the areas will be essential. The vectors of weights $\w_{i}$, $i=1,\dots,I$, are fixed as $\alr\left(\wtilde_i\right)$ (recall \eqref{eqn:alr} in the manuscript), where the transformed weights $\wtilde_i$'s are defined as
\begin{equation}
    \label{eqn:DEsimstudy_spatialweights}
    \tilde{w}_{i,1} = 3(x_{i}-\bar{x}) + 3(y_{i}-\bar{y}) \qquad
    \tilde{w}_{i,2} = -3(x_{i} - \bar{x}) - 3(y_{i} - \bar{y}),
\end{equation}
$(x_{i},y_{i})$ are the coordinates of the centre of area $i$ and $(\bar{x},\bar{y})$ the coordinates of the grid centre. It is clear that, in this simulated scenario, there is strong spatial dependence, induced by \eqref{eqn:DEsimstudy_spatialweights}, among the weights of different areas, as we can also see from \Cref{subfig:DEsimstudy_w1,subfig:DEsimstudy_w2}. We consider areas $i$ and $k$ to be geographically contiguous if they share an entire edge
(see \Cref{subfig:DEsimstudy_G}).

We fit model \eqref{eqn:data_in_clust}-\eqref{eqn:prior_H} in the manuscript to this simulated dataset, and fix the hyperparameters as follows: for $P_0$ as in \eqref{eqn:P_0} in the manuscript, we fix  $\mu_0 = 0, \lambda = 0.1, c = 2$ and $d = 2$, thus assigning  (marginal) vague priors to the means and to the variances of the mixture atoms; the prior hyperparameters associated to the across-areas variance $\sigma^2$ are $\alpha = \beta = 2$, so that we set a priori an infinite second moment, yielding to a marginal vague prior on this parameter too. Since, in this scenario, the focus is on density estimation, we assume $G$ fixed and equal to the adjacency graph. We run our reversible jump sampler for a total of $10,000$ iterations, discard the first half (i.e., the burn-in phase), and save $5,000$ draws to approximate the posterior distribution.

\begin{figure}[t]
    \centering
    \subfigure[\label{subfig:DEsimstudy_postH}]{\includegraphics[width=0.25\textwidth]{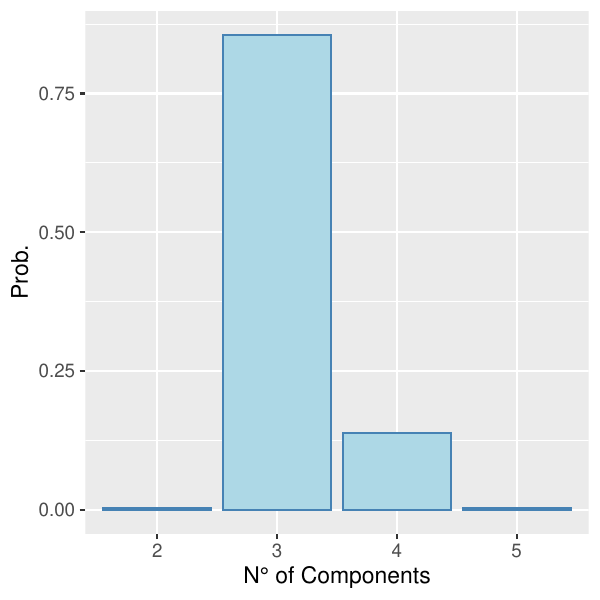}}%
    \subfigure[\label{subfig:DEsimstudy_traceH}]{\includegraphics[width=0.25\textwidth]{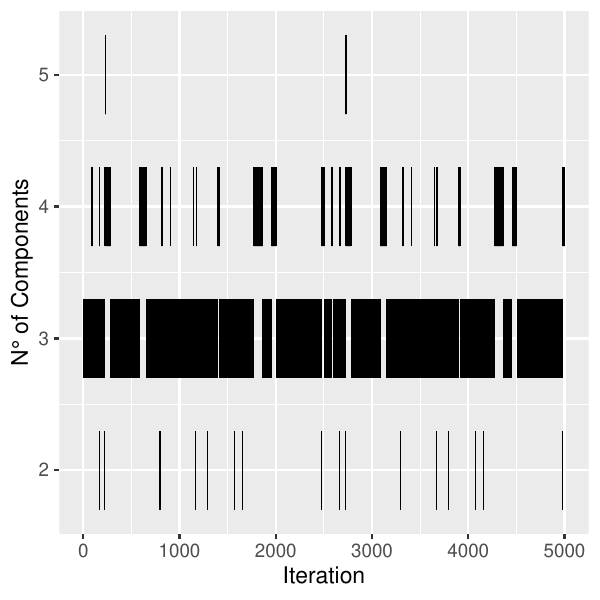}}%
    \subfigure[\label{subfig:DEsimstudy_area1}]{\includegraphics[width=0.25\textwidth]{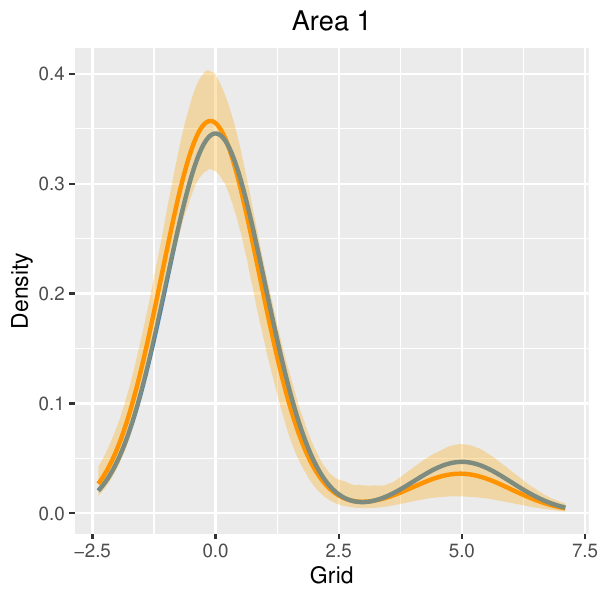}}%
    \subfigure[\label{subfig:DEsimstudy_area7}]{\includegraphics[width=0.25\textwidth]{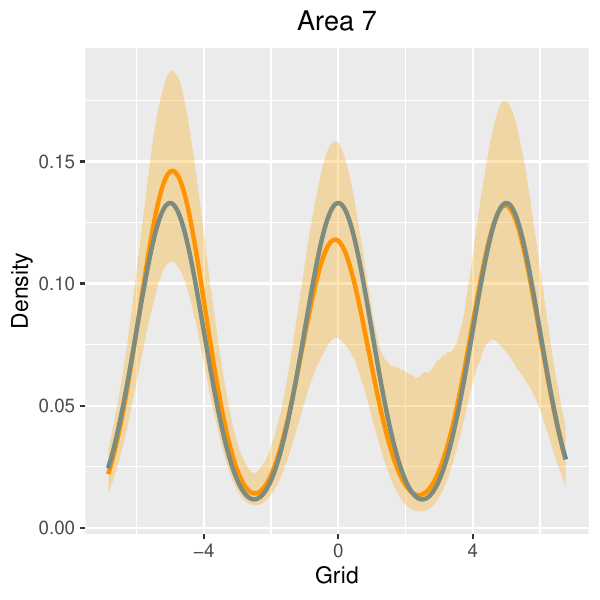}}%
    \caption{Posterior inference on the simulated dataset from spatially dependent weights in \Cref{subsec:de_simstudy} under default parameters: (a) Posterior distribution of $H$; (b) Traceplot of $H$; (c -- d) Comparison between the true (blue line) and estimated densities (orange line) in two areas. The orange ribbon represents the $95\%$ credibility band for the estimated densities.} %
    \label{fig:DEsimstudy_postInference}
\end{figure}

\Cref{subfig:DEsimstudy_postH,subfig:DEsimstudy_traceH} report the posterior distribution of $H$ and its associated traceplot. \Cref{subfig:DEsimstudy_area1,subfig:DEsimstudy_area7} display a comparison between the true (blue line) and estimated densities (orange line) in two areas. We also provide the $95\%$ credible bands (the orange shadow) for the estimated densities. Our model recovers the true number of components in the mixtures and the true densities themselves quite well.

\subsection{Simulation study - misspecified structural learning} \label{subsec:sl_simstudy}
In this section, we describe a simulation study in which we focus on the prior of the graph $G$, and we test the model for structural learning. By \virgolette{structural learning}, we mean the estimation, from the available data, of the underlying (undirected) graph that models the dependence among observations. We consider the posterior distribution of the graph $G$ in the model, with the graph density parameter $p$ fixed and with $p$ Beta-distributed. We consider $6$ different areal locations with $100$ observations each. The true graph we aim at retrieving is $G_{true} = \{(1,2), (3,4), (5,6)\}$.

We simulate data in areas $1$ and $2$ from a Student's $t$ distribution with $6$ degrees of freedom, mean $-4$ and standard deviation $1$; data in areas $3$ and $4$ are sampled from a Skew-Normal distribution of parameter $\left(\xi=4,\omega=4,\alpha=1\right)$; data in areas $5$ and $6$ come from a $\chi^2$ distribution with $3$ degrees of freedom. We run the sampler for a total of $10,000$ iterations, half of them used as the burn-in phase. Hence, the final sample size is $5,000$.
First of all, note that the model is misspecified, i.e., the true data generating distributions come from a parametric density which is not included in the likelihood \eqref{eqn:modeldata} in the manuscript.
Moreover, the true densities in areas $\{3,4,5,6\}$ are all concentrated on a partially overlapping interval of values. We then expect that the input data in these locations would make the structural learning procedure more complicated.

We consider five scenarios: $(i)$ $p$ to be fixed to values $\{0.1, 0.2, 0.3\}$; $(ii)$ $p \sim \operatorname{Beta}\left(a, b\right)$ with $\left(a, b\right) \in \{\left(1,5\right), \left(2,5\right), \left(2,2\right)\}$. Then, we focus on the sensitivity of parameters directly involved in the full conditional of $G_{i,k}$, i.e., $\rho$ and $\sigma^2$ (see \eqref{eqn:Gij_fullcond}). We set $p \sim \operatorname{Beta}\left(1, 5\right)$ and we consider: $(iii)$ $\sigma^2 \sim \invGamma\left(3,2\right)$, $\rho = \{0.90, 0.95, 0.99\}$; $(iv)$ $\rho = 0.99$, $\sigma^2 \sim \invGamma\left(\frac{1}{\nu} + 2, \frac{1}{\nu} + 1\right)$, with $\nu = \{0.5, 1, 2\}$; $(v)$ $\rho = 0.99$, $\sigma^2 \sim \invGamma(2(m^2 + 1), m(2m^2 + 1))$, with $m = 1, 2, 5$. The prior hyperparameters in $(iii)$ assess the effect of the global spatial strength $\rho$ on the posterior probabilities of edge inclusion $\mathbb{P}(G_{i,k} = 1 \mid \bm{y})$. On the other hand, prior hyperparameters in $(iv)$ and $(v)$ evaluate, respectively, the effects of $\nu$ and $m$, marginal prior variance and mean of $\sigma^2$, on $\mathbb{P}(G_{i,k} = 1 \mid \bm{y})$. In fact, notice that $\mathbb{E}\left[\sigma^2\right] = 1$, $\Var\left(\sigma^2\right) = \nu$ in $(iv)$ and $\mathbb{E}\left[\sigma^2\right] = m$, $\Var\left(\sigma^2\right) = 0.5$ in set $(v)$.

\begin{figure}[t]
    \centering
    \subfigure[\scriptsize Scenario $(i)$, $p = 0.1$]{\includegraphics[width=.3\textwidth]{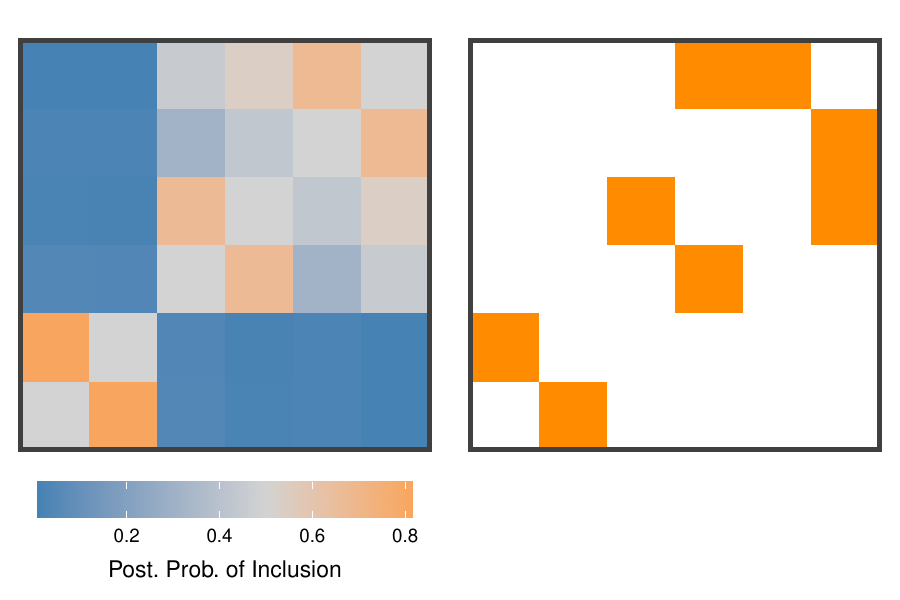}}%
    \quad
    \subfigure[\scriptsize Scenario $(i)$, $p = 0.2$]{\includegraphics[width=.3\textwidth]{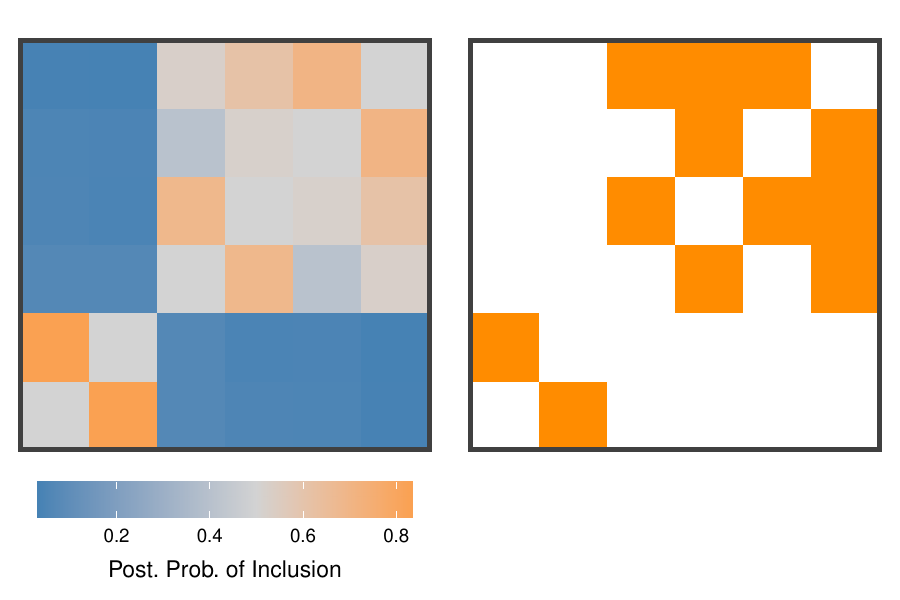}}%
    \quad
    \subfigure[\scriptsize Scenario $(i)$, $p = 0.3$]{\includegraphics[width=.3\textwidth]{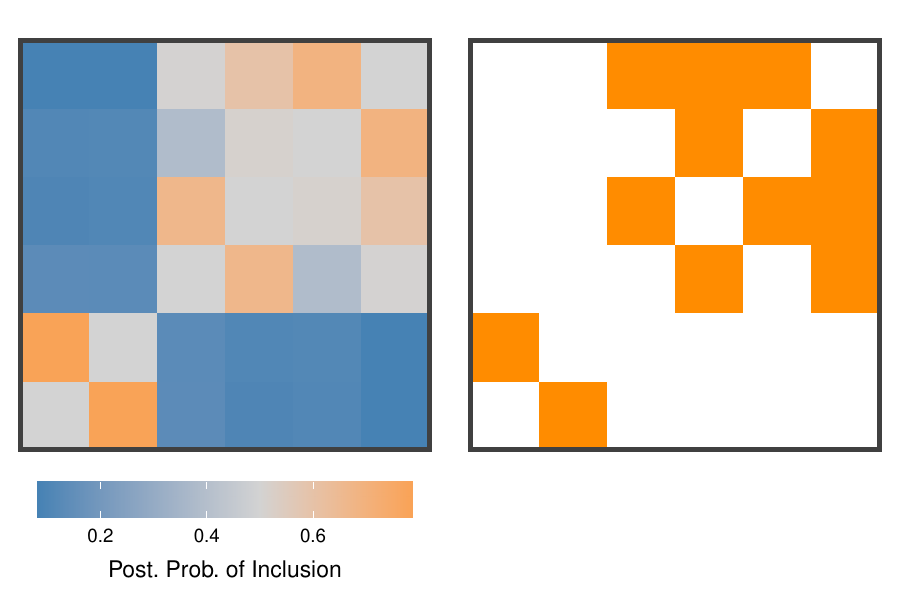}}%
    \\[-8pt]
    \subfigure[\scriptsize Scenario $(ii)$, $p \sim \operatorname{Beta}\left(1,5\right)$]{\includegraphics[width=.3\textwidth]{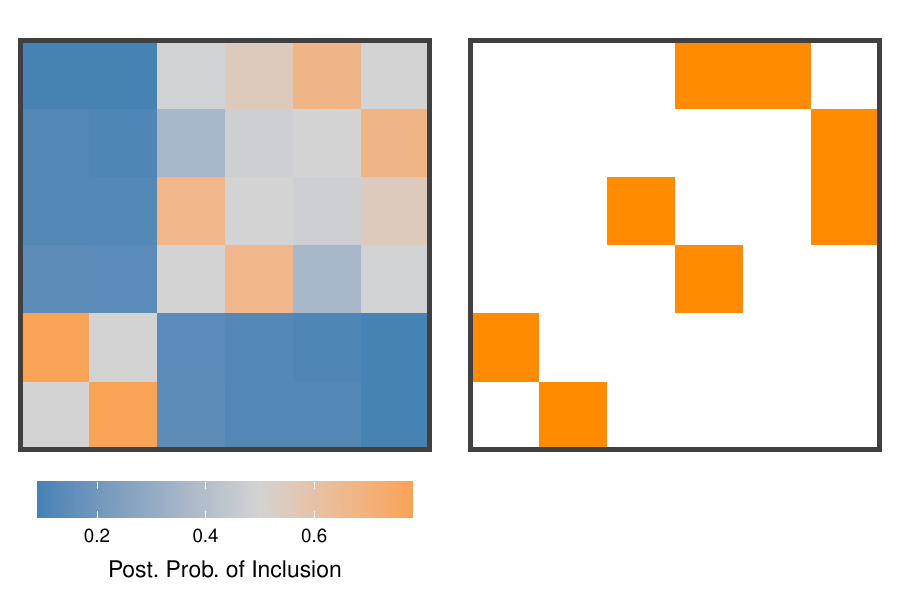}}%
    \quad
    \subfigure[\scriptsize Scenario $(ii)$, $p \sim \operatorname{Beta}\left(2,5\right)$]{\includegraphics[width=.3\textwidth]{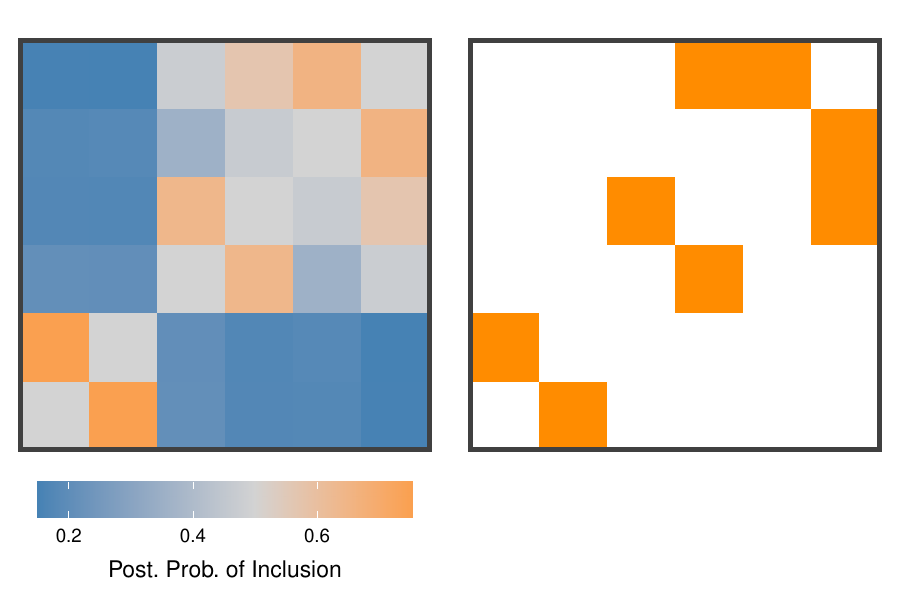}}%
    \quad
    \subfigure[\scriptsize Scenario $(ii)$, $p \sim \operatorname{Beta}\left(2,2\right)$]{\includegraphics[width=.3\textwidth]{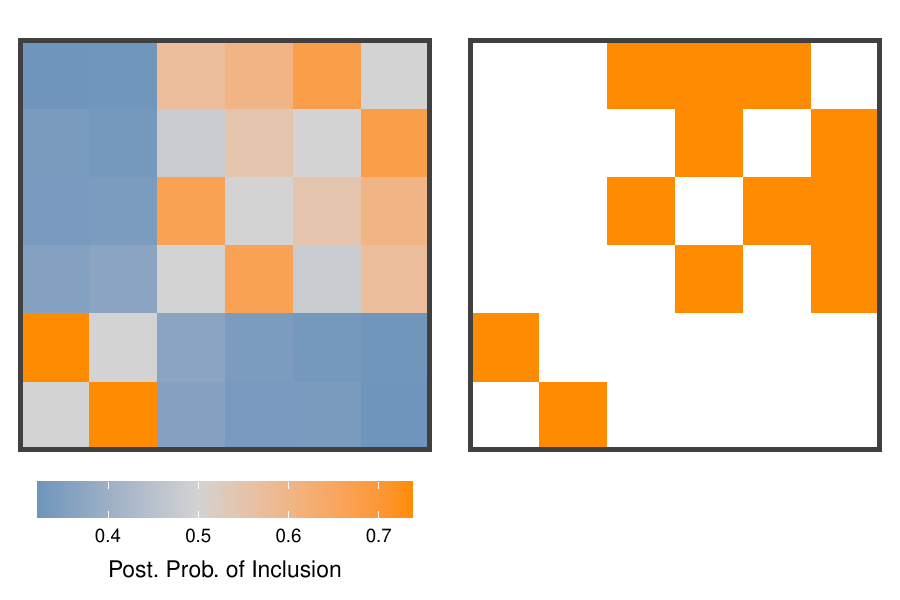}}%
    \\[-8pt]
    \subfigure[\scriptsize Scenario $(iii)$, $\rho = 0.9$]{\includegraphics[width=.3\textwidth]{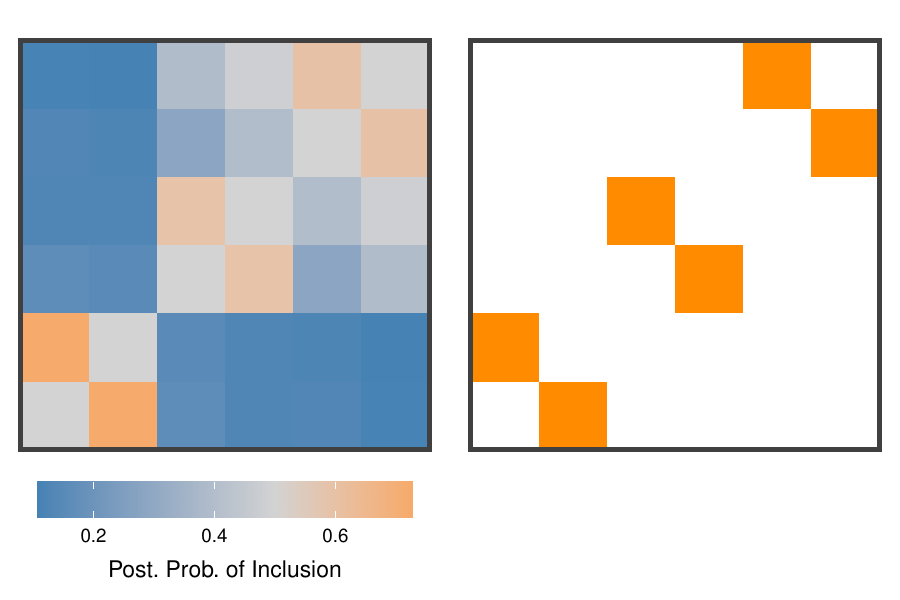}}%
    \quad
    \subfigure[\scriptsize Scenario $(iii)$, $\rho = 0.95$]{\includegraphics[width=.3\textwidth]{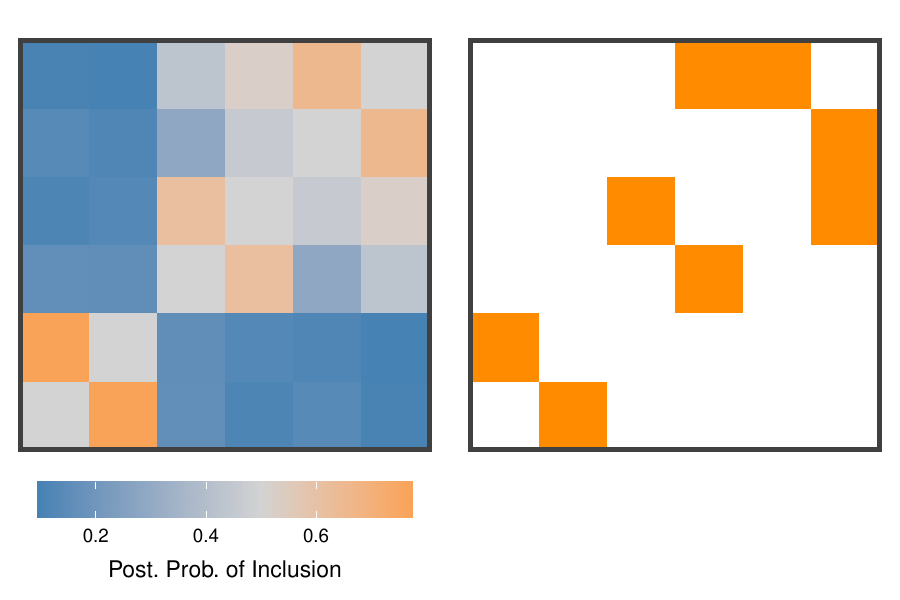}}%
    \quad
    \subfigure[\scriptsize Scenario $(iii)$, $\rho = 0.99$]{\includegraphics[width=.3\textwidth]{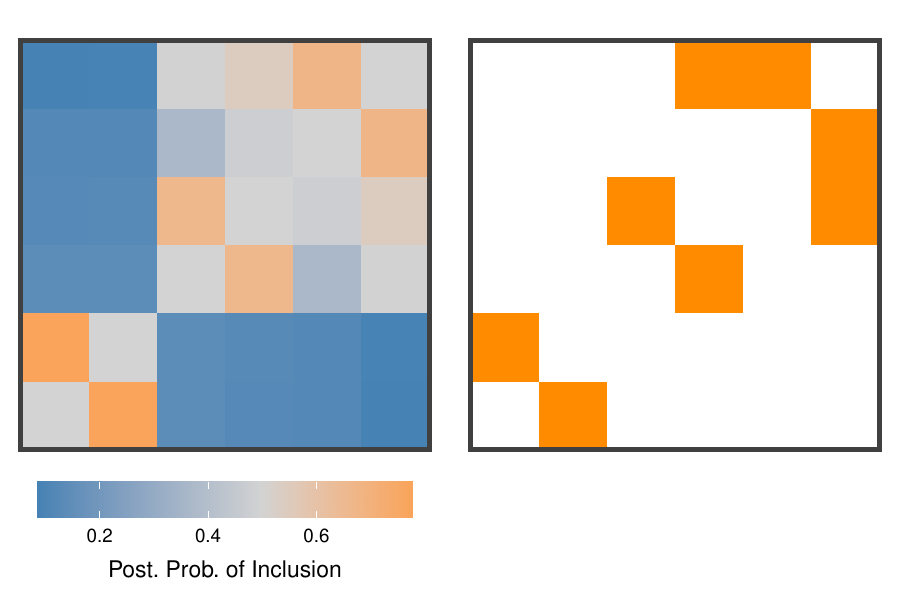}}%
    \\[-8pt]
    \subfigure[\scriptsize Scenario $(iv)$, $m = 1$]{\includegraphics[width=.3\textwidth]{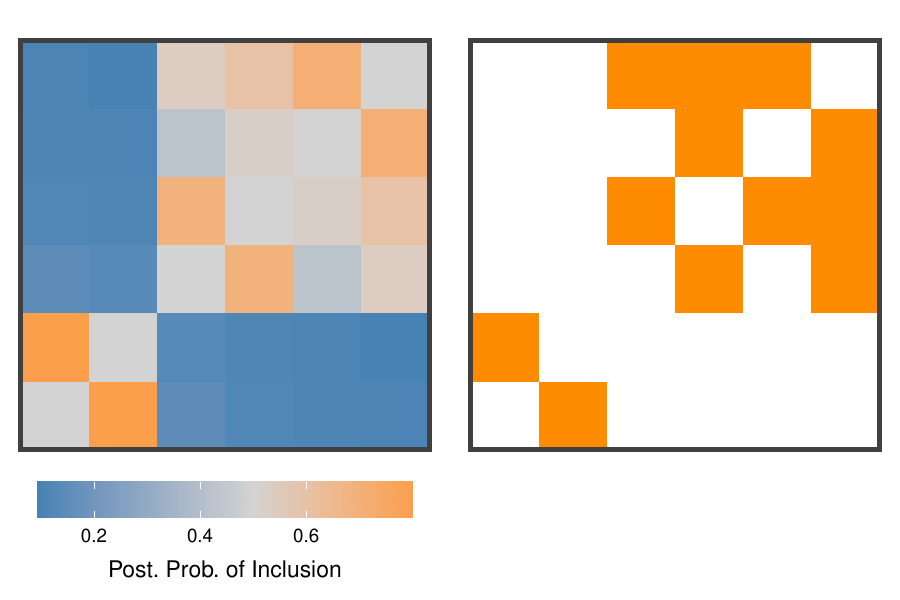}}%
    \quad
    \subfigure[\scriptsize Scenario $(iv)$, $m = 2$]{\includegraphics[width=.3\textwidth]{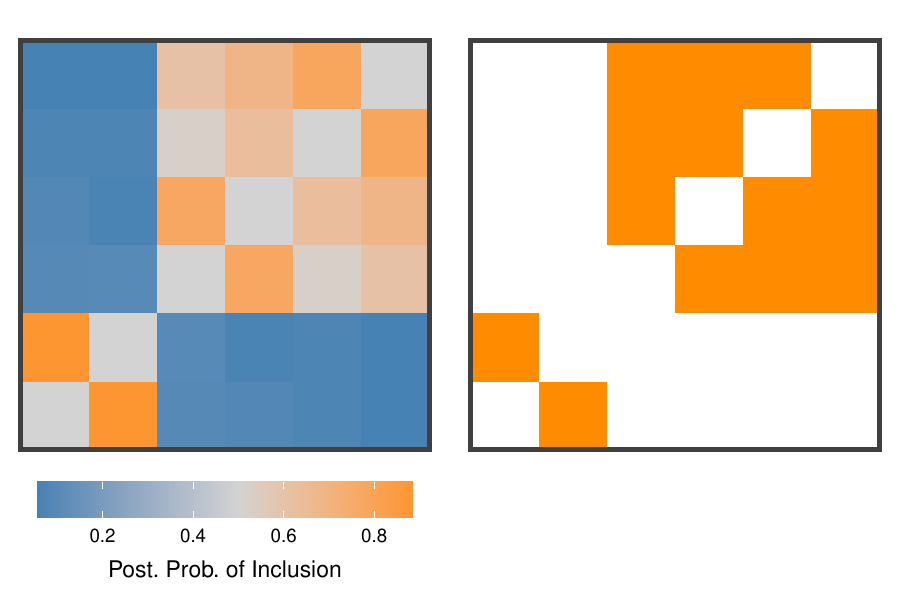}}%
    \quad
    \subfigure[\scriptsize Scenario $(iv)$, $m = 5$]{\includegraphics[width=.3\textwidth]{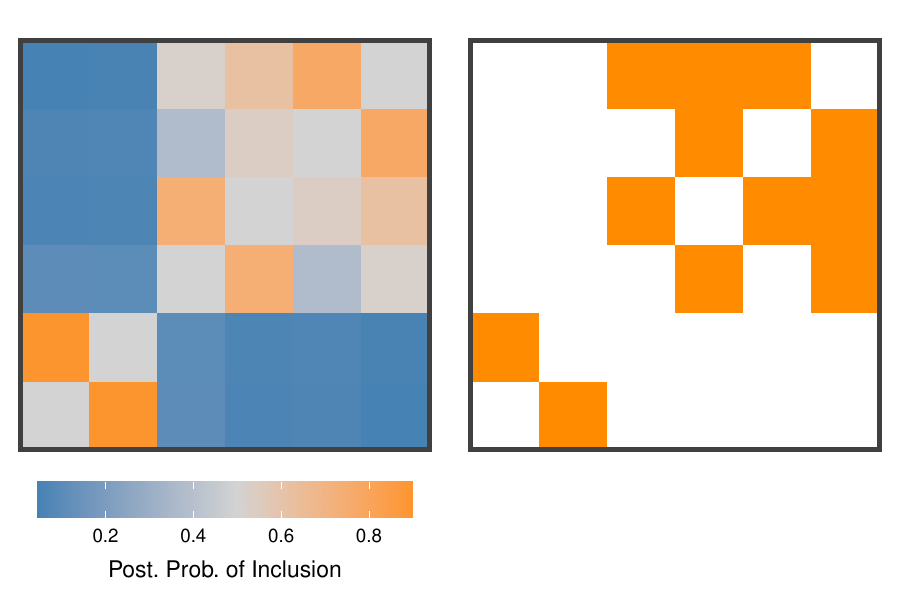}}%
    \\[-8pt]
    \subfigure[\scriptsize Scenario $(v)$, $\nu = 0.5$]{\includegraphics[width=.3\textwidth]{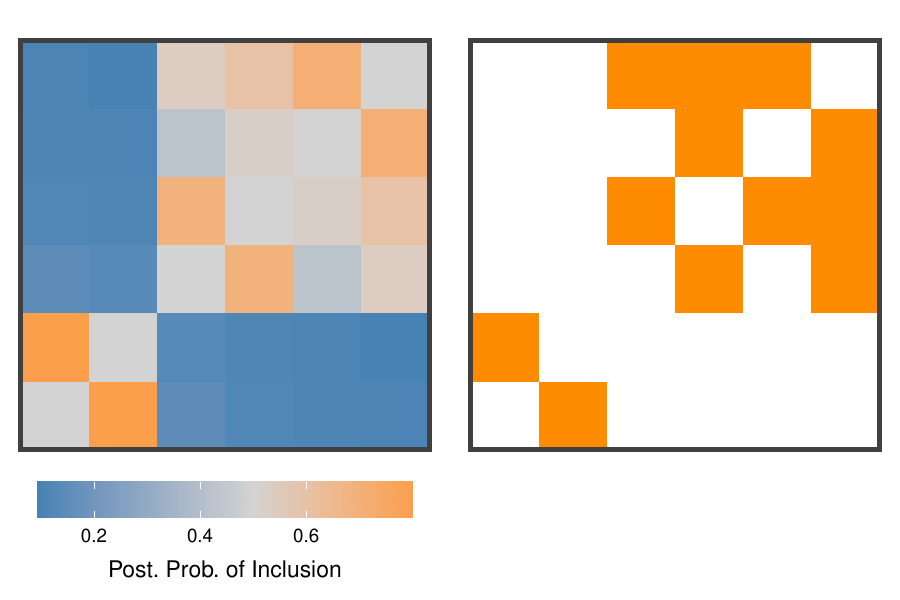}}%
    \quad
    \subfigure[\scriptsize Scenario $(v)$, $\nu = 1$]{\includegraphics[width=.3\textwidth]{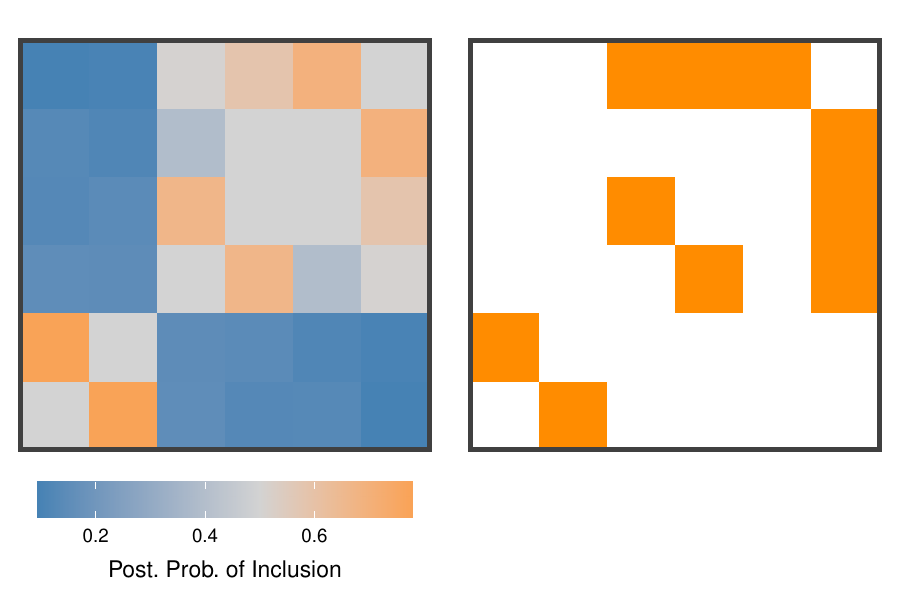}}%
    \quad
    \subfigure[\scriptsize Scenario $(v)$, $\nu = 2$]{\includegraphics[width=.3\textwidth]{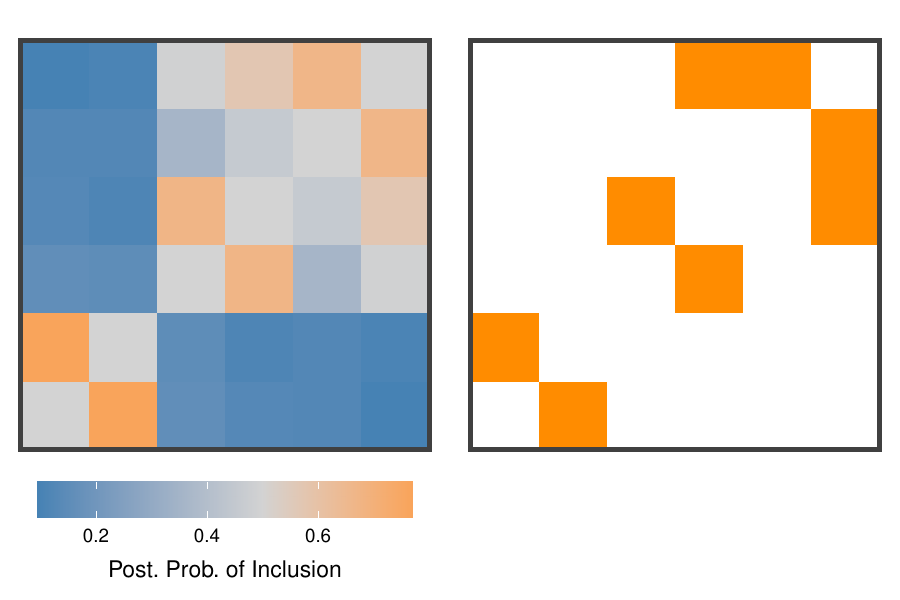}}
    \caption{Structural learning under misspecified regime (\Cref{subsec:sl_simstudy}): posterior probabilities of edge inclusion $\mathbb{P}(G_{i,k} = 1 \mid \bm{y})$ under different priors for $p$, $\rho$ and $\sigma^2$ and corresponding estimated neighbouring graph $\Ghatn$.}
    \label{fig:plt_plinks_scenario2_priorspecs}
\end{figure}

\Cref{fig:plt_plinks_scenario2_priorspecs} show the posterior probabilities of edge inclusion $\mathbb{P}(G_{i,k} = 1 \mid \bm{y})$ for each set of the prior hyperparameters as in scenarios $(i)-(v)$. Together with each posterior probability matrix, we also show the estimated neighbouring graph $\Ghatn$ with threshold $\gamma = 0.5$. We recall that the estimated neighbouring graph collects all those admissible edges $(i,k)$ for which $\mathbb{P}\left(G_{i,k} = 1 \mid \bm{y} \right) \geq \gamma$.

\subsection{Simulation Study - boundary detection under misspecification} \label{subsec:bd_simstudy}

\begin{figure}[t]
    \centering
    \subfigure[\label{subfig:scenario3_BDgroup}]{\includegraphics[width=.33\textwidth]{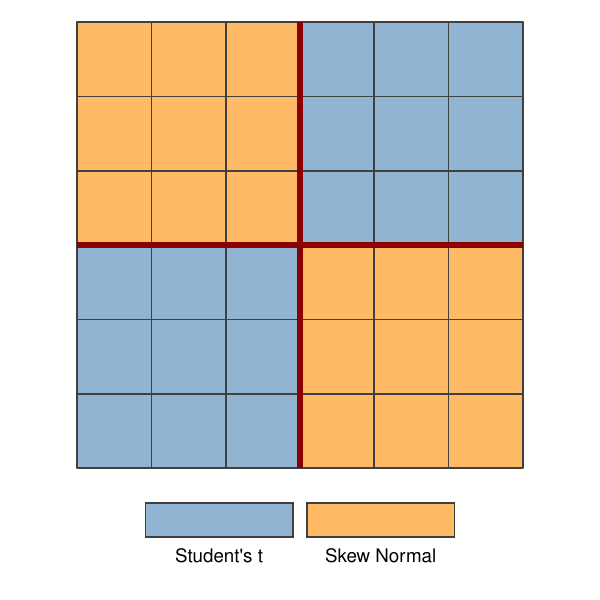}}%
    \hfill
    \subfigure[]{\includegraphics[width=.33\textwidth]{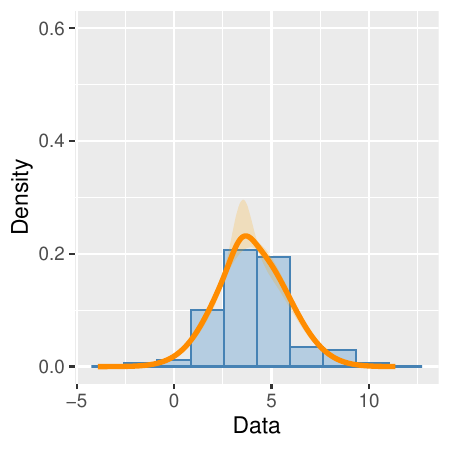}}%
    \hfill
    \subfigure[]{\includegraphics[width=.33\textwidth]{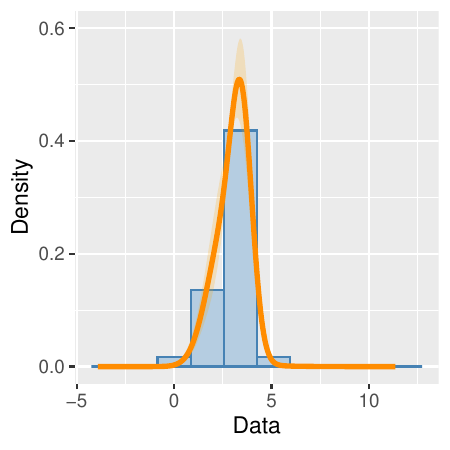}}%
    \caption{Posterior inference for the simulated dataset of \Cref{subsec:bd_simstudy}: (a) shows the spatial grid labelled according to the true data generating densities with the estimated boundary edges highlighted in red on the map; (b) and (c) report posterior estimated densities in two boundary areas, namely area 3 and area 4. The orange band represents the 95\% credible interval for the estimated density.}
    \label{fig:scenario3_postInference}
\end{figure}

We resume the analysis of the simulated dataset discussed in \Cref{sec:model} of the manuscript, where $N_i=100$ data were generated for any of $I=36$ areal units in a unit-square spatial domain. Remember that here we assume two areas as geographically contiguous if and only if they share an entire edge. As shown in \Cref{subfig:scenario3_BDgroup}, data in the blue areas were generated as i.i.d. samples from a Student's $t$ distribution, while those in the orange areas are i.i.d. simulated from a Skew Normal distribution. Note that, because the model is misspecified, i.e., data are generated from a parametric density which is not included in the likelihood of our model (see \eqref{eqn:modeldata} in the manuscript), there is no \virgolette{true} number of components $H$. For more details on the data generating densities, see \Cref{sec:model} in the manuscript. To make our conclusions more robust, we report below the inference produced on 50 replicated datasets, drawn independently according to the setting previously described. We apply our model \eqref{eqn:data_in_clust}-\eqref{eqn:prior_H} in the manuscript to each simulated dataset (generated as explained in \Cref{sec:model} of the manuscript), compute the posterior distribution via the MCMC sampler outlined in \Cref{sec:algorithm} of the manuscript and average over the 50 replicates.

Hyperparameters of $P_0$ in \eqref{eqn:P_0} of the manuscript are fixed as $\mu_0 = 0, \lambda = 0.1, c = 2, d = 2$. As the marginal prior for $\sigma^2$ in \eqref{eqn:prior_sigma} of the manuscript is concerned, we fix $\alpha = \beta = 4$, so that the prior mean is equal to $2$ and the variance is a priori infinite, yielding a vague prior for $\sigma^2$. We set the prior for $p$ in \eqref{eqn:edgeprior} in the manuscript as a $\operatorname{Beta}\left(2, I\right)$ distribution where $I = 36$, the total number of areas. This prior distribution assigns a priori a small probability of edge inclusion $p$ and, consequently, a high probability of having a boundary edge, thus inducing a sparse prior for the graph $G$.

Given the shape of the full-conditional distribution of $G_{i,k} \mid rest$, the prior for $p$ should be set with particular attention. In fact, we see from \eqref{eqn:Gij_fullcond} that the probability $\pi(G_{i,k} = 1 \mid rest)$ is (proportional to) the product of two multiplicative factors: $\operatorname{exp(logit}(p))=\displaystyle{p/(1-p)}$ and the exponential of $\rho/(2\sigma^2)$ times the scalar product of $\wtilde_{i}$ and $\wtilde_{k}$. Then, assuming a prior that assigns a high probability of edge inclusion ($p$ close to $1$) would imply no data learning since the $\operatorname{exp(logit}(p))$ were predominant w.r.t. the second term. Similarly, assuming a non-informative prior for $p$, such as the uniform density on $[0,1]$, yields that the associated prior for $p/(1-p)$ still assigns enough mass at 0 and 1, so that the $\operatorname{exp(logit}(p))$ is not defined or is equal to infinity. Note that the dimension of the space of random binary graphs the MCMC span increases with $I$. Consequently, a sparse prior, depending on $I$, will help penalizing large graphs in the MCMC. Indeed, assuming a $\operatorname{Beta}(2,I)$ prior is equivalent to a prior which concentrates mass on $2 /(2 + I)$, a small value which does take into account the size of the graph through $I$.

As discussed in \Cref{sec:model} of the manuscript, the posterior inference is sensitive to the value of $H$; this has led us to assume $H$ as random in our model. The strength of spatial association $\rho$ might play a crucial role in identifying boundaries in our model; see \eqref{eqn:logMCAR_def} in the manuscript. Moreover, we observe that in case $H$ is random, also parameter $\Lambda$, i.e., the rate of the shifted Poisson distribution in \eqref{eqn:prior_H} in the manuscript, may affect the posterior inference. Therefore, we carry out a sensitivity analysis with respect to the values of $\rho$ and $H$, also comparing how the sampler performs in case the number of components is either fixed or random. In this way, we: $(i)$ assess posterior inference and performance of the transdimensional MCMC algorithm both in terms of joint spatial density estimation and boundary detection, $(ii)$ understand the effect of these hyperparameters of the model, and $(iii)$ get more robust posterior inference by averaging over the 50 simulated datasets. For each of the simulated datasets, we fit our model with different specifications of $\rho$ and $H$. In particular, $\rho$ varies in $\{0, 0.5, 0.9, 0.95, 0.99\}$, while $H$ is assumed in $\{2,4,6,8,10\}$ or is random with prior $H - 1 \sim Poi(\Lambda)$, with $\Lambda$ in $\{1,2,5,10\}$. Each time, we run our sampler for $10,000$ iterations, discarding the first half as burn-in.

We assess boundary detection performance by comparing the posterior estimated boundary graph $\Ghatb$ (defined in \Cref{sec:model} of the manuscript) for $\gamma = 0.5$ with the true boundary graph. To this end, we define the following quantities: true positives (TP) is the number of estimated boundary edges that are also boundary edges; false positives (FP) is the number of estimated boundary edges that are, indeed, neighbouring edges; true negatives (TN) is the number estimated neighbouring edges that are also neighbouring edges; false negatives (FN) is the number of estimated neighbouring edges that are, indeed, boundary edges. We summarise boundary detection performance using several key metrics: \textit{precision}, \textit{sensitivity}, \textit{specificity} and the \textit{area under the ROC curve} (AUC), which does not depend on the cut-off threshold.

\textit{Precision} is defined as $\text{TP}/(\text{TP}+\text{FP})$ and measures the proportion of estimated boundary edges that have been correctly identified. \Cref{tab:precisions} reports mean and standard deviation of precision values across 50 simulated datasets. When $\rho = 0$, we observe poor and non-variable precision since, without spatial dependence, the posterior distribution of the $G_{i,k}$’s is dominated by the sparse marginal prior, resulting in the inclusion of nearly all admissible edges. As $\rho$ increases, precision improves for small $H$ (e.g., $H=2$), but drops rapidly to zero for $H \geq 6$ and $\rho \geq 0.9$ due to overfitting. The best precision values (with the highest means and lowest standard deviations) are achieved when $H$ is random and $\rho \geq 0.9$, corresponding to our proposed model (see \eqref{eqn:data_in_clust}–\eqref{eqn:prior_H} in the manuscript). The model is also robust to the choice of the prior for the shifted Poisson parameter $\Lambda$, with very similar precision values across settings.

\begin{table}[t]
    \centering\renewcommand{\arraystretch}{1.25}
    \resizebox{\textwidth}{!}{%
    \begin{tabular}{lccccc}
    \toprule[2pt]
    & \(\rho = 0.00\) & \(\rho = 0.50\) & \(\rho = 0.90\) & \(\rho = 0.95\) & \(\rho = 0.99\) \\\midrule[2pt]
    \(H = 2\) & 0.200 (0.000) & 0.200 (0.000) & 0.714 (0.150) & 0.896 (0.123) & 0.991 (0.029) \\ \midrule
    \(H = 4\) & 0.200 (0.000) & 0.200 (0.086) & 0.263 (0.425) & 0.339 (0.464) & 0.506 (0.487) \\ \midrule
    \(H = 6\) & 0.200 (0.000) & 0.166 (0.253) & 0.000 (0.000) & 0.000 (0.000) & 0.000 (0.000) \\ \midrule
    \(H = 8\) & 0.200 (0.000) & 0.683 (0.403) & 0.000 (0.000) & 0.000 (0.000) & 0.000 (0.000) \\ \midrule
    \(H = 10\) & 0.200 (0.000) & 0.677 (0.469) & 0.000 (0.000) & 0.000 (0.000) & 0.000 (0.000) \\ \midrule
    \(H - 1 \sim Poi(1)\) & 0.200 (0.000) & 0.200 (0.000) & 0.855 (0.157) & 0.975 (0.066) & 0.997 (0.02)  \\ \midrule
    \(H - 1 \sim Poi(2)\) & 0.200 (0.000) & 0.200 (0.000) & 0.852 (0.153) & 0.971 (0.069) & 0.997 (0.02)  \\ \midrule
    \(H - 1 \sim Poi(5)\) & 0.200 (0.000) & 0.200 (0.000) & 0.868 (0.154) & 0.982 (0.059) & 1.000 (0.000) \\ \midrule
    \(H - 1 \sim Poi(10)\) & 0.200 (0.000) & 0.200 (0.000) & 0.882 (0.146) & 0.979 (0.06)  & 1.000 (0.000) \\ \bottomrule[2pt]
    \end{tabular}}
    \caption{Precision for the estimated edges in $\Ghatb$ in the simulated scenario of \Cref{subsec:bd_simstudy} for different values of $\rho$ and $H$. All values are reported as \textnormal{mean (standard deviation)} computed over 50 simulated datasets.}
    \label{tab:precisions}
\end{table}

\begin{figure}[t]
    \centering
    \includegraphics[width=\textwidth]{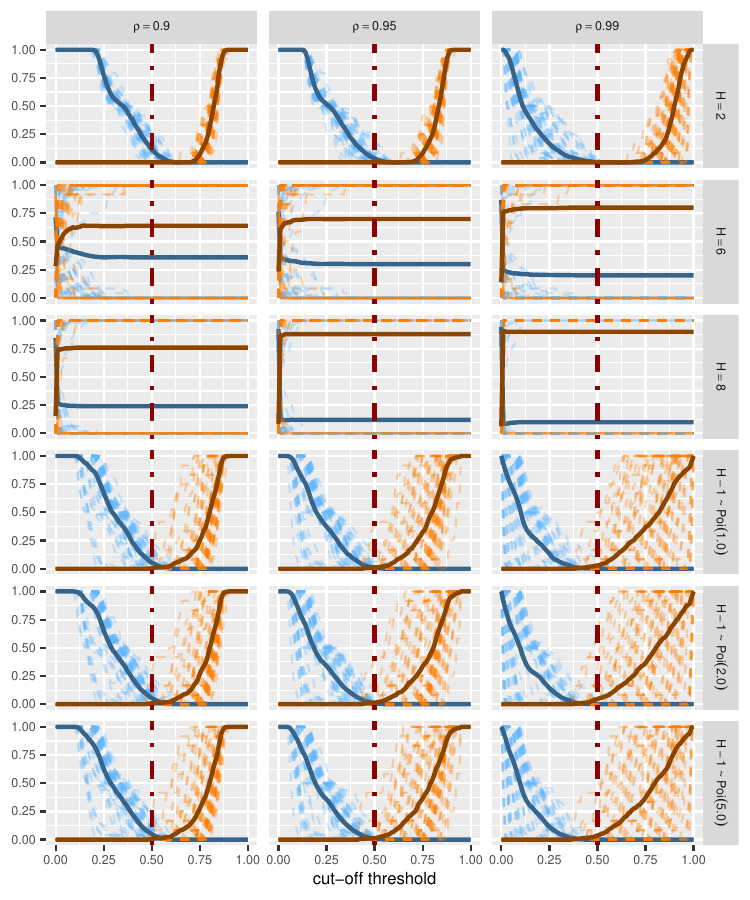}
    \caption{Mean false positive rate curve (in blue) and mean false negative rate curve (in orange) over all possible cut-off threshold for the simulated scenario in \Cref{subsec:bd_simstudy}. Values were averaged over 50 simulated datasets. The 0.5 threshold is highlighted in red.}
    \label{fig:fpr_fnr_curves}
\end{figure}

\textit{Sensitivity} ($\text{TP}/(\text{TP}+\text{FN})$) and \textit{specificity} ($\text{TN}/(\text{TN}+\text{FP})$) are reported in \Cref{tab:sens,tab:spec}, respectively. Both indices range in $(0,1)$, with higher values indicating better classification performance. The highest sensitivity and specificity values are again observed for a random number of components and strong spatial dependence (high $\rho$). The \textit{AUC} values (=the area under the ROC curve plotting TP rate against the FP rate across all possible classification thresholds $\gamma$), reported in \Cref{tab:aucs}, summarise classifier performance across all thresholds $\gamma$, thus providing a threshold-independent comparison. Consistent with other metrics, the best AUCs are obtained when $H$ is random and $\rho$ is large, confirming the superiority of our model for boundary detection in this scenario. Finally, we report in \Cref{fig:fpr_fnr_curves} the average false positive rate ($\text{FPR} = \text{FP}/(\text{FP}+\text{TN})$) and false negative rate ($\text{FNR} = \text{FN}/(\text{FN}+\text{TP})$) as a function of the threshold over a grid of values (on the x-axis); the average is taken over the 50 simulated datasets. The curves further highlight that our model with random $H$ and high $\rho$ achieves the best balance between FPR and FNR when the
threshold $\gamma = 0.5$ is selected. This fact guarantees a model selection procedure which is robust against both misclassification errors.

\begin{table}[t]
    \centering\renewcommand{\arraystretch}{1.25}
    \resizebox{\textwidth}{!}{%
    \begin{tabular}{lccccc}
    \toprule[2pt]
    & $\rho = 0.00$ & $\rho = 0.50$ & $\rho = 0.90$ & $\rho = 0.95$ & $\rho = 0.99$ \\ \midrule[2pt]
    $H = 2$ & 1.000 (0.000) & 1.000 (0.000) & 1.000 (0.000) & 1.000 (0.000) & 1.000 (0.000) \\ \midrule
    $H = 4$ & 1.000 (0.000) & 0.672 (0.375) & 0.268 (0.439) & 0.365 (0.474) & 0.557 (0.499) \\ \midrule
    $H = 6$ & 1.000 (0.000) & 0.308 (0.460) & 0.000 (0.000) & 0.000 (0.000) & 0.000 (0.000) \\ \midrule
    $H = 8$ & 1.000 (0.000) & 0.758 (0.431) & 0.000 (0.000) & 0.000 (0.000) & 0.000 (0.000) \\ \midrule
    $H = 10$ & 1.000 (0.000) & 0.673 (0.469) & 0.000 (0.000) & 0.000 (0.000) & 0.000 (0.000) \\ \midrule
    $H - 1 \sim Poi(1)$ & 1.000 (0.000) & 1.000 (0.000) & 0.997 (0.024) & 0.988 (0.063) & 0.970 (0.092) \\ \midrule
    $H - 1 \sim Poi(2)$ & 1.000 (0.000) & 1.000 (0.000) & 0.992 (0.042) & 0.987 (0.064) & 0.953 (0.124) \\ \midrule
    $H - 1 \sim Poi(5)$ & 1.000 (0.000) & 1.000 (0.000) & 0.995 (0.026) & 0.987 (0.064) & 0.963 (0.097) \\ \midrule
    $H - 1 \sim Poi(10)$ & 1.000 (0.000) & 1.000 (0.000) & 0.992 (0.059) & 0.985 (0.065) & 0.965 (0.095) \\ \bottomrule[2pt]
    \end{tabular}}\vspace{2mm}
    \caption{Estimated sensitivity in the simulated scenario in \Cref{subsec:bd_simstudy} under different value of $\rho$ and $H$. All values are reported as \textnormal{mean (standard deviation)} over 50 simulated datasets.}
    \label{tab:sens}
\end{table}

\begin{table}[t]
    \centering\renewcommand{\arraystretch}{1.25}
    \resizebox{\textwidth}{!}{%
    \begin{tabular}{lccccc}
        \toprule[2pt]
        & $\rho = 0.00$ & $\rho = 0.50$ & $\rho = 0.90$ & $\rho = 0.95$ & $\rho = 0.99$ \\ \midrule[2pt]
        $H = 2$ & 0.000 (0.000) & 0.000 (0.000) & 0.885 (0.075) & 0.965 (0.048) & 0.998 (0.008) \\ \midrule
        $H = 4$ & 0.000 (0.000) & 0.401 (0.288) & 0.970 (0.103) & 0.969 (0.120) & 0.961 (0.134) \\ \midrule
        $H = 6$ & 0.000 (0.000) & 0.897 (0.167) & 1.000 (0.000) & 1.000 (0.000) & 1.000 (0.000) \\ \midrule
        $H = 8$ & 0.000 (0.000) & 0.973 (0.048) & 1.000 (0.000) & 1.000 (0.000) & 1.000 (0.000) \\ \midrule
        $H = 10$ & 0.000 (0.000) & 0.999 (0.004) & 1.000 (0.000) & 1.000 (0.000) & 1.000 (0.000) \\ \midrule
        $H - 1 \sim Poi(1)$ & 0.000 (0.000) & 0.000 (0.000) & 0.946 (0.064) & 0.992 (0.024) & 0.999 (0.006) \\ \midrule
        $H - 1 \sim Poi(2)$& 0.000 (0.000) & 0.000 (0.000) & 0.946 (0.062) & 0.991 (0.025) & 0.999 (0.006) \\ \midrule
        $H - 1 \sim Poi(5)$& 0.000 (0.000) & 0.000 (0.000) & 0.951 (0.062) & 0.994 (0.022) & 1.000 (0.000) \\ \midrule
        $H - 1 \sim Poi(10)$ & 0.000 (0.000) & 0.000 (0.000) & 0.957 (0.058) & 0.993 (0.022) & 1.000 (0.000) \\ \bottomrule[2pt]
    \end{tabular}}\vspace{2mm}%
    \caption{Estimated specificity in the simulated scenario in \Cref{subsec:bd_simstudy} under different value of $\rho$ and $H$. All values are reported as \textnormal{mean (standard deviation)} over 50 simulated datasets.}
    \label{tab:spec}
\end{table}

\begin{table}[t]
    \centering\renewcommand{\arraystretch}{1.25}
    \resizebox{\textwidth}{!}{%
    \begin{tabular}{lccccc}
        \toprule[2pt]
        & $\rho = 0.00$ & $\rho = 0.50$ & $\rho = 0.90$ & $\rho = 0.95$ & $\rho = 0.99$ \\ \midrule[2pt]
        $H = 2$ & 0.584 (0.024) & 1.000 (0.000) & 1.000 (0.000) & 1.000 (0.000) & 1.000 (0.000) \\ \midrule
        $H = 4$ & 0.582 (0.090) & 0.753 (0.227) & 0.674 (0.214) & 0.724 (0.218) & 0.798 (0.217) \\ \midrule
        $H = 6$ & 0.571 (0.070) & 0.685 (0.218) & 0.547 (0.057) & 0.542 (0.055) & 0.522 (0.063) \\ \midrule
        $H = 8$ & 0.569 (0.068) & 0.904 (0.192) & 0.530 (0.056) & 0.523 (0.074) & 0.490 (0.063) \\ \midrule
        $H = 10$ & 0.563 (0.070) & 0.898 (0.190) & 0.480 (0.072) & 0.482 (0.057) & 0.489 (0.065) \\ \midrule
        $H - 1 \sim Poi(1)$ & 0.581 (0.047) & 1.000 (0.000) & 1.000 (0.000) & 1.000 (0.000) & 1.000 (0.000) \\ \midrule
        $H - 1 \sim Poi(2)$ & 0.586 (0.049) & 1.000 (0.000) & 1.000 (0.000) & 1.000 (0.000) & 1.000 (0.000) \\ \midrule
        $H - 1 \sim Poi(5)$ & 0.579 (0.056) & 1.000 (0.000) & 1.000 (0.000) & 1.000 (0.000) & 1.000 (0.000) \\ \midrule
        $H - 1 \sim Poi(10)$ & 0.583 (0.045) & 1.000 (0.000) & 1.000 (0.000) & 1.000 (0.000) & 1.000 (0.000) \\ \bottomrule[2pt]
    \end{tabular}} \vspace{2mm}%
    \caption{Mean area under the Receiver Operating Characteristic (ROC) curve in the simulated scenario of \Cref{subsec:bd_simstudy} for different values of $\rho$ and $H$. All values are reported as \textnormal{mean (standard deviation)} computed over 50 simulated datasets.}
    \label{tab:aucs}
\end{table}

When $H$ is random, the mode of its posterior distribution is typically $3$ across datasets. This is consistent with the known characteristics of Gaussian mixtures: when representing unimodal yet heavy-tailed distributions, they require additional components to account for skewness or asymmetry. As an illustration, \Cref{fig:scenario3_postInference} shows the estimated boundary edges (in red) and the estimated densities for two boundary areas for one simulated dataset with $H - 1 \sim \text{Poi}(1)$ and $\rho = 0.95$. Similar results are observed across all replicates.

We then evaluate density estimation accuracy using the mean $\Lone$ distance between true and posterior predictive densities and the Watanabe-Akaike Information Criterion (WAIC). For each combination of $(\rho, H)$, we compute the average $\Lone$ distance across all areas and 50 simulated datasets (see \Cref{tab:meanL1distances}). Smaller $\Lone$ distances indicate more accurate density estimates. The values show slightly smaller distances when $H$ is fixed and greater than $6$, compared to the case where $H - 1 \sim \text{Poi}(\Lambda)$, suggesting a trade-off between density estimation accuracy and boundary detection performance. Similarly, \Cref{tab:meanWAIC} reports the average WAIC (on the deviance scale) and associated standard deviations. Lower WAIC values indicate better predictive accuracy, and this is generally achieved with more mixture components. However, paired $t$-tests comparing models with $H - 1 \sim \text{Poi}(\Lambda)$ to those with fixed $H$ yield large $p$-values indicate that the observed performance differences in $\Lone$ distance and WAIC are not statistically significant. Hence, our model remains competitive for both boundary detection and density estimation.

\begin{table}[t]
    \centering\renewcommand{\arraystretch}{1.25}
    \resizebox{\textwidth}{!}{%
    \begin{tabular}{lccccc}
        \toprule[2pt]
        & $\rho = 0.00$ & $\rho = 0.50$ & $\rho = 0.90$ & $\rho = 0.95$ & $\rho = 0.99$ \\ \midrule[2pt]
        $H = 2$ & 0.175 (0.009) & 0.175 (0.009) & 0.174 (0.010) & 0.174 (0.009) & 0.172 (0.010) \\ \midrule
        $H = 4$ & 0.130 (0.016) & 0.127 (0.015) & 0.120 (0.019) & 0.115 (0.023) & 0.102 (0.028) \\ \midrule
        $H = 6$ & 0.120 (0.013) & 0.118 (0.013) & 0.120 (0.011) & 0.121 (0.012) & 0.120 (0.012) \\ \midrule
        $H = 8$ & 0.121 (0.009) & 0.114 (0.011) & 0.122 (0.010) & 0.122 (0.010) & 0.122 (0.009) \\ \midrule
        $H = 10$ & 0.122 (0.009) & 0.115 (0.011) & 0.121 (0.009) & 0.120 (0.009) & 0.120 (0.009) \\ \midrule
        $H - 1 \sim Poi(1)$ & 0.174 (0.009) & 0.171 (0.014) & 0.163 (0.017) & 0.154 (0.023) & 0.138 (0.028) \\ \midrule
        $H - 1 \sim Poi(2)$ & 0.174 (0.010) & 0.167 (0.018) & 0.161 (0.020) & 0.149 (0.026) & 0.138 (0.030) \\ \midrule
        $H - 1 \sim Poi(5)$ & 0.173 (0.010) & 0.167 (0.018) & 0.160 (0.020) & 0.151 (0.023) & 0.133 (0.027) \\ \midrule
        $H - 1 \sim Poi(10)$ & 0.173 (0.011) & 0.167 (0.018) & 0.158 (0.020) & 0.152 (0.023) & 0.136 (0.028) \\ \bottomrule[2pt]
    \end{tabular}} \vspace{2mm}%
    \caption{Estimated mean $\Lone$ distance over all areas in the simulated scenario in \Cref{subsec:bd_simstudy} under different values of $\rho$ and $H$. All values are reported as \textnormal{mean (standard deviation)} over 50 simulated datasets.}
    \label{tab:meanL1distances}
\end{table}

\begin{table}[t]
    \centering\renewcommand{\arraystretch}{1.25}
    \resizebox{\textwidth}{!}{%
    \begin{tabular}{lccccc}
        \toprule[2pt]
        & $\rho = 0.00$ & $\rho = 0.50$ & $\rho = 0.90$ & $\rho = 0.95$ & $\rho = 0.99$ \\ \midrule[2pt]
        $H = 2$ & 11778.4 (113.7) & 11778.1 (113.7) & 11775.8 (113.6) & 11776.6 (114.4) & 11780.9 (114.4) \\ \midrule
        $H = 4$ & 11541.5 (204.2) & 11508.3 (185.1) & 11460.0 (201.0) & 11446.9 (203.4) & 11396.0 (182.3) \\ \midrule
        $H = 6$ & 11437.0 (166.2) & 11396.2 (188.9) & 11469.9 (126.2) & 11463.1 (129.3) & 11462.3 (140.0) \\ \midrule
        $H = 8$ & 11208.3 (234.7) & 11190.7 (207.9) & 11438.2 (147.0) & 11432.8 (141.3) & 11438.3 (143.7) \\ \midrule
        $H = 10$ & 11087.2 (167.7) & 11142.2 (217.2) & 11417.2 (139.6) & 11416.4 (128.6) & 11415.4 (130.3) \\ \midrule
        $H - 1 \sim Poi(1)$ & 11797.7 (135.6) & 11797.6 (173.8) & 11797.9 (273.2) & 11762.7 (275.5) & 11692.2 (311.4) \\ \midrule
        $H - 1 \sim Poi(2)$ & 11797.8 (143.6) & 11768.7 (188.6) & 11780.7 (254.9) & 11751.2 (302.2) & 11681.9 (331.7) \\ \midrule
        $H - 1 \sim Poi(5)$ & 11810.4 (162.8) & 11785.4 (205.7) & 11789.2 (287.3) & 11755.4 (279.3) & 11697.4 (293.9) \\ \midrule
        $H - 1 \sim Poi(10)$ & 11810.4 (159.3) & 11785.4 (217.2) & 11770.7 (252.1) & 11761.3 (284.5) & 11690.3 (309.4) \\ \bottomrule[2pt]
    \end{tabular}} \vspace{2mm}%
    \caption{Estimated mean WAIC in the simulated scenario in \Cref{subsec:bd_simstudy} under different values of $\rho$ and $H$. All values are reported as \textnormal{mean (std. dev.)} over 50 simulated datasets.}
    \label{tab:meanWAIC}
\end{table}

\begin{figure}[!ht]
    \centering
    \subfigure{\includegraphics[width=.31\textwidth]{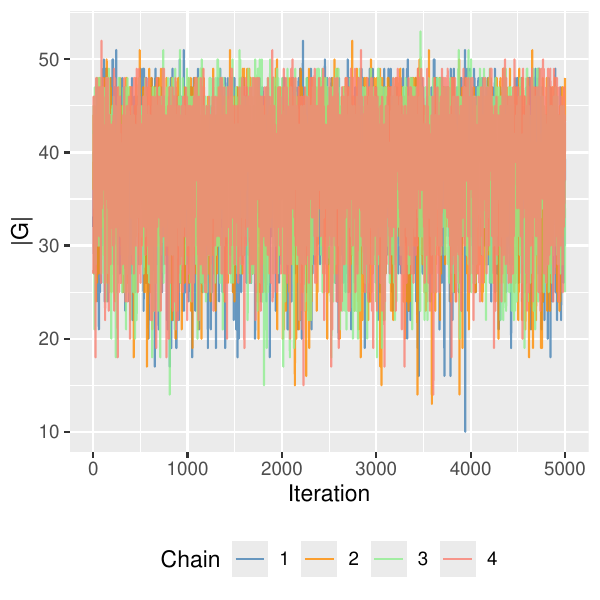}}
    \hfill
    \subfigure{\includegraphics[width=.31\textwidth]{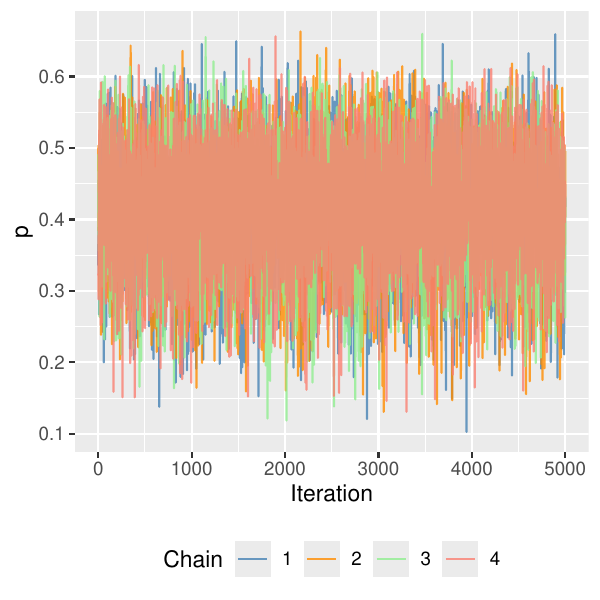}}
    \hfill
    \subfigure{\includegraphics[width=.31\textwidth]{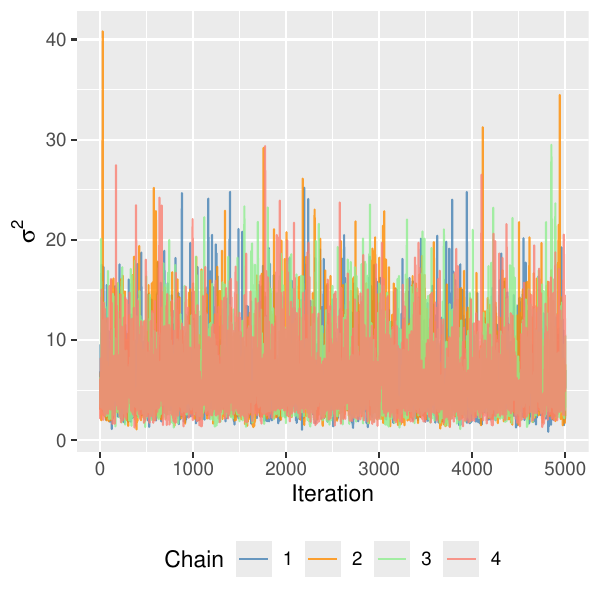}}
    \caption{MCMC diagnostics (post burn-in traceplots for different initializations of the chain) for the simulated scenario of \Cref{subsec:bd_simstudy}. This plot refers to the specification $H-1 \sim Poi(1)$, $\rho = 0.95$ and fitted using a single dataset realization.}
    \label{fig:traceplot_diagnostics}
\end{figure}

To evaluate convergence and mixing of the MCMC algorithm across the various model configurations, i.e., various couples of $(\rho, H)$, for each simulated dataset, we monitor key posterior quantities: the number of graph edges $\lvert G \rvert$, the graph sparsity parameter $p$ and the between-area variance $\sigma^2$, as they are common parameters not associated with a specific mixture component. We quantify the convergence of the MCMC chains using the Potential Scale Reduction Factor \citep[$\hat{R}$, see][]{gelman1992inference} as implemented in \cite{vehtari2021rank}. We also quantify the mixing of the MCMC chain using the Effective Sample Size (ESS) of the post burn-in iterations. \Cref{tab:Nedge_diagnostics,tab:p_diagnostics,tab:sigma_diagnostics} report the mean and standard deviation of these diagnostics across the 50 datasets. In case of random $H$ we observe an overall good performance of the chains of all parameters, with $\hat{R}$ values extremely close to $1$ and reasonable values for the ESS. In case of overfitted mixtures, i.e., when $H \geq 4$ and fixed, we observe low quality MCMC chain for $\sigma^2$, with high values of $\hat{R}$ and insufficient sample size. Finally, for the scenario and the dataset (among the 50 simulated datasets) considered for illustration in this section, \Cref{fig:traceplot_diagnostics} reports the traceplots of $\lvert G \rvert$, $p$ and $\sigma^2$ under four different chain initializations. These traceplots perfectly overlap, thus confirming a proper convergence of the MCMC even under different chain initializations.

\begin{table}[t]
    \centering\renewcommand{\arraystretch}{1.25}
    \resizebox{\textwidth}{!}{%
    \begin{tabular}{llccccc}
        \toprule[2pt]
        & & $\rho = 0.00$ & $\rho = 0.50$ & $\rho = 0.90$ & $\rho = 0.95$ & $\rho = 0.99$ \\ \midrule[2pt]
        \multirow{2}{*}{$H = 2$} & ESS & 922.8 (36.5) & 779.7 (77.2) & 2455.5 (339.4) & 1652.9 (413.1) & 267.1 (222.1) \\
        & $\hat{R}$ & 1.001 (0.000) & 1.003 (0.003) & 1.000 (0.000) & 1.001 (0.001) & 1.004 (0.005) \\ \midrule
        \multirow{2}{*}{$H = 4$} & ESS  & 1223.2 (92.4) & 996.5 (743.4) & 979.7 (1279.9) & 614.1 (1042.6) & 592.4 (1111.9) \\
        & $\hat{R}$ & 1.001 (0.001)  & 1.025 (0.070) & 1.066 (0.129) & 1.068 (0.109) & 1.122 (0.215) \\ \midrule
        \multirow{2}{*}{$H = 6$} & ESS & 1226.9 (98.8) & 1639.9 (1109.4) & 1955.6 (2029.1) & 1726.1 (2032.9) & 1797.1 (2268.3) \\
        & $\hat{R}$ & 1.001 (0.001) & 1.036 (0.164) & 1.017 (0.019) & 1.020 (0.021) & 1.034 (0.036) \\ \midrule
        \multirow{2}{*}{$H = 8$} & ESS & 1237.0 (98.0) & 775.2 (1020.7) & 3210.7 (2163.6) & 3266.1 (2093.9) & 3359.1 (2109.3) \\
        & $\hat{R}$ & 1.001 (0.001) & 1.041 (0.160) & 1.011 (0.018) & 1.011 (0.015) & 1.013 (0.028) \\ \midrule
        \multirow{2}{*}{$H = 10$} & ESS & 1193.7 (132.2) & 1082.0 (1500.4) & 4514.3 (1179.2) & 4632.1 (874.1)  & 4537.9 (1299.0) \\
        & $\hat{R}$ & 1.001 (0.001) & 1.079 (0.246) & 1.002 (0.005) & 1.002 (0.004) & 1.003 (0.006) \\ \midrule
        \multirow{2}{*}{$H - 1 \sim Pois(1)$} & ESS & 1224.4 (65.7) & 701.3 (228.4) & 1471.2 (744.6) & 889.0 (367.6) & 590.6 (344.1) \\
        & $\hat{R}$ & 1.000 (0.000) & 1.005 (0.017) & 1.006 (0.025) & 1.002 (0.004) & 1.023 (0.081) \\ \midrule
        \multirow{2}{*}{$H - 1 \sim Pois(2)$} & ESS & 1230.9 (71.3) & 681.6 (264.0) & 1419.1 (774.0) & 847.4 (399.8) & 705.4 (424.8) \\
        & $\hat{R}$ & 1.000 (0.000) & 1.006 (0.021) & 1.001 (0.003) & 1.011 (0.039) & 1.006 (0.014) \\ \midrule
        \multirow{2}{*}{$H - 1 \sim Pois(5)$} & ESS & 1220.6 (58.5) & 674.7 (278.3) & 1326.6 (758.5) & 782.7 (392.1) & 514.9 (390.3) \\
        & $\hat{R}$ & 1.000 (0.000) & 1.008 (0.022) & 1.008 (0.029) & 1.009 (0.032) & 1.031 (0.067) \\ \midrule
        \multirow{2}{*}{$H - 1 \sim Pois(10)$} & ESS & 1219.0 (66.1) & 680.4 (273.4) & 1354.7 (827.0) & 816.9 (383.7) & 733.3 (469.7) \\
        & $\hat{R}$ & 1.000 (0.001) & 1.007 (0.021) & 1.003 (0.007) & 1.008 (0.033) & 1.022 (0.061) \\ \bottomrule[2pt]
    \end{tabular}}
    \caption{MCMC diagnostics (ESS and $\hat{R}$) for $\lvert G \rvert$ for the simulated scenario of \Cref{subsec:bd_simstudy} under different values of $\rho$ and $H$. All values are reported as \textnormal{mean (std. dev.)} over 50 simulated datasets.}
    \label{tab:Nedge_diagnostics}
\end{table}

\begin{table}[t]
    \centering\renewcommand{\arraystretch}{1.25}
    \resizebox{\textwidth}{!}{%
    \begin{tabular}{llccccc}
        \toprule[2pt]
        & & $\rho = 0$ & $\rho = 0.5$ & $\rho = 0.9$ & $\rho = 0.95$ & $\rho = 0.99$ \\ \midrule[2pt]
        \multirow{2}{*}{$H = 2$} & ESS & 935.1 (34.1) & 788.5 (71.2) & 2557.7 (77.1) & 2482.6 (259.3)  & 627.7 (433.9) \\
        & $\hat{R}$ & 1.002 (0.000) & 1.003 (0.003) & 1.000 (0.000) & 1.001 (0.001) & 1.002 (0.003) \\ \midrule
        \multirow{2}{*}{$H = 4$} & ESS & 1249.4 (108.2) & 1024.4 (696.5) & 1999.0 (1401.3) & 2136.0 (1583.5) & 2668.0 (1959.0) \\
        & $\hat{R}$ & 1.001 (0.001) & 1.019 (0.053) & 1.022 (0.068) & 1.016 (0.042) & 1.034 (0.086) \\ \midrule
        \multirow{2}{*}{$H = 6$} & ESS & 1242.0 (103.9) & 1828.9 (958.8) & 4348.7 (644.5) & 4494.5 (677.8)  & 4563.7 (526.4) \\
        & $\hat{R}$ & 1.001 (0.001) & 1.028 (0.136) & 1.001 (0.002) & 1.001 (0.002) & 1.001 (0.002) \\ \midrule
        \multirow{2}{*}{$H = 8$} & ESS & 1249.4 (115.3) & 1087.9 (1140.8) & 4824.7 (266.1) & 4882.3 (237.9) & 4935.1 (222.5) \\
        & $\hat{R}$ & 1.001 (0.001) & 1.032 (0.134) & 1.000 (0.001) & 1.000 (0.001) & 1.000 (0.000) \\ \midrule
        \multirow{2}{*}{$H = 10$} & ESS & 1222.2 (114.8) & 1269.9 (1442.0) & 4904.7 (191.5) & 4891.5 (193.2) & 4922.9 (201.8) \\
        & $\hat{R}$ & 1.001 (0.001) & 1.065 (0.209) & 1.000 (0.000) & 1.000 (0.000) & 1.000 (0.000) \\ \midrule
        \multirow{2}{*}{$H - 1 \sim Poi(1)$}  & ESS  & 1223.5 (71.7)  & 712.2 (223.3)   & 1936.5 (865.5)  & 1741.4 (769.2)  & 1576.6 (1312.5) \\
        & $\hat{R}$ & 1.000 (0.000) & 1.005 (0.016) & 1.004 (0.014) & 1.001 (0.002) & 1.009 (0.029) \\ \midrule
        \multirow{2}{*}{$H - 1 \sim Poi(2)$} & ESS  & 1232.0 (79.5) & 693.0 (256.3) & 1894.9 (907.3) & 1693.6 (864.1) & 1752.9 (1355.4) \\
        & $\hat{R}$ & 1.000 (0.000) & 1.006 (0.018) & 1.001 (0.002) & 1.006 (0.022) & 1.002 (0.006) \\ \midrule
        \multirow{2}{*}{$H - 1 \sim Poi(5)$} & ESS  & 1224.3 (68.9) & 682.4 (273.0) & 1809.1 (989.5) & 1483.1 (841.2) & 1318.8 (1242.7) \\
        & $\hat{R}$ & 1.000 (0.000) & 1.007 (0.019) & 1.005 (0.020) & 1.005 (0.017) & 1.013 (0.026) \\ \midrule
        \multirow{2}{*}{$H - 1 \sim Poi(10)$} & ESS & 1220.1 (72.4) & 688.3 (266.1) & 1817.5 (983.2) & 1527.2 (826.3) & 1621.9 (1304.6) \\
        & $\hat{R}$ & 1.000 (0.000) & 1.006 (0.018) & 1.002 (0.005) & 1.005 (0.017) & 1.008 (0.024) \\ \bottomrule[2pt]
    \end{tabular}}
    \caption{MCMC diagnostics (ESS and $\hat{R}$) for $p$ for the simulated scenario of \Cref{subsec:bd_simstudy} under different values of $\rho$ and $H$. All values are reported as \textnormal{mean (std. dev.)} over 50 simulated datasets.}
    \label{tab:p_diagnostics}
\end{table}

\begin{table}[t]
    \centering\renewcommand{\arraystretch}{1.25}
    \resizebox{\textwidth}{!}{%
    \begin{tabular}{llccccc}
        \toprule[2pt]
        & & $\rho = 0$ & $\rho = 0.5$ & $\rho = 0.9$ & $\rho = 0.95$ & $\rho = 0.99$ \\ \midrule[2pt]
        \multirow{2}{*}{$H = 2$} & ESS & 702.3 (164.3) & 663.6 (145.6) & 1327.5 (326.6) & 1049.5 (191.8) & 198.1 (112.8) \\
        & $\hat{R}$ & 1.004 (0.003) & 1.002 (0.002) & 1.001 (0.002)  & 1.001 (0.001) & 1.004 (0.005) \\ \midrule
        \multirow{2}{*}{$H = 4$} & ESS  & 5.1 (14.6) & 29.8 (63.0) & 36.1 (83.5) & 31.9 (61.2)    & 35.4 (76.5) \\
        & $\hat{R}$ & 1.936 (0.291) & 1.658 (0.476) & 1.799 (0.411) & 1.711 (0.477) & 1.667 (0.494) \\ \midrule
        \multirow{2}{*}{$H = 6$} & ESS  & 8.9 (25.5) & 28.4 (50.8) & 1.3 (0.01) & 1.3 (0.01) & 1.3 (0.01) \\
        & $\hat{R}$ & 1.934 (0.336) & 1.747 (0.483) & 2.061 (0.023) & 2.061 (0.028) & 2.075 (0.026) \\ \midrule
        \multirow{2}{*}{$H = 8$} & ESS  & 64.4 (63.0) & 62.4 (54.4) & 1.3 (0.01) & 1.3 (0.01)     & 1.3 (0.01) \\
        & $\hat{R}$ & 1.378 (0.515) & 1.282 (0.463) & 2.084 (0.013)  & 2.084 (0.014) & 2.095 (0.015) \\ \midrule
        \multirow{2}{*}{$H = 10$} & ESS  & 83.2 (50.1) & 65.3 (63.7) & 1.3 (0.01) & 1.3 (0.01)     & 1.3 (0.01) \\
        & $\hat{R}$ & 1.087 (0.257) & 1.364 (0.506) & 2.089 (0.008) & 2.089 (0.011) & 2.100 (0.009) \\ \midrule
        \multirow{2}{*}{$H - 1 \sim Poi(1)$} & ESS & 401.3 (126.3) & 456.9 (188.5) & 1086.4 (557.8) & 697.2 (369.4) & 381.7 (391.2) \\
        & $\hat{R}$ & 1.002 (0.003) & 1.003 (0.008) & 1.001 (0.002)  & 1.003 (0.005)  & 1.021 (0.069) \\ \midrule
        \multirow{2}{*}{$H - 1 \sim Poi(2)$} & ESS  & 385.6 (128.8) & 453.0 (223.2) & 1053.0 (598.3) & 554.7 (327.3) & 416.2 (408.8) \\
        & $\hat{R}$ & 1.002 (0.003) & 1.004 (0.010) & 1.003 (0.009) & 1.010 (0.030) & 1.009 (0.019) \\ \midrule
        \multirow{2}{*}{$H - 1 \sim Poi(5)$} & ESS & 384.8 (130.5) & 441.7 (222.1) & 1026.3 (600.7) & 631.2 (369.5)  & 312.9 (377.9) \\
        & $\hat{R}$ & 1.002 (0.003) & 1.004 (0.011) & 1.003 (0.009)  & 1.010 (0.036)  & 1.032 (0.070) \\ \midrule
        \multirow{2}{*}{$H - 1 \sim Poi(10)$} & ESS & 385.1 (121.3) & 448.1 (218.2) & 1057.8 (648.4) & 652.6 (366.8)  & 433.4 (477.0) \\
        & $\hat{R}$ & 1.002 (0.002) & 1.005 (0.011) & 1.002 (0.002)  & 1.009 (0.036)  & 1.019 (0.049) \\ \bottomrule[2pt]
    \end{tabular}}
    \caption{MCMC diagnostics (ESS and $\hat{R}$) for $\sigma^2$ for the simulated scenario of \Cref{subsec:bd_simstudy} under different values of $\rho$ and $H$. All values are reported as \textnormal{mean (std. dev.)} over 50 simulated datasets.}
    \label{tab:sigma_diagnostics}
\end{table}

Note that, in the design of our RJ-MCMC algorithm, particular care has been taken to account for the sampling efficiency. From a computational standpoint, the main bottleneck is the reversible-jump step used to select the number of components. When $H$ is fixed, the number of MCMC iterations per second decreases as $H$ increases (from about 838 iterations/s for $H=2$ to 306 iterations/s for $H=10$). When $H$ is random, the rate drops to approximately 11 iterations/s, corresponding to a runtime of roughly 15 minutes for 10,000 iterations using our \texttt{C++} implementation. All computational times have been obtained on a machine with an Intel i7-1255U @ 4.700 GHz processor and 32 GB RAM.

\section{Stress-testing our model for boundary detection}
\label{sec:stress_tests}
In this section, we assess the robustness of
boundary detection provided by our model for two demanding stress tests. Specifically, in \Cref{sec:bd_simstudy_samemedian} we test if \textit{SPMIX} is able to detect boundaries between geographically continuous areas in case the distributional discrepancies exist only in the tails.
Then, in \Cref{subsec:bd_simstudy-highdim} we evaluate the scalability of \textit{SPMIX} and the associated MCMC algorithm in a large-sample-size regime involving over 100,000 observations and local data perturbations.

\subsection{True area-specific densities are different only in the tails}
\label{sec:bd_simstudy_samemedian}
In this section, we evaluate the sensitivity of our model to distributional discrepancies that occur outside the bulk of the distribution. We stress-test our model's ability to identify boundaries when neighbouring areas have identical medians but differ in their tail behaviour and skewness.
To this end, we consider the same scenario presented in \Cref{subsec:bd_simstudy} using a slightly different data generating process. As before, we consider $I = 36$ areas in a unit squared domain and, in each area, we simulate $100$ i.i.d. data points either $(i)$ from a Student's $t$ distribution with $6$ degrees of freedom
and with standard deviation equal to 1.5 or $(ii)$ from a Skew Normal distribution with location $\xi = 4$, scale  $\omega = 1.3$ and shape $\alpha = -3$.
To assume identical central tendencies between the data generating distributions in all areas, we numerically calibrated the location of the Student's $t$ distribution to $\mu_0 \approx 3.12$. We choose this specific value to ensure that both the Student's $t$ and the Skew-Normal distributions share the same theoretical median. In this way, we force the boundary detection mechanism to rely exclusively on higher-order moments and tail characteristics.
See \Cref{subfig:scenario3_BDgroup} to see in which areas data are simulated from $(i)$ and from $(ii)$.

\begin{figure}[t]
    \centering
    \subfigure[]{\includegraphics[width=0.32\textwidth]{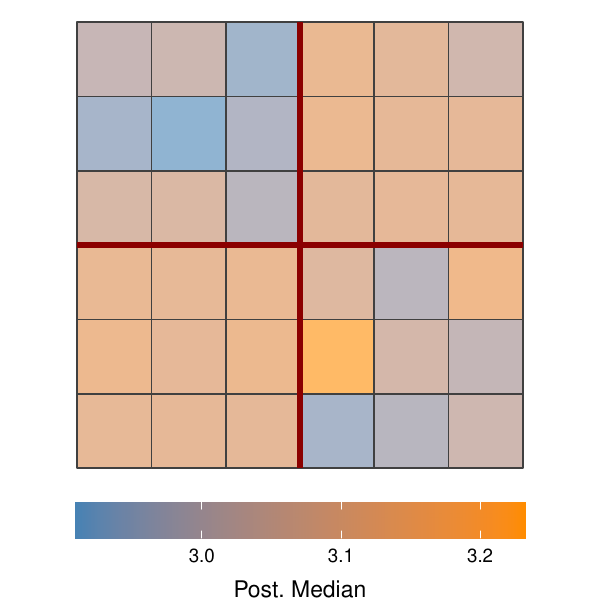}}
    \hfill
    \subfigure[]{\includegraphics[width=0.32\textwidth]{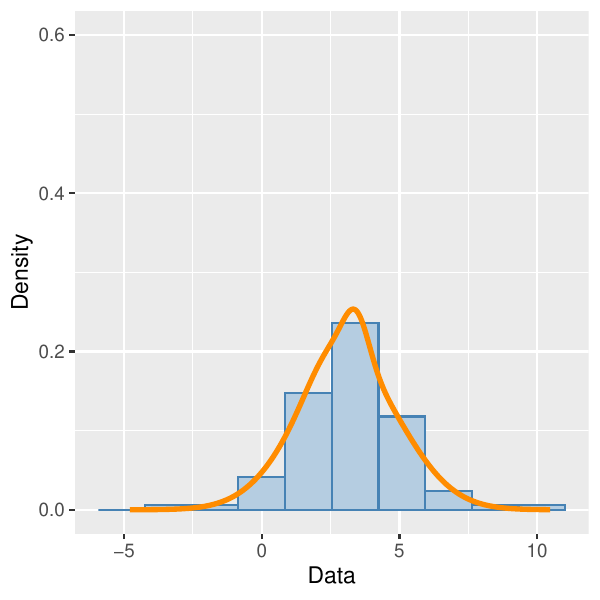}}
    \hfill
    \subfigure[]{\includegraphics[width=0.32\textwidth]{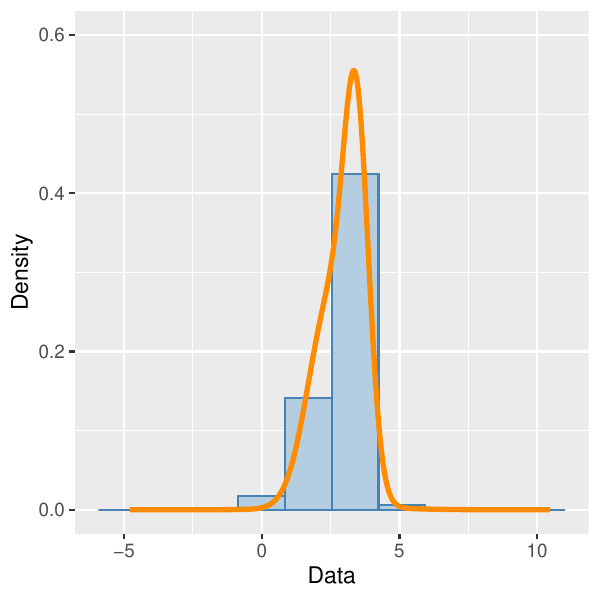}}
    \caption{Posterior inference for the simulated scenario in \Cref{sec:bd_simstudy_samemedian}; panel (a) displays the posterior medians of the estimated densities on the lattice with estimated boundaries in red; (b) and (c) report posterior estimated densities in two boundary areas.}
    \label{fig:BD_Scenario1_SameMedian_postInference}
\end{figure}

We apply model \eqref{eqn:data_in_clust}-\eqref{eqn:prior_H} in the manuscript to the dataset, with prior hyperparameters as in the simulated scenario in \Cref{subsec:bd_simstudy}, with $\rho=0.95$. We run the MCMC sampler for a total of 10,000 iterations, discarding the first half as burn-in. \Cref{fig:BD_Scenario1_SameMedian_postInference} reports the estimated boundaries in red over the spatial grid coloured according to the value of the associated posterior median (left panel); we also show the estimated densities in two boundary areas (central and right panels).
Despite the identical local medians across the domain, the model successfully identifies the true boundaries only through the tail behaviour of the distributions.

\subsection{Large-sample-size regimes}
\label{subsec:bd_simstudy-highdim}
In this section, we stress-test our model by fitting it to a simulated scenario with a very large number of observational units per area. Moreover, we introduce noise into the data-generating process so that data from geographically contiguous areas may come from similar (but not identical) distributions. To this end, we still consider $I = 36$ areas in a square unit spatial domain, as in \Cref{subsec:bd_simstudy}. In each area, we simulate $N_i = 3000$ i.i.d observations either from a Student's $t$ distribution or from a Skew Normal distribution. We consider a slightly different scenario from the one in the previous section. In particular, we generate data in the blue areas of \Cref{subfig:scenario3_BDgroup} from a Student's $t$ distribution with 6 degrees of freedom, with standard deviation equal to 1.5 and now with a random mean $\mu \sim \mathcal{N}(4, 0.12^2)$. The data in the orange areas are generated from a Skew Normal distribution with location $\xi = 4$, scale $\omega = 1.3$ and shape $\alpha = -3$ as in the previous section.

Since the simulated scenario is similar to the one discussed in \Cref{subsec:bd_simstudy}, the prior hyperparameters for model \eqref{eqn:data_in_clust}-\eqref{eqn:prior_H} in the manuscript are set as in the previous section. As before, we set the prior for $p$ in \eqref{eqn:edgeprior} in the manuscript as a $\operatorname{Beta}\left(2, I\right)$ distribution where $I = 36$, the total number of areas. Since the total number of observations is $36 \times 3,000 = 108,000$, we expect the MCMC algorithm to require a large number of iterations to properly explore the posterior distribution. Specifically, for the scenario analyzed here, the MCMC routine required approximately four days to complete $40,000$ iterations on a machine equipped with an Intel i7-1255U @ 4.700 GHz processor and 32 GB RAM.

\begin{figure}[t]
    \centering
    \subfigure[]{\includegraphics[width=.3\textwidth]{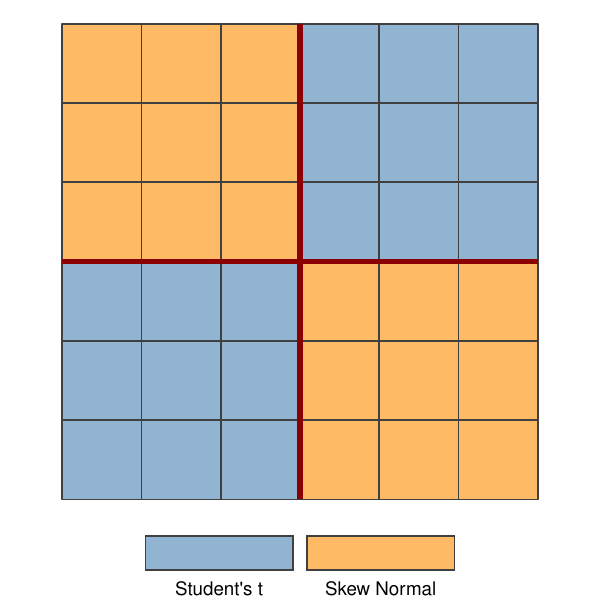}\label{subfig:BD_highdim-boundaries}}
    \hfill
    \subfigure[]{\includegraphics[width=.3\textwidth]{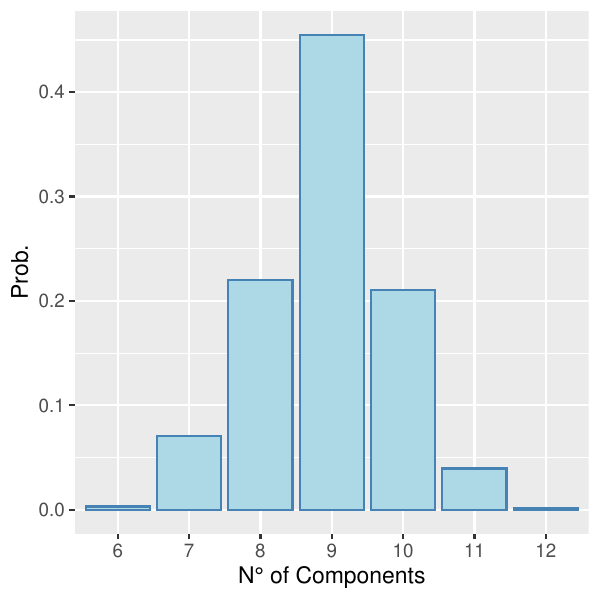}}
    \hfill
    \subfigure[]{\includegraphics[width=.3\textwidth]{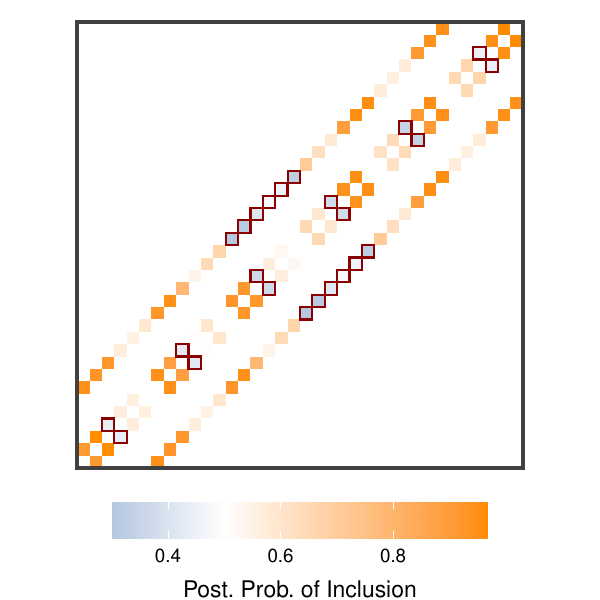}}
    \caption{Posterior inference on the simulated scenario in \Cref{subsec:bd_simstudy-highdim}: spatial domain coloured according to the data generating process (left); posterior distribution of the number of mixture components $H$ (centre); posterior probabilities $\mathbb{P}(G_{i,k} = 1 \mid \bm{y})$ of edge inclusion (right). The estimated boundary graph is highlighted in red.}
    \label{fig:BD_highdim-postinference}
\end{figure}

\Cref{fig:BD_highdim-postinference} (left panel) displays the spatial domain coloured according to the data generating process, with the estimated boundary edges highlighted in red, and while the right panel reports the matrix of posterior probabilities $\mathbb{P}(G_{i,k} = 1 \mid \bm{y})$ of edge inclusion. Even if the posterior mode of $H$ is set to 9 (see the central panel of \Cref{fig:BD_highdim-postinference}), we observe that our model correctly identifies an edge as a boundary based on non-trivial distributional discrepancies: it remains robust by avoiding the detection of boundaries between areas where the data are generated from slightly perturbed Student's $t$ distributions, while successfully detecting all boundaries between geographically contiguous areas with different data-generating distributions. The higher number of components is a direct consequence of the higher number of observations we have simulated in each area.

\section{Comparison with competitor models}\label{sec:comparisons}
In this section, we compare our model (referred to as \textit{SPMIX} in the following) with other boundary detection models in the literature. To the best of our knowledge, existing models and algorithms perform boundary detection only in the case of a single response per area, while our model achieves the same goal for multiple responses in each geographical unit. For this reason, we compute summary statistics of the data in each area. As in the manuscript, we denote by $y_{i, j}$  observation $j$ in area $i$, for  $j = 1,2,\dots,N_i$ and $i = 1,\ldots, I$. As summary statistics of the data, we consider the five-dimensional vector of empirical quantiles for area $i$ $(q_{i, \alpha_1}, \ldots, q_{i, \alpha_5}) \in \mathbb{R}^5$, where $q_{i, \alpha}$ is the empirical quantile of order $\alpha$ for $y_{i, 1}, \ldots, y_{i, N_i}$ and $(\alpha_1, \ldots, \alpha_5) = (0.05, 0.25, 0.5, 0.75, 0.95)$. Moreover, let $\bm{q}_{\alpha} = (q_{1,\alpha}, \ldots q_{I, \alpha})$ be the vector of $\alpha$-quantiles for all the areas, for any $\alpha$. In the following, we use the notation $\bm{1}_p$ for the unit vector in $\mathbb{R}^p$, and $\bm{I}_p$ is the $p \times p$ identity matrix.

\subsection{Competitor models}\label{subsec:competitor_models}
We first describe the competitor models against which we benchmark the proposed \textit{SPMIX}. These approaches represent different strategies for boundary detection or related tasks.

\subsubsection{Naive MCAR model for boundary detection}\label{sec:BDSumStats}
We consider an naive alternative MCAR model for boundary detection, based on empirical quantiles of the empirical distribution.
The model for boundary detection based on the empirical quantiles assumes
\begin{equation}
    \label{eqn:MCAR-likelihood}
    \bm{q}_{\alpha} \mid \bm{\psi}_{\alpha}, \tau^2 \ind \mathcal{N}\left(\bm{\psi}_{\alpha}, \tau^2\bm{I}_{I}\right) \qquad \alpha \in \{0.05, 0.25, 0.5 0.75, 0.95\},
\end{equation}
where $\bm{\psi}_{\alpha}$ is the $I$-dimensional vector of the spatial random effect associated to the $\alpha$-quantiles. We jointly model the vector of spatial random effects $vec(\bm{\psi}_{0.05}, \bm{\psi}_{0.25}, \bm{\psi}_{0.5},\-\bm{\psi}_{0.75},\-\bm{\psi}_{0.95})$, conditionally on the graph $G$ and the local variance $\sigma^2$, as follows:
\begin{equation}
    \label{eqn:MCAR-psi_prior}
    \scalebox{0.95}{$vec\left(\bm{\psi}_{0.05}, \bm{\psi}_{0.25}, \bm{\psi}_{0.5}, \bm{\psi}_{0.75}, \bm{\psi}_{0.95}\right) \mid \sigma^2, G \sim \operatorname{MCAR}\left(\bm{1}_5 \otimes \tilde{\bm{m}}, \sigma^2\bm{I}_5 \otimes (F - \rho G)^{-1}\right).$}
\end{equation}
Note that we assume the variance $\sigma^2$ to be constant over the areas since, in our examples, the variability associated to each vector $\bm{q}_\alpha$ is quite similar. We complete the model as follows:
\begin{equation}
    \begin{split}
        G_{i,k} \mid p &\iid \operatorname{Be}(p) \ \text{for all } (i,k) \in \Eadj \\
        p &\sim \operatorname{Beta}\left(a, b\right), \ a,b>0 \\
        \sigma^2 &\sim \operatorname{InvGamma}(\alpha_s, \beta_s), \ \alpha_s,\beta_s >0 \\
        \tau^2 &\sim \operatorname{InvGamma}(\alpha_t, \beta_t), \ \alpha_t,\beta_t >0. 
        \label{eqn:MCAR-all_priors}
    \end{split}
\end{equation}
In particular, note that the prior for the graph $G$ matches the prior used in our spatial mixture model.

Observe that model \eqref{eqn:MCAR-likelihood}-\eqref{eqn:MCAR-psi_prior} is not fully adequate to model a vector of quantiles, as it disregards the fact that, in each area, $(q_{i, 0.05}, \ldots, q_{i, 0.95})$ is a nondecreasing sequence. However, such a model can perform boundary detection based on the differences between the vector of empirical quantiles in geographically contiguous areas, as the full conditional of $G_{i,k}$, $(i, k) \in \Eadj$, is
\begin{equation*}
    \pi\left(G_{i,k} = 1 \mid rest \right) \propto \exp \left\{ \log\left(\frac{p}{1-p}\right) - \frac{\rho}{2\sigma^2}\| \Psi_i - \Psi_k\|^2 \right\}, \ \pi\left(G_{i,k} = 0 \mid rest\right) \propto 1,
\end{equation*}
where $\Psi_{i} = (\psi_{i, 0.05}, \ldots, \psi_{i, 0.95})$. That is, the probability of detecting an edge is directly proportional to the $L_2$ distance between the random effects associated with the quantiles in each area. In particular, such a distance can also be regarded as a discretization of the Wasserstein distance between random effects, therefore justifying the use of model \eqref{eqn:MCAR-likelihood}-\eqref{eqn:MCAR-all_priors} for boundary detection. Since this model detects boundaries using the same approach followed by \textit{SPMIX}, an admissible edge $(i,k)$ is an estimated boundary edge if $\mathbb{P}(G_{i,k} = 1 \mid \bm{y}) < \gamma$, with $\gamma = 0.5$. For notational convenience, we refer to \eqref{eqn:MCAR-likelihood}-\eqref{eqn:MCAR-all_priors} as the \emph{naive MCAR} in the rest of this section.

\subsubsection{SKATER algorithm for regionalisation}\label{sec:skater}
SKATER \citep[Spatial `K'luster Analysis by Tree Edge Removal,][]{assunccao2006efficient} is a regionalisation method based on graph partitioning. \textit{Regionalisation} refers to the process of grouping a set of areal units (e.g., census tracts, municipalities, or districts) into a smaller number of contiguous regions that are internally homogeneous based on specific attributes. This procedure typically ensures that the resulting regions are spatially contiguous (i.e., they share boundaries and form a connected area). This definition makes it clear that regionalisation is more closely related to clustering than boundary detection. However, one could reasonably interpret the borders between different regions as boundaries. Although this interpretation slightly alters the traditional definition of a boundary, it allows us to compare the posterior inference from our model with the output from the SKATER algorithm. This algorithm first constructs a minimum spanning tree (MST) from the adjacency graph that encodes the spatial structure of the data. Then, regionalisation is achieved by optimally pruning the MST.

We apply the SKATER algorithm as implemented in the \texttt{R} package \texttt{spdep} \citep{bivand2022r} using as observation in each area $i$ the vector of empirical quantiles $(q_{i, 0.05}, q_{i, 0.25},\-q_{i, 0.5}$, $q_{i, 0.75}, q_{i, 0.95}) \in \mathbb R^5$. While constructing the minimum spanning tree, SKATER requires the computation of the adjacency graph, which is typically undirected and can be either weighted or unweighted. The \textit{weight} or \textit{cost} associated with each edge is usually determined by covariate-based dissimilarity metrics between geographically contiguous regions, with a higher cost reflecting a greater dissimilarity. For a fair comparison with our model that does not make use of covariates, we exclude dissimilarity metrics in the adjacency graph computation. In this context, boundaries are defined as the edges between the $K+1$ identified regions, where $K$ is the number of tree prunings, a user-defined parameter. We refer to the SKATER algorithm as \textit{SKATER} in the rest of this section.

\subsubsection{CARBayes model for boundary detection}\label{sec:CARBayes}
The \texttt{CARBayes} package \citep{lee2013carbayes} is a well-known \texttt{R} package for modelling areal data using conditional autoregressive (CAR) priors within a Bayesian framework via MCMC simulation. It supports a broad range of models for areal data, handling both univariate and multivariate response variables that follow binomial, Gaussian, multinomial, Poisson, or zero-inflated Poisson distributions. We use \texttt{CARBayes} to fit the following model:
\begin{align*}
    y_i \mid \psi_{i}, \tau^2 &\ind \mathcal{N}(\psi_i, \tau^2) \qquad i = 1, \dots, I, \\ %
    \tau^2 &\sim \invGamma(a, b);
\end{align*}
where $y_i$ represents the observed data in area $i$. The model can be applied only to cases with one datapoint per each area. The spatial random effect is denoted as $\bm{\psi} = (\psi_1, \dots, \psi_I)$, and it is modelled using the class of CAR priors. In particular, we consider the boundary detection model proposed by \cite{lee.mitchell2012}, which uses the CAR prior from \cite{leroux2000estimation}, defined as:
\begin{equation*}
    \psi_i \mid \psi_{-i}, G, \rho, m, \sigma^2 \sim \mathcal{N}\left( \frac{\rho \sum_{k} G_{i,k}\psi_k + (1- \rho)m}{\rho\sum_{k} G_{i,k} + (1-\rho)}, \frac{\sigma^2}{\rho\sum_{k} G_{i,k} + (1-\rho)} \right),
\end{equation*}
and achieves boundary detection using covariate-based dissimilarity metrics. The model detects \textit{boundaries}  between areas with distinct characteristics. The model assumes that each admissible edge $G_{i,k}$ is  a binary random variable, modelled as $G_{i,k}(\bm{\alpha})$ according to:
\begin{align*}
    G_{i,k}(\bm{\alpha}) &=
    \begin{cases}
        1 & \text{if } \exp\left(-\sum_{l = 1}^{q} z_{l,i,k}\,\alpha_l\right) \geq 0.5 \text{ and } i \sim k \\
        0 & \text{otherwise}
    \end{cases}; \\
    \alpha_l &\ind \mathcal{U}(0, M_i) \qquad l = 1, \dots, q,
\end{align*}
In the above equation, $z_{l,i,k}$ represents the $l$-dissimilarity metric between areas $i$ and $k$, $\bm{\alpha} = (\alpha_1, \dots, \alpha_q)$ is the regression parameter vector that controls the influence of the dissimilarity metrics, and $q$ is the number of dissimilarities used. Typically,  dissimilarities $z_{l,i,k}$, $l=1,\ldots,q$, between areas $i$ and $k$ are assumed as the difference in the values of a vector of $q$ areal covariates $\bm x_i$ and $\bm x_k$. 
This model is implemented in the \texttt{CARBayes} package via the \texttt{S.CARdissimilarity()} function. In this model, a boundary is identified if the posterior mean of the adjacency weight $w_{ij}$ between two areas is less than $0.5$. This threshold serves as the decision rule for edge exclusion, providing a direct comparison to the $\gamma = 0.5$ threshold used in our \textit{SPMIX} model. In the remainder of this section, we refer to this specific model as \textit{CARBayes}.

It is worth noticing that a comparison between \textit{SPMIX} and \textit{CARBayes} is not straightforward. Our model does not use covariates or dissimilarity metrics and is designed for multiple observations per area, while the \textit{CARBayes} model requires a single, univariate response variable per area and depends on dissimilarity metrics. We choose to fix the \textit{CARBayes} model, using the empirical median $q_{0.5,i}$ as the response variable in each area, for $i = 1, \dots, I$. We then consider a set of $q = 4$ dissimilarity metrics, each corresponding, respectively, to the absolute difference $z_{l,i,k} = \lvert q_{l,i} - q_{l,k} \rvert$ for all $(i,k)$ such that $i \sim k$ and for $l = \{0.05, 0.25, 0.75, 0.95\}$.

\subsection{Model comparison on a simulated dataset}\label{subsec:comparison_synthetic}
\begin{figure}[t]
    \centering
    \subfigure[\textit{Truth}]{\includegraphics[width=.3\textwidth]{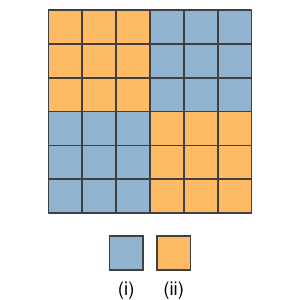}}
    \subfigure[\textit{SPMIX}]{\includegraphics[width=.3\textwidth]{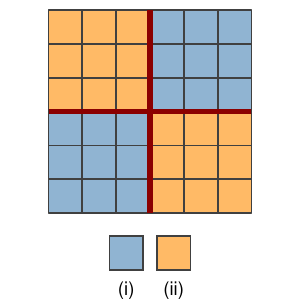}}
    \subfigure[\textit{naive MCAR}]{\includegraphics[width=.3\textwidth]{img/Comparison/SyntheticData/plt_boundaries-naiveMCAR.pdf}} \\
    \subfigure[\textit{SKATER}]{\includegraphics[width=.3\textwidth]{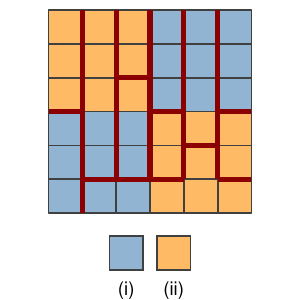}}
    \subfigure[\textit{CARBayes}]{\includegraphics[width=.3\textwidth]{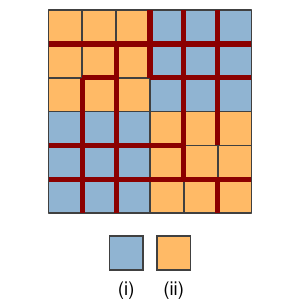}}
    \caption{Comparative study on the simulated dataset in \Cref{subsec:comparison_synthetic}: the top left panel reports the regular lattice with each area coloured according to the data generating distribution. Each other panel also highlights in red the the estimated boundary graph $\Ghatb$ according to each model or algorithm considered in the comparative study.}
    \label{fig:Comparison-SyntheticData}
\end{figure}

In this section, we compare \textit{SPMIX} with all competitor models in a controlled simulation study. In this simulated scenario, we consider $I = 36$ areas in a unit-squared domain, as in all other simulated scenarios we have considered. According to the area $i$, we simulate $N_i=200$ i.i.d. data points  $(y_{i,j})$,  for  $j = 1,2,\dots,N_i$,  either $(i)$ from a mixture of two Gaussian distributions with equal weights ($=0.5$), unit variances and centred in $-2$ and $2$ or $(ii)$ from a Gaussian distribution with zero mean and variance equal to $5$. In this way, datapoints across all areas have the same theoretical mean and variance despite being simulated from two very different distributions. Please, refer to \Cref{fig:Comparison-SyntheticData}(a) to see in which areas the data are simulated according to $(i)$ or according to $(ii)$.

Hyperparameters for the \textit{SPMIX} model are fixed as follows: $\mu_0 = 0, \lambda = 0.1, c = 2$ and $d = 2$ in $P_0$. As the marginal prior for $\sigma^2$, the across-area variance, we fix $\alpha = \beta = 2$. We set the prior for $p$, the probability of edge inclusion, equal to a $\operatorname{Beta}(2,I)$ distribution, where $I = 36$, total number of areas. Finally, we set $\rho=0.95$ to encourage a strong global spatial association. Recall that \textit{SPMIX} model is fit to the original simulated data $(y_{i,j})$, for  $j = 1,2,\dots,N_i$, $i=1,\ldots,I$, while the other competitor models or algorithms are fitted to empirical quantiles $\bm{q}_\alpha$ of the data in each area.

The hyperparameters of the \emph{naive MCAR} model are chosen as follows: we set $\bm{\tilde{m}} = m_0\bm{1}_I$, where $m_0 \in \mathbb R$ is equal to the empirical mean of $vec(\bm{q}_{0.05}, \dots, \bm{q}_{0.95})$. For $\sigma^2$ and $\tau^2$, we fix $\alpha_s = \beta_s = \alpha_t = \beta_t = 2$, which is a common choice in the literature. The prior for $G$ matches that for our spatial mixture model, namely $p \sim \mathrm{Beta}(2, I)$, $I = 36$ being the number of areas; in addition, the global spatial association parameter $\rho$ is set equal to $0.95$. The MCMC algorithm for posterior computation for the naive MCAR has been implemented in JAGS \citep{plummer2003jags} via the \texttt{rjags} package, storing a total of $1,000$ Monte Carlo samples from the posterior distribution.

As long as the \textit{SKATER} algorithm is concerned, we can fix either the number of tree prunings $K$, which will split the map into $K+1$ regions, or constraints on the minimum and maximum number of areas in each region. We choose to consider a total of $10$ tree prunings, with at least two observations in each region. We choose the default vague prior as prior for the \textit{CARBayes} model. For further details about the default prior specification, see \cite{lee2013carbayes}.

\Cref{fig:Comparison-SyntheticData} (b)-(e) report
the detected boundaries (in red) according to each model or algorithm we have considered in this comparative study. It is clear from the figure that our model is able to identify the true boundaries by detecting very different (estimated) densities. The \textit{naive MCAR} model, instead, is not capable of detecting any boundary by fitting only summary statistics of the data, while both \textit{SKATER} and \textit{CARBayes} detect some boundaries, even if many of these boundaries appear between areas with the same data-generating process. This clearly entails that our model outperforms competitors and detects boundaries between geographically contiguous yet different areal densities.

\subsection{Model comparison on the California census income dataset} \label{subsec:comparison_calicensusdata}
In this section, we compare \textit{SPMIX} with all competitor models using the California census income dataset presented in \Cref{subsec:data_exploration} of the manuscript. The available software to fit the alternative methods is not designed for large datasets (about 80,000 individual observations). For this reason,  we limit the comparison to the case of a reduced dataset. We randomly select 100 observations from each PUMA, resulting in a total sample size of 9,300. Hyperparameters for the \textit{SPMIX} model are selected following empirical estimates: namely, we set the hyperparameters of the base measure $P_0$ as $\mu_0 = 10, \lambda = 0.1, c = 6, d = 4$. As the marginal prior for $\sigma^2$, the across-area variance, we fix $\alpha = \beta = 2$. We set the prior for $p$, the probability of edge inclusion, equal to a $\operatorname{Beta}(2,I)$ distribution, where $I = 93$ is the total number of PUMAs. Finally, we set $\rho=0.95$ to encourage a strong global spatial association.

The hyperparameters of the \emph{naive MCAR} model are chosen as follows: we set $\bm{\tilde{m}} = m_0\bm{1}_I$, where $m_0 \in \mathbb R$ is equal to the empirical mean of $vec(\bm{q}_{0.05}, \dots, \bm{q}_{0.95})$. For $\sigma^2$ and $\tau^2$, we fix $\alpha_s = \alpha_t = 3$ and $\beta_s = \beta_t = 2$, which lead to a prior distribution for the variances centred in $1$ and with unit variance. The prior for $G$ matches that for our spatial mixture model, namely $p \sim \mathrm{Beta}(2, I)$; in addition, the global spatial association parameter $\rho$ is set equal to $0.95$. The MCMC algorithm for posterior computation for the naive MCAR has been implemented in JAGS \citep{plummer2003jags} via the \texttt{rjags} package, storing a total of $1,000$ Monte Carlo samples from the posterior distribution.

As long as the \textit{SKATER} algorithm is concerned, 
we choose to consider a total of $10$ tree prunings, with at least two observations in each region. Finally, also in this case, we choose the default, vague prior as prior for the \textit{CARBayes} model. 

\begin{figure}
    \centering
    \subfigure[\textit{naive MCAR}]{\includegraphics[width=.48\textwidth]{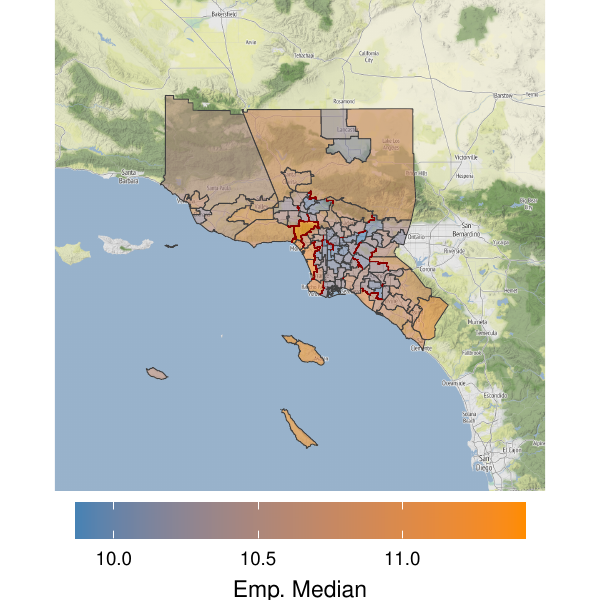}}
    \hfill
    \subfigure[\textit{SPMIX}]{\includegraphics[width=.48\textwidth]{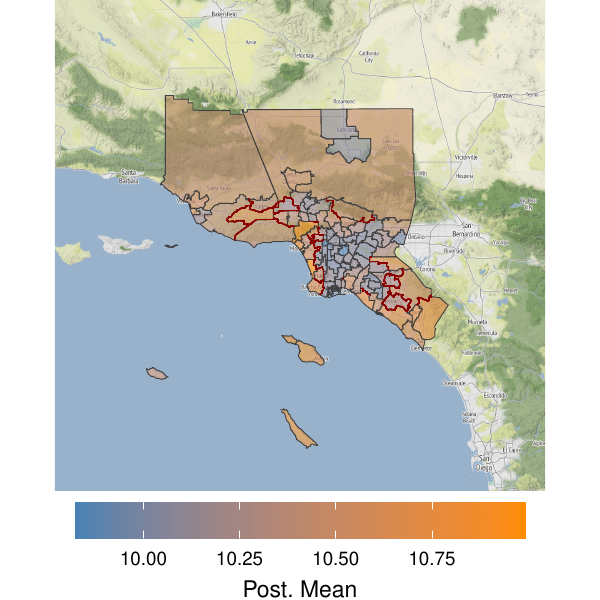}}
    \caption{Boundary detection under the \textit{naive MCAR}  (panel (a)) and \textit{SPMIX} (panel (b)) models for the California census income dataset: empirical median of the data in each areal unit. The estimated boundary graph $\Ghatb$ is highlighted in red.}
    \label{fig:SPMIXvsMCAR-CaliCensusData}
\end{figure}

\begin{figure}[t]
    \centering
    \subfigure[\label{subfig:SPMIXvsMCAR-boxplots_DistDens}]{\includegraphics[width=.48\textwidth]{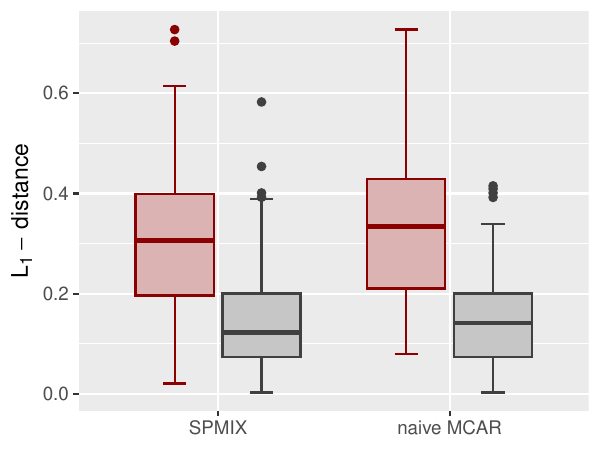}}
    \hfill
    \subfigure[\label{subfig:SPMIXvsMCAR-boxplots_DistVect}]{\includegraphics[width=.48\textwidth]{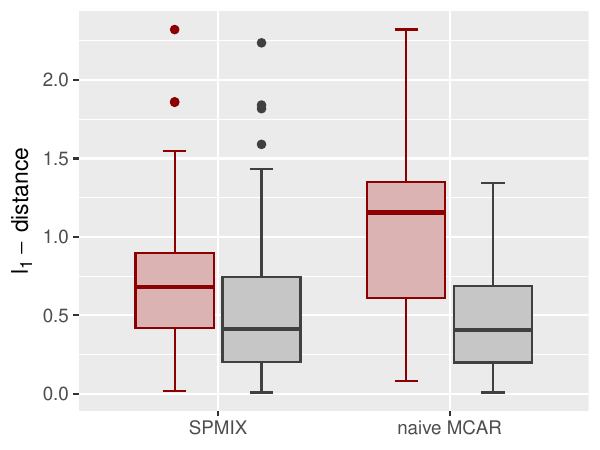}}
    \caption{Boxplots of $\Lone$ distances between estimated densities (a) and $l_1$ distances between vectors of empirical quantiles (b) over the sets $\BEhat_{\mbox{\tiny{SPMIX}}}$, $\BEhat_{\mbox{\tiny{MCAR}}}$ (in red), $\NEhat_{\mbox{\tiny{SPMIX}}}$, $\NEhat_{\mbox{\tiny{MCAR}}}$ (in gray) for the California census income dataset.}
    \label{fig:SPMIXvsMCAR-boxplots-CaliCensusData}
\end{figure}

\paragraph{\textit{SPMIX} vs \textit{naive MCAR}}
\Cref{fig:SPMIXvsMCAR-CaliCensusData} displays the detected boundaries using the \textit{naive MCAR} (panel (a)) and \textit{SPMIX} (panel (b)) on the map. \textit{SPMIX} identifies 52 boundaries, while \textit{naive MCAR} only 46; 20 boundaries are identified by both. The Jaccard similarity index between the boundaries estimated by the two models is $20 / (52+46-20) \simeq 0.256$. To validate the boundaries detected by the two models, as detailed in \Cref{subsec:post_inf_cali} of the manuscript, we compute the $\Lone$ distance between all couples of boundary and all couples of neighbouring densities,  detected by \textit{SPMIX} and \textit{naive MCAR} models. Let $\NEhat_{\mbox{\tiny{SPMIX}}}$ and $\NEhat_{\mbox{\tiny{MCAR}}}$ be the sets of neighbouring edges detected by \textit{SPMIX} and \textit{naive MCAR}, respectively. Similarly, $\BEhat_{\mbox{\tiny{SPMIX}}}$ and $\BEhat_{\mbox{\tiny{MCAR}}}$ denote the sets of boundary edges identified by each model. \Cref{subfig:SPMIXvsMCAR-boxplots_DistDens} displays the boxplots of the $\Lone$ distances between couples of estimated densities over these sets: red boxplots correspond to boundary edges while grey ones correspond to neighbouring edges. Clearly, both models are able to identify boundary edges associated to larger values of the $\Lone$ distance. We also compute the $l_1$ distance between couples of vectors of empirical quantiles for the same edge sets, shown in \Cref{subfig:SPMIXvsMCAR-boxplots_DistVect}.

\begin{figure}[t]
    \centering
    \includegraphics[width=\textwidth]{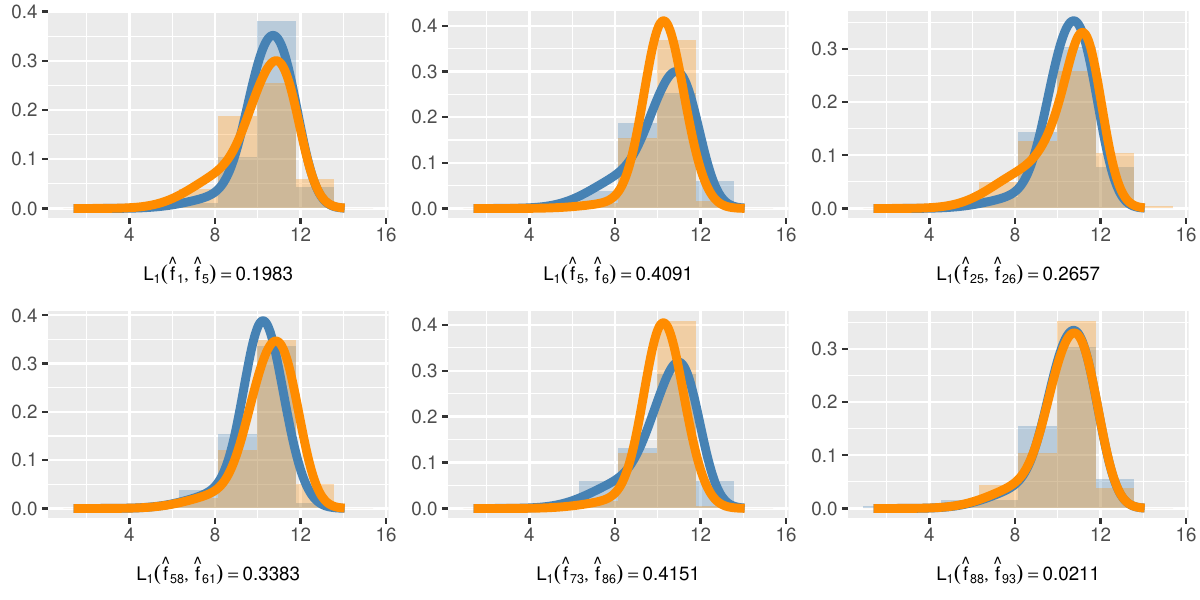}
    \caption{Empirical histograms and associated posterior estimated densities for couples of boundary areas detected by \textit{SPMIX} but not by \textit{naive MCAR} for the California census income dataset. The value of the $\Lone$ distance is reported at the bottom of each panel. Histograms and estimated densities of the couples of areas are depicted in blue and orange.}
    \label{fig:SPMIXvsMCAR-DensityComparison}
\end{figure}

While the \textit{naive MCAR} model clearly separates boundary and neighbouring edges, the two boxplots associated to \textit{SPMIX} are more similar. This indicates that the \textit{SPMIX} model can detect boundaries even when the $l_1$ distances between the vectors of empirical quantiles are small. Keeping in mind the limitations of the \textit{naive MCAR} highlighted above, we expect that the boundaries detected by \textit{SPMIX} but not by the \textit{naive MCAR} correspond to areas with relatively different densities but similar summary statistics. \Cref{fig:SPMIXvsMCAR-DensityComparison} compares empirical density histograms and posterior estimated densities for several couples of boundary areas detected by \textit{SPMIX} but not by the \textit{naive MCAR}; each panel also reports the $\Lone$ distance between estimated densities. Since these edges are not identified as boundaries by the \textit{naive MCAR} model, their associated empirical quantiles are similar. Except for the right-bottom panel, all the couples of estimated densities are different (in the location, skewness or tails). Such differences cannot be captured only through quantiles or other summary statistics, which is why \textit{SPMIX} can detect boundaries not identified by the \textit{naive MCAR} model.

\begin{figure}[t]
    \centering
    \subfigure[$K = 5$]{\includegraphics[width=0.48\textwidth]{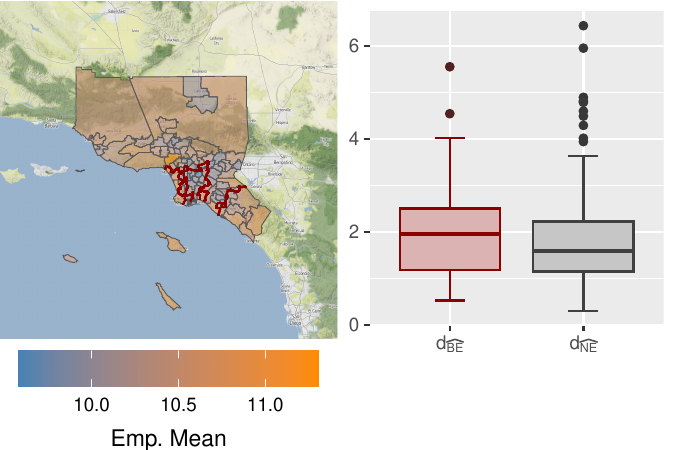}}
    \hfill
    \subfigure[$K = 10$]{\includegraphics[width=0.48\textwidth]{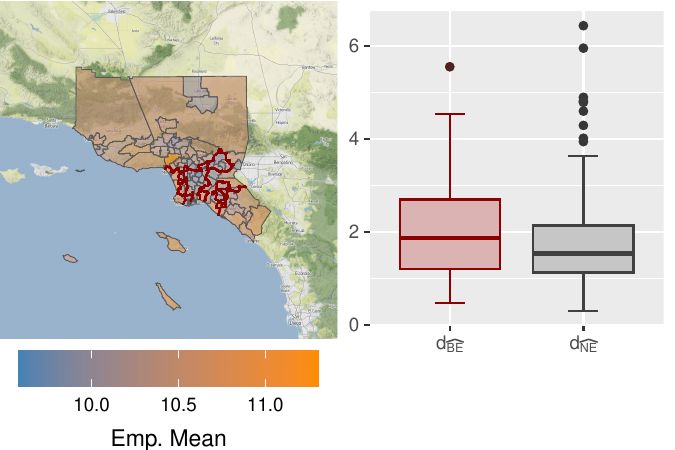}}
    \caption{Heatmap of the empirical means of the 
    California census income data in each area, with estimated boundaries (by SKATER) marked in red (left); associated boxplots (right) between $l_1$ distance of the data in neighbouring and boundary couples of areas. Panels (a) and (b) refer to $K=5,10$ pruned edges, respectively.}
    \label{fig:SKATER_inference}
\end{figure}

\paragraph{\textit{SPMIX} vs. \textit{SKATER}}
In \Cref{fig:SKATER_inference}, we display the boundaries detected by SKATER, defined as borders between areas assigned to different regions, for various edge-pruning specifications. It is important to note that pruning $K$ edges results in the partition of the areal units into $K+1$ regions.

There are similarities between the boundaries detected by our model and those identified by the SKATER algorithm. Both approaches tend to distinguish the central part of Los Angeles from the wealthier Bay Area, with some boundaries also observed in Orange County. However, a key limitation of the SKATER algorithm is that the number of tree pruning must be specified manually and cannot be learned from the data. Consequently, any value between $0$ and the total number of areas can yield different results. The algorithm also allows the user to specify the minimum and maximum number of areas to estimate in each region, making the inference outcome even more sensitive to the input parameters. Moreover, the minimum spanning tree required by the algorithm is usually computed using a greedy algorithm that finds a local, but generally sub-optimal, solution using a node as a starting point. This entails that, according to the starting node, the MST can be different, and since pruning happens only between branches of the MST, the regions identified by SKATER can differ.

To compare our model with the boundaries detected using the SKATER algorithm, we replicate the global comparison described in \Cref{subsec:post_inf_cali} of the manuscript, where we have computed the $\Lone$ distances between posterior densities for the set of neighbouring areas and the set of boundary areas. Since the output of the SKATER algorithm is a list of labels assigning each area to its corresponding region, in this case, we can only compute distances between the input data for the set of neighbouring areas and the set of boundary areas. We use the $l_1$ distance between vectors, i.e., given two vectors $\bm{x}$ and $\bm{x}'$ in $\mathbb{R}^q$, $d_{l_1}(\bm{x}, \bm{x'}) = \sum_{l = 1}^{q}\lvert x_l - x'_l \rvert$. \Cref{fig:SKATER_inference} shows the boxplots of the $l_1$ distances in the set of neighbouring and boundary areas when the number of tree prunings $K$ is equal to $5$ and $10$. In both cases, the boxes of the two boxplots span over the same values, suggesting that the algorithm has found no significant difference between the neighbouring and boundary areas; compare, instead, with \Cref{subfig:L1_distances_local,subfig:L1_distances_global} in the manuscript, where the $\Lone$ distances between our estimated densities in the set of neighbouring areas are smaller than those between boundary areas.

\paragraph[SPMIX vs. CARBayes]{\textit{SPMIX} vs. \textit{CARBayes}}
As previously mentioned, the comparison between \textit{SPMIX} and \textit{CARBayes} is not straightforward. Our model does not use covariates nor dissimilarity metrics and is designed for multiple observations per area, while the \textit{CARBayes} model requires a single, univariate response variable per area and depends on dissimilarity metrics. To address this, we assume the response variable $y_i$ as the empirical median $q_{i,0.5}$ in each area, and we augment our dataset with a list of dissimilarity metrics $\{\bm{Z}^l\}_{l=1}^{q}$, where each $\bm{Z}^l$ is a $I \times I$ dissimilarity matrix, i.e., $\bm{Z}^l=[z^l_{i,k}]_{i,k}$. We specify two different sets of dissimilarity metrics, using the exogenous variables introduced later in \Cref{sec:interpret_boundaries}, to interpret the boundaries estimated by \textit{SPMIX}. Focusing solely on LA County, we calculate, for each PUMA: $(a)$ the number of recorded crimes in 2020, and define the dissimilarity metric $\bm{Z}^a$ by setting $z^a_{i,k}$ equal to the difference, in absolute value,  of the number of crimes between areas $i$ and $k$, for all $(i,k) \in \Eadj$; $(b)$ the percentage of people without health insurance in 2020, and define the dissimilarity metric $\bm{Z}^b$ by setting $z^b_{i,k}$ equal to the difference, in absolute value, of the percentage of uninsured people between areas $i$ and $k$, for all admissible edges $(i,k)$.

\begin{figure}[t]
    \centering
    \subfigure[Number of crimes]{\includegraphics[width=0.48\linewidth]{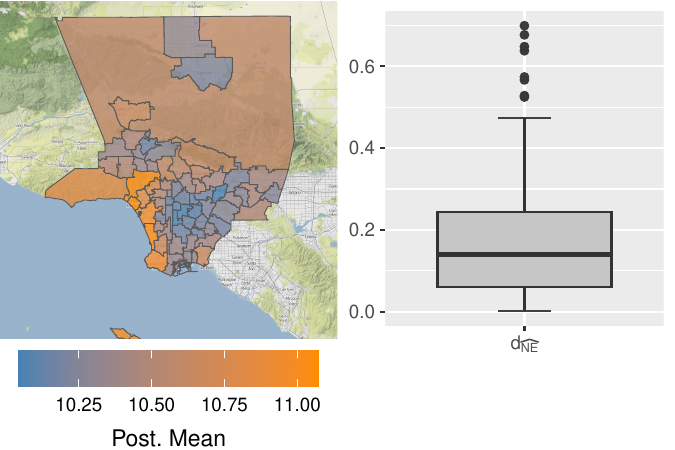}}
    \hfill
    \subfigure[\% of people w/o health insurance \label{subfig:CARBayes_boundaries_HealthInsurance}]{\includegraphics[width=0.48\textwidth]{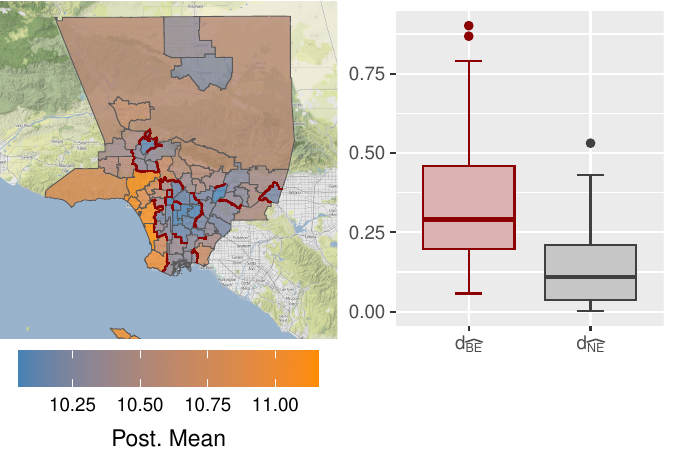}} \\
    \caption{Posterior means in each LA County PUMA displayed on the map with estimated boundaries highlighted in red. For each plot, we report the boxplot comparisons between neighbouring and boundary couples of areas using the posterior mean predictive value in each area. The boundaries are estimated via the \texttt{S.CARdissimilarity()} function in the \texttt{CARBayes} package in \texttt{R}.}
    \label{fig:CARBayes_boundaries}
\end{figure}
 
Since the output of \texttt{S.CARdissimilarity()} does not include estimated densities, we resort to a different global comparison between neighbouring and boundary areas, as we have done in the case of the comparison with SKATER. To compute distances, we use the estimated posterior mean in each area that we get as an output of the \texttt{S.CARdissimilarity()} function and compute the absolute values of their differences over the set of neighbouring and boundary areas. Each panel of \Cref{fig:CARBayes_boundaries} displays the estimated boundaries (highlighted in red) over the map, with each area coloured according to the value of the posterior mean and the boxplots of the absolute differences over the set of boundary and neighbouring areas for both dissimilarity metrics $(a)$ and $(b)$.
In case $(a)$, the set of neighboring edges coincides with $\Eadj$, which represents a trivial estimate.
In case $(b)$, \textit{CARBayes} estimates a non-trivial boundary graph. Moreover, we see that the differences of the posterior means over the set of boundary areas assume larger values than those computed over the set of neighbouring areas; see \Cref{subfig:CARBayes_boundaries_HealthInsurance}.

\begin{figure}[t]
    \centering
    \subfigure[\textit{CARBayes}]{\includegraphics[width=0.48\textwidth]{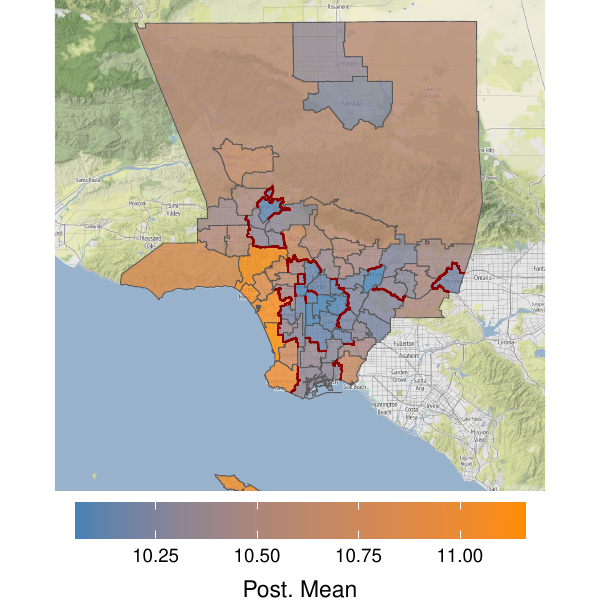}}
    \hfill
    \subfigure[\textit{SPMIX}]{\includegraphics[width=0.48\textwidth]{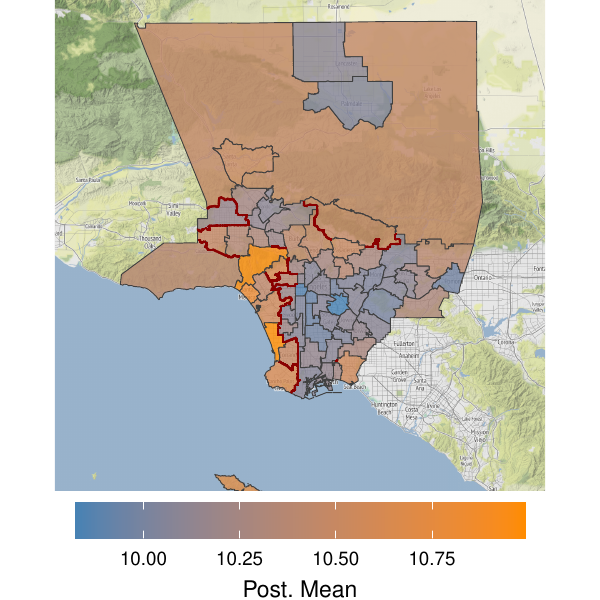}}%
    \caption{Boundary detection under the \textit{CARBayes} and \textit{SPMIX} models for the California census income dataset in LA County: posterior mean of the spatial random effect in each PUMA as an output of S.CARdissimilarity() (a); posterior mean of the estimated density in each PUMA by \textit{SPMIX} (b). Estimated boundary edges are highlighted in red.}
    \label{fig:SPMIXvsCARBayes-boundaries}
\end{figure}

For this reason, here we shed light on the differences between the boundary detection we achieve by \textit{SPMIX} and \textit{CARBayes}, respectively. We then consider the MCMC chain obtained by \textit{SPMIX} (when the data are only those from the LA County)
and the MCMC chain of \textit{CARBayes} that uses as dissimilarity metric the difference in the percentage of the population without health insurance (case $(b)$). \Cref{fig:SPMIXvsCARBayes-boundaries} displays the detected boundaries according to \textit{CARBayes} (panel (a)) and \textit{SPMIX} (panel (b)) on the map. In the LA County, \textit{SPMIX} identifies $31$ boundaries, while \textit{CARBayes} $47$; $14$ boundaries are identified by both. The Jaccard similarity index between the boundaries estimated by the two models is $14 / (31 + 47 - 14) \simeq 0.219$, indicating that the boundaries detected by the two models differ.

\begin{figure}[t]
    \centering
    \includegraphics[width=0.75\textwidth]{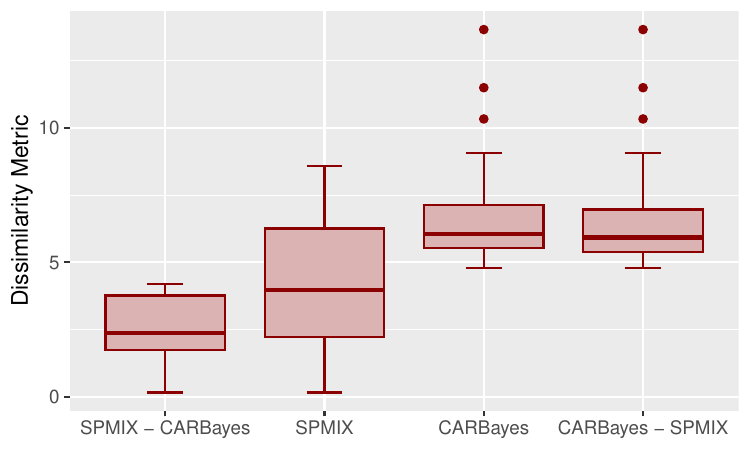}
    \caption{Boxplots of the values of the dissimilarity metric based on percentage of population without health insurance over the sets $\BEhat_{\mbox{\tiny{SPMIX - CARBayes}}}$, $\BEhat_{\mbox{\tiny{SPMIX}}}$, $\BEhat_{\mbox{\tiny{CARBayes}}}$ and $\BEhat_{\mbox{\tiny{CARBayes - SPMIX}}}$ for the California census income dataset in LA County.}
    \label{fig:SPMIXvsCARBayes-DissMetricBoxplots}
\end{figure}

Similarly as in the comparison with \textit{naive MCAR}, we define $\BEhat_{\mbox{\tiny{SPMIX}}}$ and $\NEhat_{\mbox{\tiny{SPMIX}}}$ as the sets of boundary and neighbouring edges detected by \textit{SPMIX}, and  $\BEhat_{\mbox{\tiny{CARBayes}}}$ and $\NEhat_{\mbox{\tiny{CARBayes}}}$ are as the sets of boundary and neighbouring edges detected by \textit{CARBayes}. Since the boundary detection via the \textit{CARBayes} model is driven by dissimilarity metrics, we display in \Cref{fig:SPMIXvsCARBayes-DissMetricBoxplots} the boxplots of the values of the dissimilarity metrics over the set of boundary edges detected by \textit{SPMIX} and \textit{CARBayes}. Moreover, we report the boxplot of the values of the dissimilarity metric over the subset of boundary edges that are detected by \textit{CARBayes} but not by \textit{SPMIX} (we denote such a subset as $\BEhat_{\mbox{\tiny{CARBayes - SPMIX}}}$) and vice versa ($\BEhat_{\mbox{\tiny{SPMIX - CARBayes}}}$).

The dissimilarity metrics over the sets $\BEhat_{\mbox{\tiny{CARBayes}}}$ and $\BEhat_{\mbox{\tiny{CARBayes - SPMIX}}}$ assume similar ranges of values; see \Cref{fig:SPMIXvsCARBayes-DissMetricBoxplots}. This suggests that the boundaries identified exclusively by \textit{CARBayes} are primarily influenced by the dissimilarity metric, which is coherent with the definition of a boundary as proposed by \cite{lee.mitchell2012}. In contrast, the boundary detection by \textit{SPMIX} is not influenced by the dissimilarity metric, since our approach is not driven by covariates. As a consequence, the associated boxplot includes smaller values of the dissimilarity metric. The distinction becomes even clearer when focusing on the boundaries detected by \textit{SPMIX} but not by \textit{CARBayes} (left-most boxplot in \Cref{fig:SPMIXvsCARBayes-DissMetricBoxplots}), as this boxplot and those corresponding to $\BEhat_{\mbox{\tiny{CARBayes}}}$ and $\BEhat_{\mbox{\tiny{CARBayes - SPMIX}}}$ span different ranges of values.

\begin{figure}[!ht]
    \centering
    \includegraphics[width=\textwidth]{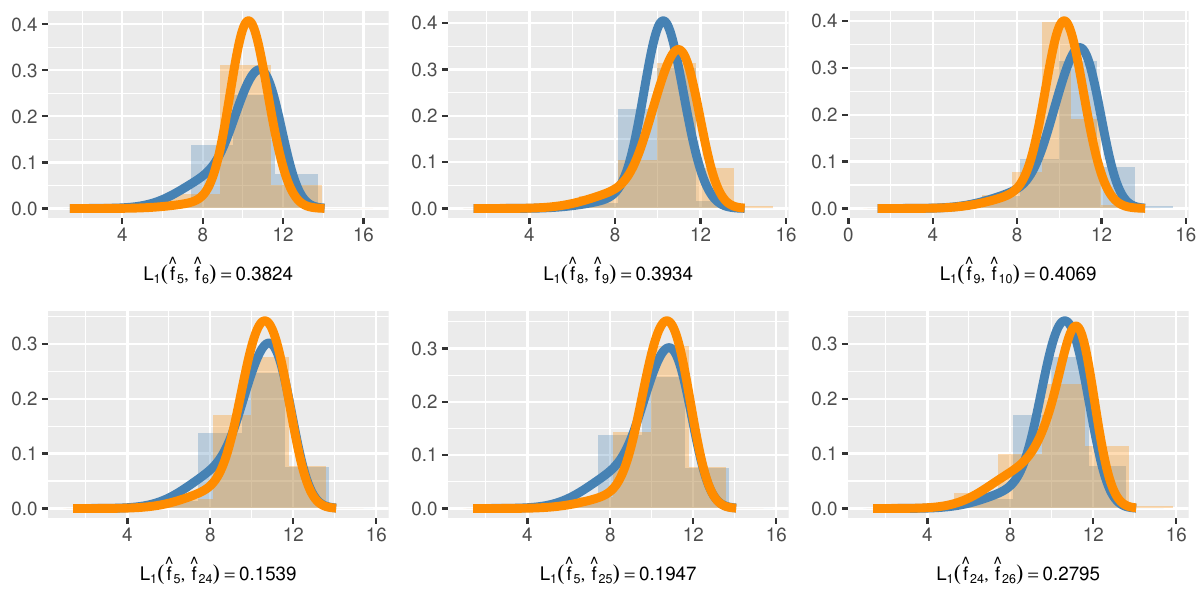}
    \caption{Empirical histograms and associated posterior estimated densities for couples of boundary areas detected by \textit{SPMIX} but not by \textit{CARBayes} for the California census income dataset. The value of the $\Lone$ distance is reported at the bottom of each panel. Histograms and estimated densities of the couples of areas are depicted in blue and orange.}
    \label{fig:SPMIXvsCARBayes-DensityComparison}
\end{figure}

We then focus on the boundaries detected by \textit{SPMIX} and not by \textit{CARBayes} to show that, despite the values of the associated dissimilarity metric are small, the log-income distribution in those areas is different and explains why \textit{SPMIX} detects a boundary. In \Cref{fig:SPMIXvsCARBayes-DensityComparison}, we compare empirical density histograms and posterior estimated densities for several couples of boundary areas detected by \textit{SPMIX} but not by the \textit{CARBayes}; each panel also reports the $\Lone$ distance between estimated densities. All the couples of estimated densities are different (in the location, in the skewness or in the tails). Such differences cannot be captured through the empirical median in each area and its outcome is not influenced by the metric adopted. If compared to \textit{CARBayes}, \textit{SPMIX} detects boundaries without summarising the distribution of the data in each area without being influenced by the dissimilarity metric.

\paragraph{\textit{SPMIX} vs. \textit{SPMIX}: sensitivity w.r.t. the global spatial parameter}
\label{subsec:rho_sens}
In this paragraph, we provide sensitivity analysis with respect to the global spatial parameter \scalebox{0.94}{$\rho\in\{0, 0.5, 0.9, 0.95, 0.99\}$}; see \eqref{eqn:logMCAR_def} in the manuscript. When $\rho=0$,  we assume no spatial dependence in our model, while $\rho = 0.5$ corresponds to a moderate spatial dependence. The other three values are typically used in boundary detection problems to encourage global spatial association and foster the spatial process to be learned locally via $G$ (see \Cref{sec:model} of the manuscript).

\begin{figure}[t]
    \centering
    \subfigure[$\rho = 0.00$]{\includegraphics[width=0.4\textwidth]{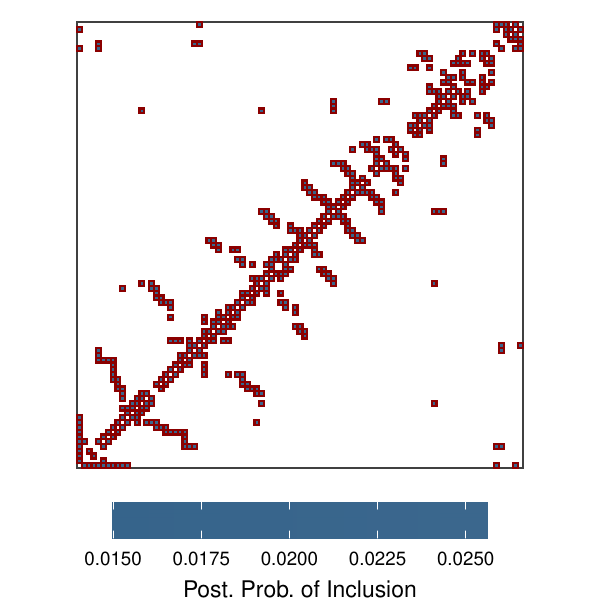}}%
    \subfigure[$\rho = 0.50$]{\includegraphics[width=0.4\textwidth]{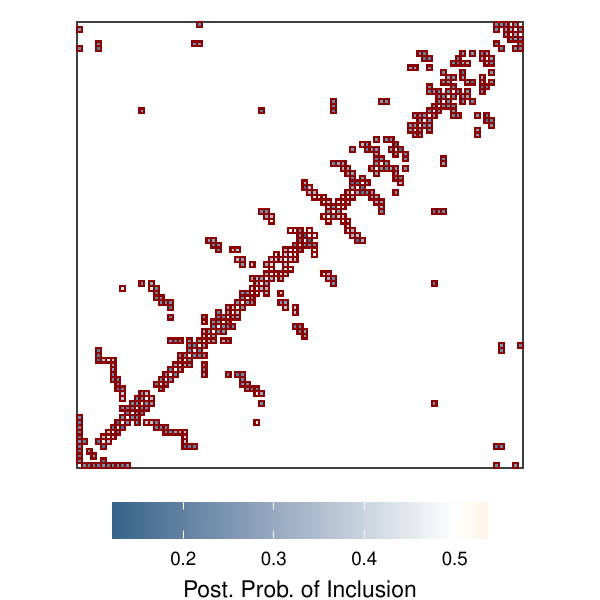}}\\
    \subfigure[$\rho = 0.90$]{\includegraphics[width=0.4\textwidth]{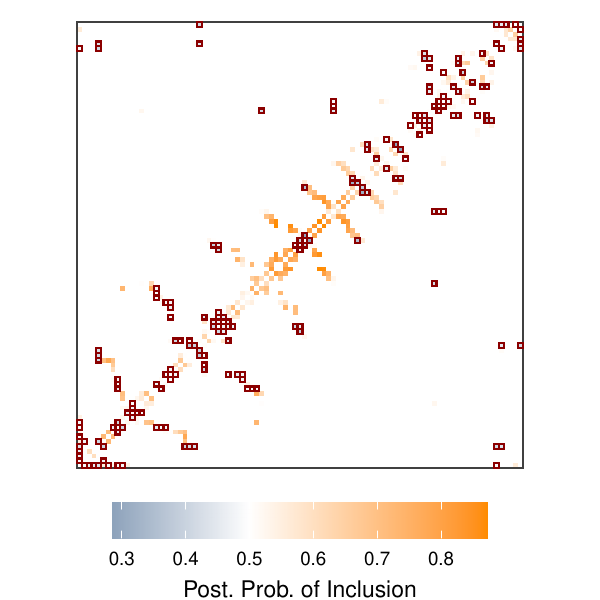}}%
    \subfigure[$\rho = 0.99$]{\includegraphics[width=0.4\textwidth]{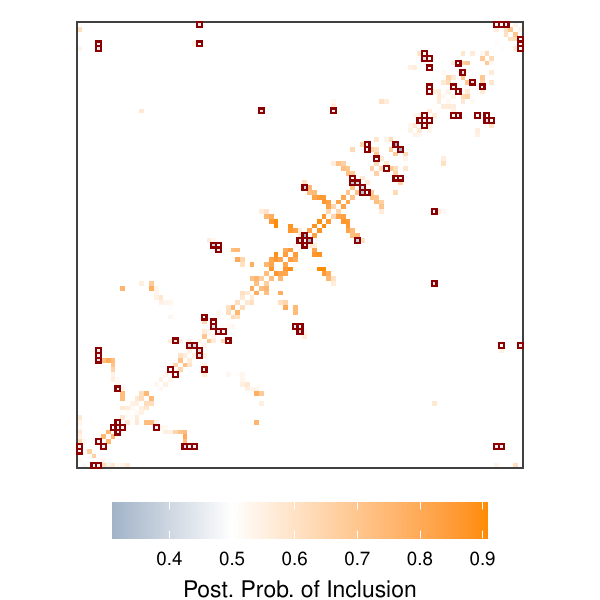}}
    \caption{Posterior probabilities of edge inclusion $\mathbb{P}(G_{i,k} = 1 \mid \bm{y})$ for different values of $\rho$ for the California census income dataset in \Cref{sec:real_case_scenario} of the manuscript. The corresponding estimated boundary graph $\hat{G}^{\text{b}}$ with threshold $\gamma=0.5$ is highlighted in red.}
    \label{fig:CaliCensusData_rho_sens}
\end{figure}

\Cref{fig:CaliCensusData_rho_sens} shows posterior probabilities of edge inclusion $\mathbb{P}(G_{i,k} = 1 \mid \bm{y})$ for each value of $\rho$. In the case $\rho=\{0, 0.5\}$, the estimated boundary graph is meaningless since all entries in the posterior probability matrix are close to $0$, i.e. the model estimates a boundary between any couple of geographically contiguous areas. In the other three cases, when we specify a high value for $\rho$, we see that the number of boundaries detected decreases as $\rho$ increases.

For completeness, we also compare the averages (over all the areal units) of $\Lone$ distances between the posterior estimated densities for $\rho=\{0, 0.5, 0.9, 0.99\}$ and the estimated densities for $\rho = 0.95$, that is the case we have discussed in great detail in \Cref{sec:real_case_scenario} in the manuscript; see \Cref{tab:CaliCensusData_meanL1}. When $\rho$ is greater or equal to 0.9,  these distances are below $0.006$, while we get a six times larger value in the case of null or moderate values of $\rho$. Finally, we also report in \Cref{tab:CaliCensusData_WAIC} the WAIC obtained for each value of $\rho$ in the deviance scale. The smallest value is obtained in case $\rho = 0.95$, the value we used to perform posterior inference in \Cref{sec:real_case_scenario} in the manuscript.

\begin{table}[t]
    \centering\renewcommand{\arraystretch}{1.25}
    \begin{tabular}{ccccc}
      \toprule[2pt]
      $\rho = 0.00$ & $\rho = 0.50$ & $\rho = 0.90$ & $\rho = 0.95$ & $\rho = 0.99$ \\ \midrule[2pt]
      $0.039$ & $0.028$ & $0.008$ & $0.000$ & $0.007$ \\ \bottomrule[2pt]
    \end{tabular}\vspace{2mm}
    \caption{Estimated mean $\Lone$ distance over all areas in the California census income dataset discussed in \Cref{sec:real_case_scenario} of the manuscript for different values of $\rho$. The $\Lone$ distances are computed between the estimated densities for the values of $\rho$ in the table and the estimated densities for the case $\rho = 0.95$, with all other hyperparameters left unchanged.}
    \label{tab:CaliCensusData_meanL1}
\end{table}

\begin{table}[t]
    \centering
    \renewcommand{\arraystretch}{1.25}
    \begin{tabular}{ccccc}
      \toprule[2pt]
      $\rho = 0.00$ & $\rho = 0.50$ & $\rho = 0.90$ & $\rho = 0.95$ & $\rho = 0.99$ \\ \midrule[2pt]
      $28155.66$ & $28114.31$ & $28030.91$ & $28015.95$ & $28042.50$ \\ \bottomrule[2pt]
    \end{tabular}\vspace{2mm}
    \caption{Estimated WAIC index in the California census income dataset discussed in \Cref{sec:real_case_scenario} of the manuscript for different values of $\rho$. WAIC index is in deviance scale, hence a smaller value means a better fit.}
    \label{tab:CaliCensusData_WAIC}
\end{table}

\section{Understanding the estimated boundary of the California census income dataset}
\label{sec:interpret_boundaries}
 
As mentioned since the Introduction, our model analyses i.i.d. individual log-income data from areal units, with the aim of detecting boundaries among the units themselves. In our framework, a \textit{boundary} is detected
if two geographically contiguous areal units show negligible spatial dependence between the corresponding income densities, thus producing borders when the annual income distributions are estimated as different. Intuitively, the estimated boundary underlines where the income gap, as represented by a random density, is more marked.  We do not include extra information in the model, such as dissimilarity metrics based on area-specific covariates.

However, if we focus only on the LA county PUMAs, extra information for each area can be collected. In particular, we consider the total number of crimes and the percentage of the population without health insurance. From the literature, it is known that lower income and social inequity are related to crimes \citep{hipp2007income} and that income and healthcare are strongly associated \citep{braveman2010socioeconomic}. Monthly crime statistics in the LA county are openly available on the official Los Angeles County Sheriff's Department site. We consider all the crimes recorded during 2020 (the same year as the income data). The proportion of the population without health insurance in LA is available on the Los Angeles County open data portal \citep{lac.opendata-portal}. In this section, we investigate whether areas separated by a boundary in our model show relevant differences in these two additional variables. We consider the estimated boundaries detected by our model between PUMAs of the LA county.

\begin{figure}[t]
    \centering
    \subfigure{\includegraphics[width=0.5\textwidth]{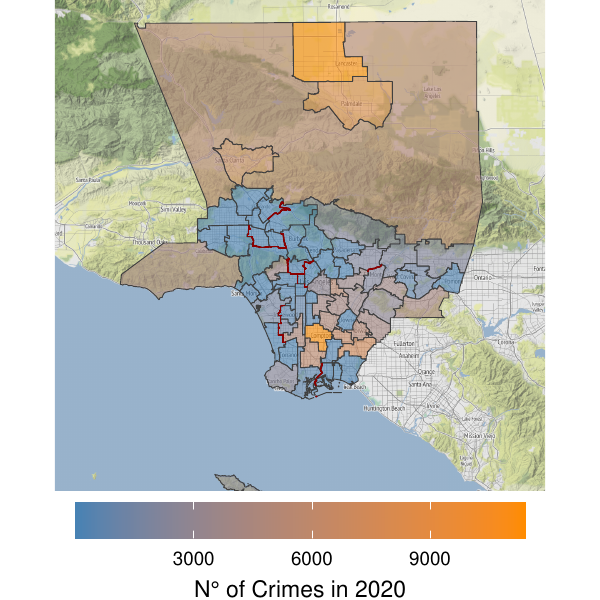}}%
    \subfigure{\includegraphics[width=0.5\textwidth]{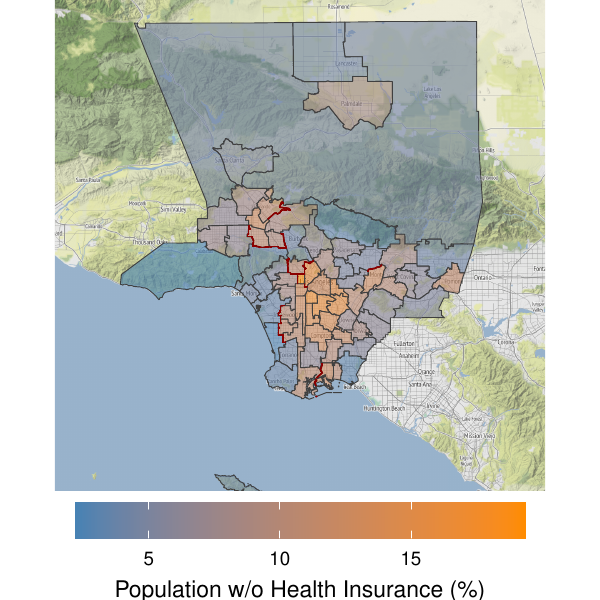}}%
    \caption{Number of all crimes recorded in 2020 in LA County per PUMA (left); percentage of population without health insurance in LA County per PUMA (right). Estimated boundary edges according to \textit{SPMIX} are highlighted in red.}
    \label{fig:CaliCensusData_covariates}
\end{figure}

\Cref{fig:CaliCensusData_covariates} (left panel) shows the heatmap of the total number of crimes grouped by PUMA, with the boundaries detected by our model in red.  There is no relevant difference in this variable's values across PUMAs estimated as boundary areas, suggesting that the number of crimes might not be associated with economic inequality, unlike expected; see \cite{hipp2007income} and references therein. \Cref{fig:CaliCensusData_covariates} (right panel) shows the percentage of people without health insurance for each PUMA in LA County, along with the boundaries detected by our model. From the figure, we see that the areas with a higher percentage of citizens without medical insurance are located in the centre and south of LA, as expected. Moreover, the estimated boundaries separate areas where this variable is different. Of course, this is expected since access to health insurance is highly correlated to a high income. Summing up, we found that the estimated boundary areas can be \textit{explained} in terms of the percentage of population without health insurance. However, the total number of crimes does not seem to \textit{explain} the estimated boundary areas we found.

\section{California census income dataset - additional plots and tables}\label{sec:extra_plots_and_tables}

\begin{figure}[t]
    \centering
    \includegraphics[width=0.5\textwidth]{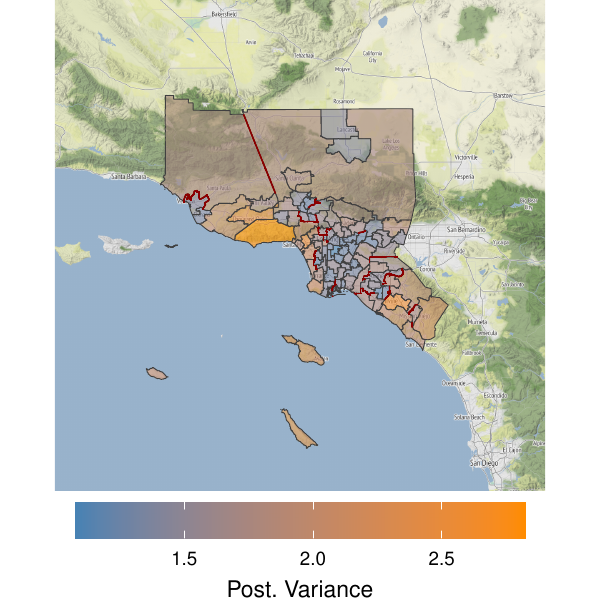}
    \caption{Posterior variances of the estimated densities on the map with estimated boundaries in red for the California census income dataset in \Cref{sec:real_case_scenario} of the manuscript.}
    \label{fig:CaliCensusData_BDresults-supp}
\end{figure}

\begin{figure}[t]
    \centering
    \subfigure{\includegraphics[width=.31\textwidth]{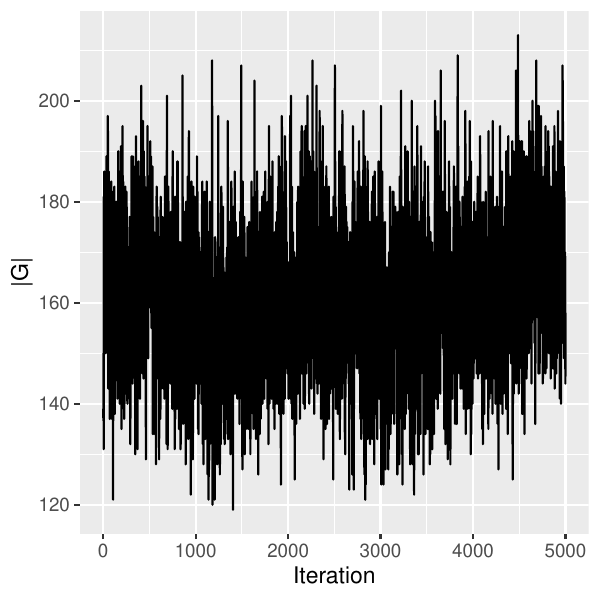}}
    \hfill
    \subfigure{\includegraphics[width=.31\textwidth]{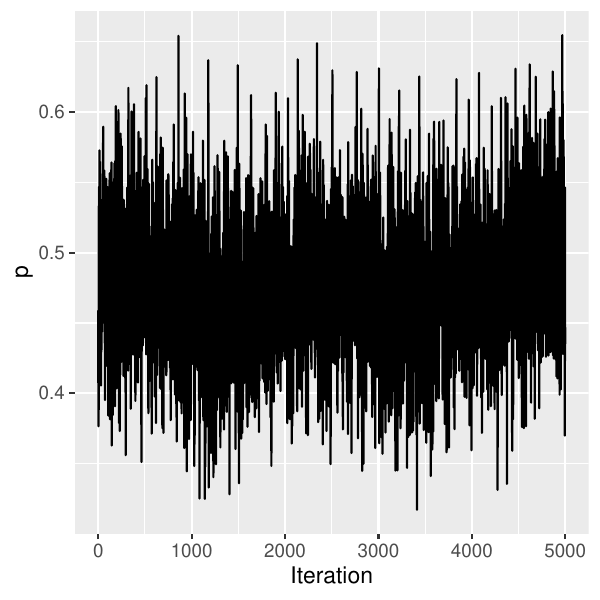}}
    \hfill
    \subfigure{\includegraphics[width=.31\textwidth]{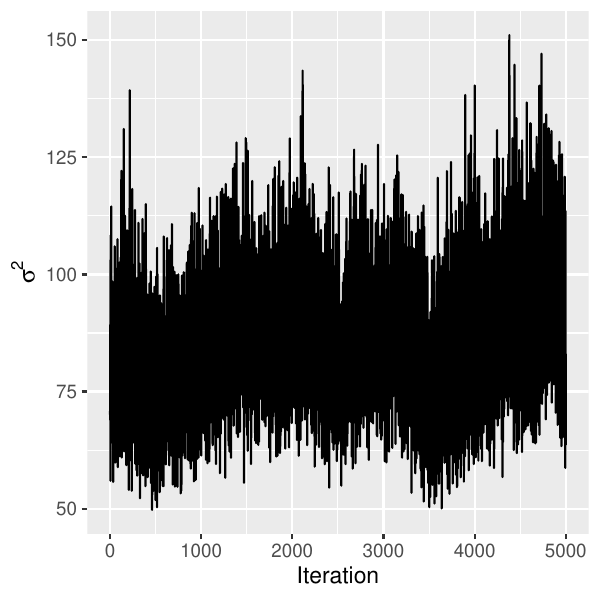}}
    \caption{MCMC diagnostics (post burn-in traceplots) for the California census income dataset in \Cref{sec:real_case_scenario} of the manuscript: $\text{ESS}(\lvert G \rvert) = 436.9$, $\text{ESS}(p) = 1458.8$, $\text{ESS}(\sigma^2) = 180.7$.}
    \label{fig:CaliCensusData_traceplot_diagnostics}
\end{figure}



\end{document}